\documentclass[11pt,a4paper,oneside]{article}
\usepackage[english]{babel}
\usepackage[utf8]{inputenc}
\usepackage[T1]{fontenc}
\ifdefined\pdfgentounicode
  \input{glyphtounicode}
\fi
\usepackage{lmodern,microtype}
\usepackage{xcolor}
\usepackage{amsmath,amssymb,amsthm,mathtools}
\usepackage{graphicx,booktabs,array,tabularx,longtable}
\usepackage{caption}
\newcolumntype{Y}{>{\raggedright\arraybackslash}X}
\usepackage[left=25mm,right=25mm,top=25mm,bottom=25mm]{geometry}
\usepackage[sort]{natbib}
\usepackage{xurl}
\usepackage{hyperref}
\hypersetup{
  colorlinks=true,
  citecolor=blue,
  urlcolor=blue,
  linkcolor=blue,
  pdftitle={On the closed-loop Volterra method for analyzing and forecasting time series},
  bookmarksnumbered=true,
  pdfdisplaydoctitle=true
}
\usepackage{orcidlink}
\usepackage{needspace,float}
\makeatletter
\renewcommand{\NAT@nmfmt}[1]{{\color{blue}#1}}
\makeatother
\theoremstyle{plain}
\newtheorem{theorem}{Theorem}[section]
\newtheorem{lemma}[theorem]{Lemma}
\newtheorem{proposition}[theorem]{Proposition}
\newtheorem{corollary}[theorem]{Corollary}
\newtheorem*{riskratecorollary}{Corollary}
\theoremstyle{definition}

\newtheorem{assumption}[theorem]{Assumption}

\theoremstyle{remark}
\newtheorem{remark}[theorem]{Remark}
\newcommand{\E}{\mathbb E}
\newcommand{\Pbb}{\mathbb P}
\newcommand{\Rbb}{\mathbb R}

\newcommand{\cH}{\mathcal H}
\newcommand{\cF}{\mathcal F}

\newcommand{\argmin}{\operatorname*{arg\,min}}

\numberwithin{equation}{section}
\title{On the closed-loop Volterra method for analyzing and forecasting time series}
\author{Nabaneet Das\orcidlink{0000-0002-2605-626X},
  Maryam Movahedifar\orcidlink{0000-0003-0283-7560}
 and Thorsten Dickhaus\orcidlink{0000-0003-3084-3036}\thanks{Correspondence to: Thorsten Dickhaus, Institute for Statistics, University of Bremen, Germany. Email: dickhaus@uni-bremen.de}}
\date{
	\small{\textit{Institute for Statistics, University of Bremen, Germany}}
}
\begin{document}
\maketitle
\begin{abstract}
This paper studies the closed-loop Volterra method for analysing and predicting stationary time series. The method represents the dependence of a current observation on past observations through polynomial terms, including powers of individual lags and interactions between different lags. Since the number of coefficients increases rapidly with the memory length and polynomial degree, we use the theory of reproducing kernel Hilbert spaces to estimate these coefficients by regularised least squares through kernel evaluations. We establish conditions on marginal moments, temporal dependence, model complexity and regularisation under which the prediction risk converges to the minimum mean squared prediction error as the sample size increases. We also consider data-driven selection of tuning parameters by a rolling-origin procedure based on one-step-ahead prediction errors, and show that the resulting selector asymptotically attains the prediction risk of the best candidate in the considered grid. Simulation studies and a real-data application examine the finite-sample performance of the method and its comparison with linear and nonlinear prediction methods.\\
Keywords: Time series analysis; Volterra series; closed-loop method;
kernel ridge regression; prediction.\\
\textbf{MSC Classification}: 37M10, 62M10, 62M20.
\end{abstract}

\section{Introduction}\label{section.Intro}
Nonlinear systems with memory are frequently encountered in time series analysis,
where the current output depends on preceding inputs. Learning this
relationship from observations provides a basis for describing the system
and predicting its subsequent behaviour.\\[0.05 in]
Volterra's approach extends the convolution representation of linear systems
through polynomial integral operators of increasing order \citep{RN19}.
The first-order term describes the linear response, while higher-order terms
represent products of delayed inputs. This polynomial expansion with memory
has been developed in systems theory, signal processing and nonlinear
time-series analysis \citep{RN14,RN15,RN16}. \citet{RN5} established
uniform approximation of fading-memory, time-invariant operators by
finite Volterra series under suitable assumptions. More recently,
\citet{ortega2025fading} characterize fading-memory properties and convolution
representations for linear filters and derive associated kernel-based approximation
results, while \citet{gonon2025reservoir} construct a Volterra reservoir kernel
for causal, time-invariant fading-memory filters.\\[0.05 in]
Consider a discrete-time system with input $x_t$ and output $y_t$. A Volterra
representation with memory $m$ and degree $p$ approximates the output by a
polynomial in $(x_t,x_{t-1},\ldots,x_{t-m+1})^\top$. The memory determines
which past inputs enter the model, and the degree determines the highest
order of their products.\\[0.05 in]
Setting $x_t=y_{t-1}$ feeds the preceding output back as the current input;
this is a closed-loop system, whose finite-memory Volterra predictor
is a polynomial nonlinear autoregression \citep{RN4}. It takes the form
\begin{equation}\label{V_m,p}
\widehat y_t=\sum_{\substack{\boldsymbol{\alpha}\in\mathbb N_0^m\\
|\boldsymbol{\alpha}|\leq p}}a_{\boldsymbol{\alpha}}
\prod_{j=1}^m y_{t-j}^{\alpha_j},\qquad t=m+1,\ldots,T.
\end{equation}
Here $|\boldsymbol{\alpha}|=\sum_{j=1}^m\alpha_j$. These monomials form a basis for the space of polynomials in $m$ variables
of total degree at most $p$, whose dimension is $D_{m,p}=\binom{m+p}{p}$.
The basis includes the constant, linear effects, powers of individual lags
and interactions between lags.\\[0.05 in]
The dimension grows rapidly when both memory and degree increase:
$D_{10,10}=184756$. Kernel methods address the computational cost of this
representation by evaluating polynomial inner products without constructing
all monomials. \citet{RN7} apply reproducing-kernel methods to Volterra series estimation,
and \citet{Franz} relate regularised Volterra and Wiener representations to
polynomial kernel regression. \citet{dallalibera2021volterra} develop
multiplicative polynomial kernels for regularised Volterra identification.
More recent work includes kernel-based regularised identification of nonlinear
dynamical systems through polynomials of Hilbert--Schmidt operators
\citep{cacace2024} and kernel design for regularised Volterra identification
of polynomial Wiener--Hammerstein systems \citep{xu2025kernel}.\\[0.05 in]
Choosing the memory, degree and regularisation remains a statistical problem.
Adaptive selection of model complexity for dependent time series
is studied by \citet{Meir2000}; related approaches to regularised regression
and forecast selection are developed by
\citet{farahmand2012regularized,AlquierWintenberger2012,mcdonald2017risk}.
Rolling-origin evaluation assesses forecasts in their temporal order
\citep{Dawid1984,Tashman2000}. Online cross-validation and rolling validation
provide related procedures for model or tuning-parameter selection with
sequential data \citep{Benkeser2018,ZhangLei2025}, while
\citet{biau2010sequential} study nonparametric sequential prediction of
time series. More generally, \citet{steinwartHushScovel2009} establish
consistency results for kernel learning under dependent observations,
\citet{steinwartAnghel2009} study forecasting of ergodic dynamical systems,
and \citet{hangSteinwart2014} derive learning bounds under mixing.\\[0.05 in]
For closed-loop Volterra prediction, important questions are whether
increasing the memory and polynomial degree improves approximation of the
prediction target, and whether the resulting predictors can be estimated
consistently under temporal dependence. We establish polynomial approximation
under a marginal Carleman condition, finite-sample risk bounds, and risk
consistency for growing memory and degree. We also establish oracle
comparisons for rolling-origin selection over fixed and growing grids.
To the best of our knowledge, this is the first rigorous statistical
analysis of these properties for the closed-loop Volterra method.\\[0.05 in]
The rest of the manuscript is organized as follows.
Section~\ref{volterra} introduces the Volterra representation and its
discrete closed-loop form. Section~\ref{RKHS} develops the RKHS formulation
and regularised estimation. Section~\ref{sec:theoretical-results} establishes
the risk and tuning-parameter selection results. Section~\ref{sim} presents
simulated and real-data applications, and Section~\ref{conclusion} concludes.

\subsection{Notation and conventions}
The notation and conventions used throughout the manuscript are stated as follows:
Scalars are written in nonbold type, vectors in bold italic type, and
matrices in bold upright capitals, as in $Y_t$, $\boldsymbol{X}$ and
$\mathbf{K}$, respectively. For scalars and vectors, uppercase data symbols
denote random quantities and lowercase symbols their observed values, as in
$Y_t,y_t$ and $\boldsymbol{X},\boldsymbol{x}$. Vector and matrix entries are
nonbold. Parameters, functions, dimensions and population moments retain
their stated notation; a hat denotes an estimator or its realised estimate
according to the context. The superscript $\top$ denotes transpose, and
$\mathbf{I}_d$ is the $d\times d$ identity matrix; its subscript is omitted
when the dimension is clear. For real symmetric matrices of the same size,
$\mathbf{A}\preceq\mathbf{B}$ means that $\mathbf{B}-\mathbf{A}$ is
positive semidefinite.\\[0.05 in]
For a real-valued random variable $Z$ and $1\leq q<\infty$, we write
$\|Z\|_q=(\E|Z|^q)^{1/q}$ for its $L^q$ norm. For a vector
$\boldsymbol{u}\in\Rbb^d$, $\|\boldsymbol{u}\|_2$ denotes the Euclidean norm.
For matrices of the same size, we write
$\langle\mathbf A,\mathbf B\rangle_{\mathrm F}
=\operatorname{tr}(\mathbf A^\top\mathbf B)$ and
$\|\mathbf A\|_{\mathrm F}
=\langle\mathbf A,\mathbf A\rangle_{\mathrm F}^{1/2}$
for the Frobenius inner product and norm, respectively
\citep[Section~5.7]{hornjohnson2012matrix}. For a real matrix
$\mathbf A\in\Rbb^{r\times s}$, its operator norm induced by the Euclidean norm is
\begin{equation}\label{eq:operator-norm}
\|\mathbf A\|_{\mathrm{op}}
=\sup_{\substack{\boldsymbol u\in\Rbb^s\\\|\boldsymbol u\|_2=1}}
\|\mathbf A\boldsymbol u\|_2.
\end{equation}
For a symmetric matrix, the operator norm equals its largest absolute
eigenvalue.

\section{Volterra theory of stationary nonlinear systems}\label{volterra}

Consider a discrete-time closed-loop model in which the current output $Y_t$
depends on the $m$ preceding outputs and an additive noise term:
\begin{equation}\label{io}
Y_t=f(Y_{t-1},\ldots,Y_{t-m})+E_t.
\end{equation}
Here $f$ is a system operator that maps the delayed-input vector to the
corresponding output, and $E_t$ represents an additive noise term. Equation~\eqref{io}
is an illustrative finite-memory model; the theory of
Section~\ref{sec:theoretical-results} does not require such a model.
The input--output rule is time invariant when the same function
$f$ is used at every time point. This property of the system operator is
distinct from strict stationarity of the stochastic process $\{Y_t\}$,
which is imposed in Assumption~\ref{ass:stationarity and causality} of
Section~\ref{sec:theoretical-results}.\\[0.05 in]
Following the definitions in \citet{RN5}, an input--output operator is
causal if the output at a given time depends only on present and past inputs,
and time invariant if a time shift of the input produces the same time shift
of the output. For a linear causal, time-invariant operator, the input--output
relation is represented by a convolution. For a continuous-time input $u(s)$
and output $v(s)$, this takes the form
\begin{equation}\label{eq:continuous-linear-response}
v(s)=H_1[u](s)=\int_0^\infty h^{(1)}(\tau)u(s-\tau)\,d\tau,
\end{equation}
where $h^{(1)}$ is the linear kernel or impulse response. An extension of
this convolution expression is the Volterra series expansion
\begin{equation}\label{eq:continuous-volterra-expansion}
v(s)=V[u](s)=H_0[u](s)+H_1[u](s)+H_2[u](s)+\cdots,
\end{equation}
whenever the integrals and series converge. The zero-order term
$H_0[u](s)=h^{(0)}$ is the constant response. The $n$th-order Volterra
operator is defined by
\begin{equation}\label{eq:continuous-volterra-component}
H_n[u](s)=\int_0^\infty\!\cdots\!\int_0^\infty
h^{(n)}(\tau_1,\ldots,\tau_n)
\prod_{j=1}^n u(s-\tau_j)\,d\tau_1\cdots d\tau_n.
\end{equation}
The integral kernels $h^{(n)}$ are the Volterra kernels. Depending on the
system, the integration domain may be finite or infinite. The memory of
the system determines their support, specifying the interval over which
past inputs can affect the current output. Thus a Volterra series extends
the linear convolution representation by adding higher-order integral
operators whose kernels act jointly on several delayed input values
\citep{RN14,RN16,bib.2}.

\subsection{Discrete Volterra system}
In applied signal processing, a discrete form of the Volterra system is
used for a finite sample of observations. Consider the nonlinear,
discrete-time, time-invariant relationship in \eqref{io}.
The observed input vector is
$\boldsymbol{x}_t=(y_{t-1},\ldots,y_{t-m})^\top$. The positive-degree
components of its Volterra polynomial are
\begin{equation}\label{hn}
H_n(\boldsymbol{x})=\sum_{i_1=1}^{m}\cdots\sum_{i_n=1}^{m}
h^{(n)}_{i_1,\ldots,i_n}x_{i_1}\cdots x_{i_n},\qquad 1\leq n\leq p.
\end{equation}
Define the degree-zero component by $H_0(\boldsymbol{x})=h^{(0)}$.
The truncated expansion is
\begin{equation}\label{fullvolterra}
v_{m,p}(\boldsymbol{x})=\sum_{n=0}^{p}H_n(\boldsymbol{x})
=h^{(0)}+\sum_{n=1}^{p}H_n(\boldsymbol{x}).
\end{equation}
For $p=0$, the expansion consists only of $H_0$.
Each ordered coefficient tensor may be symmetrised without
changing the function, because multiplication is invariant under
permutations of the indices \citep{RN1,RN14}.\\[0.05 in]
There are $m^n$ ordered tuples at degree $n$, but only
$\binom{m+n-1}{n}$ distinct monomials. Including the intercept, the
polynomial space therefore has dimension
\begin{equation}\label{mdim}
D_{m,p}=1+\sum_{n=1}^{p}\binom{m+n-1}{n}
=\binom{m+p}{p}=\binom{m+p}{m}.
\end{equation}
For fixed $m$ this grows polynomially in $p$, and for fixed $p$ it grows
polynomially in $m$; simultaneous growth can nevertheless be rapid.
For example, $D_{10,10}=184756$. The kernel representation in the next
section permits regularised estimation of these coefficients through inner
products of the lag vectors.

\section{Estimating closed-loop Volterra series using linear regression in RKHS}
\label{RKHS}
Direct estimation of a closed-loop Volterra series becomes increasingly
challenging as the memory length ($m$) and the degree of nonlinearity ($p$)
increase,
since the number of coefficients in the higher-order Volterra kernels
grows rapidly. We therefore recast the closed-loop Volterra series in a
reproducing kernel Hilbert space (RKHS). For a lag vector of length $m$,
evaluating the degree-$p$ polynomial kernel takes $O(m+p)$ operations.
Regularised estimation then uses an $n_T\times n_T$ kernel matrix rather
than explicitly constructing $D_{m,p}$ polynomial features.

\Needspace{9\baselineskip}
\subsection{Problem formulation}
Suppose that a real-valued time series
$\boldsymbol{Y}_{1:T}=(Y_1,\ldots,Y_T)^\top$ of length $T$ is observed,
with realised values $\boldsymbol{y}_{1:T}=(y_1,\ldots,y_T)^\top$. At
time $t$, prediction is based on the available history
$(Y_{t-1},\ldots,Y_1)$. Accordingly, we consider predictors
$f(Y_{t-1},\ldots,Y_1)$ that are
$\sigma(Y_{t-1},\ldots,Y_1)$-measurable and minimise the squared-error risk
\begin{equation}\label{eq:finite-history-risk}
R_t^{\mathrm{hist}}(f)=\E\left[\{Y_t-f(Y_{t-1},\ldots,Y_1)\}^2\right].
\end{equation}
Assuming $\E(Y_t^2)<\infty$, the conditional mean
\begin{equation}\label{eq:finite-history-target}
f_t^\star=\E[Y_t\mid Y_{t-1},\ldots,Y_1]
\end{equation}
minimises $R_t^{\mathrm{hist}}(f)$ over all square-integrable
$\sigma(Y_{t-1},\ldots,Y_1)$-measurable predictors. This is the
finite-history target. Under stationarity, its optimal risk equals the
risk using $t-1$ lags and converges, as $t\to\infty$, to the
infinite-past optimal risk $R(f^\star)$ defined in \eqref{eq:prediction-targets}
by Lemma~\ref{lem:finite-past-l2}.
The objective is to identify conditions under which the prediction problem in \eqref{eq:finite-history-risk} can be
approximated by a finite-memory Volterra series and estimated by regularised
regression using the polynomial Volterra kernel defined in \eqref{e2}.
For a fixed memory length $m\geq1$, define the lag vector
\begin{equation}\label{eq:lag-vector}
\boldsymbol{X}_t^{(m)}=(Y_{t-1},\ldots,Y_{t-m})^\top\in\Rbb^m.
\end{equation}
Its observed value is
$\boldsymbol{x}_t^{(m)}=(y_{t-1},\ldots,y_{t-m})^\top$.
By \eqref{V_m,p}, a finite-memory Volterra predictor with memory $m$ and
degree of nonlinearity at most $p$ can be written as
\begin{equation}\label{eq:finite-volterra-polynomial}
v_{m,p}(\boldsymbol{x})=
\sum_{|\boldsymbol{\alpha}|\leq p}a_{\boldsymbol{\alpha}}
\boldsymbol{x}^{\boldsymbol{\alpha}},\qquad
\boldsymbol{x}=(x_1,\ldots,x_m)^\top,
\end{equation}
where $\boldsymbol{\alpha}=(\alpha_1,\ldots,\alpha_m)^\top\in\mathbb N_0^m$,
$|\boldsymbol{\alpha}|=\sum_{j=1}^m\alpha_j$, and
$\boldsymbol{x}^{\boldsymbol{\alpha}}=\prod_{j=1}^m x_j^{\alpha_j}$.

\subsection{Volterra feature maps and kernels}
For $n\geq1$, let $\boldsymbol{\phi}_n(\boldsymbol{x})\in\Rbb^{m^n}$ contain
all ordered products $x_{i_1}\cdots x_{i_n}$, and put
$\boldsymbol{\phi}_0(\boldsymbol{x})=(1)$. The $n$th-degree component satisfies
\begin{equation}\label{eq:nth-order-volterra-feature}
H_n(\boldsymbol{x})=\boldsymbol{\eta}_n^\top\boldsymbol{\phi}_n(\boldsymbol{x}),
\qquad
\boldsymbol{\phi}_n(\boldsymbol{x})^\top\boldsymbol{\phi}_n(\boldsymbol{z})
=(\boldsymbol{x}^\top\boldsymbol{z})^n.
\end{equation}
Here $\boldsymbol{\eta}_n$ contains the symmetric ordered coefficients in
\eqref{hn}, and $\eta_0=h^{(0)}$. Put
$\boldsymbol{\eta}^{(p)}=(\eta_0,\boldsymbol{\eta}_1^\top,\ldots,
\boldsymbol{\eta}_p^\top)^\top$. Stacking the feature maps gives
\begin{equation}\label{eq:stacked-feature-map}
\boldsymbol{\phi}^{(p)}(\boldsymbol{x})
=(1,\boldsymbol{\phi}_1(\boldsymbol{x})^\top,\ldots,
\boldsymbol{\phi}_p(\boldsymbol{x})^\top)^\top,
\qquad
v_{m,p}(\boldsymbol{x})=\{\boldsymbol{\eta}^{(p)}\}^\top
\boldsymbol{\phi}^{(p)}(\boldsymbol{x}).
\end{equation}
The associated inner product is the original Volterra kernel
\citep{RN17,Franz}:
\begin{equation}\label{e2}
k^{(p)}(\boldsymbol{x},\boldsymbol{z})
=\boldsymbol{\phi}^{(p)}(\boldsymbol{x})^\top\boldsymbol{\phi}^{(p)}(\boldsymbol{z})
=\sum_{n=0}^{p}(\boldsymbol{x}^\top\boldsymbol{z})^n.
\end{equation}
Let $\cH_{m,p}$ denote its RKHS. Since $k^{(p)}$ is positive definite, the
Moore--Aronszajn theorem implies that this RKHS is uniquely determined by
$k^{(p)}$ and satisfies the reproducing property \citep{Aronszajn1950}:
$$
f(\boldsymbol{x})=\langle f,k^{(p)}(\,\cdot\,,\boldsymbol{x})\rangle_{\cH_{m,p}}.
$$
To specify its norm without repeated monomials, let
$\mathcal A_{m,p}=\{\boldsymbol{\alpha}\in\mathbb N_0^m:
|\boldsymbol{\alpha}|\leq p\}$ and define
\begin{equation}\label{eq:scaled-monomials}
c_{\boldsymbol{\alpha},m,p}=\frac{|\boldsymbol{\alpha}|!}{\alpha_1!\cdots\alpha_m!},
\qquad
\psi_{\boldsymbol{\alpha},m,p}(\boldsymbol{x})
=\sqrt{c_{\boldsymbol{\alpha},m,p}}\,\boldsymbol{x}^{\boldsymbol{\alpha}}.
\end{equation}
Define the feature map by
$$
\boldsymbol{\psi}_{m,p}(\boldsymbol{x})
=
\left(\psi_{\boldsymbol{\alpha},m,p}(\boldsymbol{x})\right)_{
\boldsymbol{\alpha}\in\mathcal A_{m,p}}
\in\Rbb^{D_{m,p}},
$$
where the coordinates are arranged in a fixed order. The coordinate indexed
by $\boldsymbol{\alpha}=\boldsymbol{0}$ equals
$\psi_{\boldsymbol{0},m,p}(\boldsymbol{x})=1$.

\Needspace{13\baselineskip}
\begin{lemma}[Finite feature representation]
\label{lem:feature-representation}
For every $m,p$ and every
$f_{\boldsymbol{\theta}},f_{\boldsymbol{\beta}}\in\cH_{m,p}$,
\begin{equation}\label{eq:finite-feature-representation}
\cH_{m,p}
=\{f_{\boldsymbol{\theta}}:
 f_{\boldsymbol{\theta}}(\boldsymbol{x})
 =\boldsymbol{\theta}^{\top}\boldsymbol{\psi}_{m,p}(\boldsymbol{x}),\ 
 \boldsymbol{\theta}\in\Rbb^{D_{m,p}}\},
\qquad
\langle f_{\boldsymbol{\theta}},f_{\boldsymbol{\beta}}\rangle_{\cH_{m,p}}
=\boldsymbol{\theta}^{\top}\boldsymbol{\beta}.
\end{equation}
Hence $\|f_{\boldsymbol{\theta}}\|_{\cH_{m,p}}=\|\boldsymbol{\theta}\|_2$.
In particular,
\begin{equation}
\|\boldsymbol{\psi}_{m,p}(\boldsymbol{x})\|_2^2
=k^{(p)}(\boldsymbol{x},\boldsymbol{x})
=\sum_{n=0}^{p}\|\boldsymbol{x}\|_2^{2n}.
\label{eq:feature-norm}
\end{equation}
\end{lemma}
\begin{proof}
The multinomial theorem gives
$k^{(p)}(\boldsymbol{x},\boldsymbol{z})
=\sum_{\boldsymbol{\alpha}\in\mathcal A_{m,p}}
c_{\boldsymbol{\alpha},m,p}\boldsymbol{x}^{\boldsymbol{\alpha}}
\boldsymbol{z}^{\boldsymbol{\alpha}}
=\boldsymbol{\psi}_{m,p}(\boldsymbol{x})^\top\boldsymbol{\psi}_{m,p}(\boldsymbol{z})$.
Since the functions $\psi_{\boldsymbol{\alpha},m,p}$ are linearly
independent, each function
$f_{\boldsymbol{\theta}}(\boldsymbol{x})
=\boldsymbol{\theta}^\top\boldsymbol{\psi}_{m,p}(\boldsymbol{x})$
has a unique coefficient vector $\boldsymbol{\theta}\in\Rbb^{D_{m,p}}$.
The RKHS associated with a finite feature map has the Euclidean
coefficient inner product when the features are linearly independent.
Thus the kernel identity above gives
$\langle f_{\boldsymbol{\theta}},f_{\boldsymbol{\beta}}\rangle_{\cH_{m,p}}
=\boldsymbol{\theta}^\top\boldsymbol{\beta}$ and hence
$\|f_{\boldsymbol{\theta}}\|_{\cH_{m,p}}=\|\boldsymbol{\theta}\|_2$,
which proves \eqref{eq:finite-feature-representation}. Setting
$\boldsymbol{z}=\boldsymbol{x}$ proves \eqref{eq:feature-norm}.
\end{proof}

The ordinary coefficients in \eqref{eq:finite-volterra-polynomial} satisfy
$a_{\boldsymbol{\alpha}}=\sqrt{c_{\boldsymbol{\alpha},m,p}}\theta_{\boldsymbol{\alpha}}$.
Consequently,
\begin{equation}\label{eq:volterra-coefficient-norm}
\|v_{m,p}\|_{\cH_{m,p}}^2
=\sum_{\boldsymbol{\alpha}\in\mathcal A_{m,p}}
\frac{a_{\boldsymbol{\alpha}}^2}{c_{\boldsymbol{\alpha},m,p}}.
\end{equation}
The factor $c_{\boldsymbol{\alpha},m,p}$ counts the ordered tuples producing
that monomial. Each corresponding symmetric ordered coefficient is
$a_{\boldsymbol{\alpha}}/c_{\boldsymbol{\alpha},m,p}$. Thus
\begin{equation}\label{eq:ordered-volterra-norm}
\|v_{m,p}\|_{\cH_{m,p}}^2
=\eta_0^2+\sum_{n=1}^{p}\|\boldsymbol{\eta}_n\|_2^2
=\|\boldsymbol{\eta}^{(p)}\|_2^2.
\end{equation}
To verify \eqref{eq:ordered-volterra-norm}, fix $1\leq n\leq p$.
For each $\boldsymbol{\alpha}$ with $|\boldsymbol{\alpha}|=n$, the
$c_{\boldsymbol{\alpha},m,p}$ ordered terms associated with the monomial
$\boldsymbol{x}^{\boldsymbol{\alpha}}$ all have coefficient
$a_{\boldsymbol{\alpha}}/c_{\boldsymbol{\alpha},m,p}$. Hence
$$
\|\boldsymbol{\eta}_n\|_2^2
=\sum_{|\boldsymbol{\alpha}|=n}
 c_{\boldsymbol{\alpha},m,p}
 \left(\frac{a_{\boldsymbol{\alpha}}}{c_{\boldsymbol{\alpha},m,p}}\right)^2
=\sum_{|\boldsymbol{\alpha}|=n}
 \frac{a_{\boldsymbol{\alpha}}^2}{c_{\boldsymbol{\alpha},m,p}}.
$$
Adding the squared constant coefficient and summing over
$n=1,\ldots,p$ gives \eqref{eq:ordered-volterra-norm} from
\eqref{eq:volterra-coefficient-norm}. Thus the squared RKHS norm is the
sum of squared symmetric ordered Volterra coefficients.\\[0.05 in]
A commonly used alternative is the inhomogeneous polynomial kernel
\citep{RN7},
\begin{equation}\label{K_m,p}
K_{m,p}(\boldsymbol{x},\boldsymbol{z})
=(1+\boldsymbol{x}^\top\boldsymbol{z})^p
=\sum_{n=0}^{p}\binom pn(\boldsymbol{x}^\top\boldsymbol{z})^n.
\end{equation}
It spans the same polynomial space, but scales each degree's ordered features
by the square root of its binomial weight in \eqref{K_m,p}. This changes the
RKHS norm. An infinite-degree alternative, obtained by factorial weighting, is
\begin{equation}\label{eq:infinite-volterra-kernel}
k^{(\infty)}(\boldsymbol{x},\boldsymbol{z})
=\exp(\boldsymbol{x}^\top\boldsymbol{z})
=\sum_{n=0}^{\infty}\frac{(\boldsymbol{x}^\top\boldsymbol{z})^n}{n!}.
\end{equation}
Finite and infinite discrete-time Volterra representations can therefore be
expressed through feature maps in suitable RKHSs; see \citet{Franz} for a
detailed treatment. For a representation with symmetric ordered
coefficients $\boldsymbol{\eta}_n$, membership in the infinite-degree
RKHS requires
$\sum_{n=0}^\infty n!\|\boldsymbol\eta_n\|_2^2<\infty$.
Throughout the estimation and theoretical developments,
we use \eqref{e2} to retain the coefficient penalty in
\eqref{eq:ordered-volterra-norm}.

\subsection{Regularised estimation}\label{subsec:ridge-estimation}
For integers $1\leq m_T<T$ and $p_T\geq0$, put
$\cH_T=\cH_{m_T,p_T}$, $\boldsymbol{X}_t^{(T)}=\boldsymbol{X}_t^{(m_T)}$,
and $n_T=T-m_T$. The available regression sample is
$$
\mathcal D_T=\{(\boldsymbol{X}_t^{(T)},Y_t):t=m_T+1,\ldots,T\}.
$$
The mean squared training error is
$$
\widehat R_T(f)=\frac1{n_T}\sum_{t=m_T+1}^T
\{Y_t-f(\boldsymbol{X}_t^{(T)})\}^2.
$$
We combine this error with the coefficient norm to obtain the regularised
Volterra estimator
\begin{equation}\label{eq:ridge-estimator}
\widehat f_T=\argmin_{f\in\cH_T}
\{\widehat R_T(f)+\lambda_T\|f\|_{\cH_T}^2\},
\qquad \lambda_T>0.
\end{equation}
By \eqref{eq:finite-feature-representation}, the objective in
\eqref{eq:ridge-estimator} is a strictly convex quadratic function of
$\boldsymbol{\theta}$ when $\lambda_T>0$, and therefore has a unique
minimiser.\\[0.05 in]
For a general RKHS $\mathbb F$ with kernel $k$, a loss function
$c:\Rbb\times\Rbb\to\Rbb$, and a strictly increasing penalty
$\Omega:[0,\infty)\to\Rbb$, the representer theorem states that any minimiser,
when one exists, has the form
\begin{equation}\label{eq:representer-general}
\begin{aligned}
\widehat f&\in\argmin_{f\in\mathbb F}
\left\{\sum_{i=1}^N c(y_i,f(\boldsymbol{x}_i))+\Omega(\|f\|_{\mathbb F})\right\},\\
\widehat f(\boldsymbol{x})&=\sum_{i=1}^N\gamma_i k(\boldsymbol{x},\boldsymbol{x}_i),
\qquad \gamma_i\in\Rbb,
\end{aligned}
\end{equation}
for the data $(\boldsymbol{x}_i,y_i)_{i=1}^N$
\citep[Section~11.6]{RN18}. Taking squared loss and
$\Omega(r)=n_T\lambda_T r^2$ gives \eqref{eq:ridge-estimator}, after division
of the objective by $n_T$.
Let $\boldsymbol{y}_T=(y_{m_T+1},\ldots,y_T)^\top$ be the
training-response vector. Define the Gram matrix
$\mathbf{K}_T\in\Rbb^{n_T\times n_T}$ and, for
$\boldsymbol{x}\in\Rbb^{m_T}$, the kernel-evaluation vector
$\boldsymbol{\kappa}_T(\boldsymbol{x})\in\Rbb^{n_T}$ by
$$
(\mathbf{K}_T)_{ij}
=k^{(p_T)}\!\left(\boldsymbol{x}_{m_T+i}^{(m_T)},
\boldsymbol{x}_{m_T+j}^{(m_T)}\right),\qquad
[\boldsymbol{\kappa}_T(\boldsymbol{x})]_i
=k^{(p_T)}\!\left(\boldsymbol{x}_{m_T+i}^{(m_T)},\boldsymbol{x}\right),
$$
for $1\leq i,j\leq n_T$.
Thus $\mathbf{K}_T$ contains the pairwise kernel evaluations among the
training lag vectors, while $\boldsymbol{\kappa}_T(\boldsymbol{x})$
contains their kernel evaluations with $\boldsymbol{x}$.
The realised fitted predictor is
\begin{equation}\label{eq:regularised-kernel-solution}
\widehat f_T(\boldsymbol{x})=\boldsymbol{\kappa}_T(\boldsymbol{x})^\top
(\mathbf{K}_T+n_T\lambda_T \mathbf{I}_{n_T})^{-1}\boldsymbol{y}_T.
\end{equation}
The factor $n_T$ in the diagonal addition follows from averaging the training
loss. Thus the fit can be computed directly from the kernel, and the
regularisation makes the linear system positive definite.
Selection of $m_T,p_T,\lambda_T$ is developed in
Section~\ref{subsec:tuning-parameter-selection}, after the analysis of a
prespecified sequence of models.

\begin{remark}[Intercept penalisation and rescaling]\label{rem:scale-penalty}
The RKHS penalty in \eqref{eq:ridge-estimator} penalises the constant
coefficient together with the positive-degree coefficients. For a
constant-only model, the fitted constant is
$\overline Y/(1+\lambda)$, where $\overline Y$ is the mean of the
training responses. If a constant $c$ is added to every response, the fitted
constant becomes $(\overline Y+c)/(1+\lambda)$, which is not equal to
$\overline Y/(1+\lambda)+c$ when $\lambda>0$. Thus adding $c$ to the data
does not simply add $c$ to the fitted values; the estimator is not location
equivariant when the intercept is penalised. For a closed-loop predictor,
shifting the series also shifts its lagged inputs. The results of
Section~\ref{sec:theoretical-results} concern the series in its stated
units and do not use location equivariance. Rescaling the data also
changes the relative penalty applied to coefficients of different
degrees. If the series is divided by $s>0$ before fitting, its fitted
predictor $f$ corresponds in the original units to
$g(\boldsymbol{x})=s f(\boldsymbol{x}/s)
=\sum_{\boldsymbol{\alpha}}b_{\boldsymbol{\alpha}}
\boldsymbol{x}^{\boldsymbol{\alpha}}$. Multiplying the objective for the
scaled data by $s^2$ returns the squared loss to the original units and
changes the penalty to
$\lambda\sum_{|\boldsymbol{\alpha}|\leq p}
s^{2|\boldsymbol{\alpha}|}b_{\boldsymbol{\alpha}}^2/
c_{\boldsymbol{\alpha},m,p}$. Hence rescaling changes the relative
shrinkage applied to coefficients of different polynomial degrees.
\end{remark}

\section{Theoretical results}\label{sec:theoretical-results}
We establish conditions under which regularised Volterra estimation
approaches the minimum squared-error prediction risk. The analysis first
approximates the conditional mean by a finite-memory polynomial and then
controls the error from estimating its coefficients.
Section~\ref{subsec:tuning-parameter-selection} uses these ideas to analyse
data-driven selection of memory, degree and penalty.

\begin{assumption}[Stationarity and square integrability]
\label{ass:stationarity and causality}
The process $\{Y_t:t\in\mathbb Z\}$ is strictly stationary and
$\mu_2=\E Y_0^2<\infty$.
\end{assumption}
Stationarity makes the distribution of a response and its lag vector
independent of the prediction time. For a Borel-measurable function
$f:\Rbb^m\to\Rbb$ with finite squared-error risk, define
\begin{equation}\label{eq:population-risk}
R_m(f)=\E\{Y_0-f(\boldsymbol{X}_0^{(m)})\}^2,
\qquad R_T(f)=R_{m_T}(f)\quad(f\in\cH_T).
\end{equation}
For a fitted function $\widehat f$, $R_m(\widehat f)$ denotes the
population risk of the realised fitted function.\\[0.05 in]
Write $\cF_m=\sigma(Y_{-1},\ldots,Y_{-m})$ and
$\cF_\infty=\sigma(Y_{-1},Y_{-2},\ldots)$, and define
\begin{equation}\label{eq:prediction-targets}
f_m=\E[Y_0\mid\cF_m],\qquad f^\star=\E[Y_0\mid\cF_\infty],
\qquad R(f^\star)=\E(Y_0-f^\star)^2.
\end{equation}
Let $P_m$ be the distribution of $\boldsymbol{X}_0^{(m)}$.
To obtain polynomial approximation, we impose the classical moment condition
of \citet{carleman1922moments} on the marginal distribution of the observations.

\begin{assumption}[Marginal Carleman condition]\label{ass:carleman}
The random variable $Y_0$ has finite moments of all orders, that is,
$\E|Y_0|^r<\infty$ for every $r>0$, and
\begin{equation}\label{eq:carleman}
\sum_{j=1}^{\infty}\{\E|Y_0|^{2j}\}^{-1/(2j)}=\infty,
\end{equation}
where a zero moment contributes $+\infty$.
\end{assumption}

\begin{remark}[Interpretation of the moment condition]\label{rem:carleman-context}
Condition~\eqref{eq:carleman} restricts the growth of the even moments. It is
an established condition in the moment problem and polynomial approximation;
see \citet[Section~3]{berg1996moment}. The extended Carleman theorem of
\citet[Theorem~2.3]{de2003determinate} implies the polynomial-density result
below. Because the process is strictly stationary, the criterion can be
checked from the marginal distribution of $Y_0$.
\end{remark}

\begin{remark}[A sufficient condition for the Carleman condition]
If $\E e^{a|Y_0|}<\infty$ for some $a>0$, then
Assumption~\ref{ass:carleman} is satisfied. Indeed, with
$C_1=\E e^{a|Y_0|}$,
$$
\E|Y_0|^{2j}\leq C_1(2j)!a^{-2j},\qquad
\{\E|Y_0|^{2j}\}^{-1/(2j)}\geq\frac{a}{2j}C_1^{-1/(2j)},
$$
so the series in \eqref{eq:carleman} diverges. In particular, bounded and
Gaussian marginal distributions satisfy Assumption~\ref{ass:carleman}.
\end{remark}

\begin{lemma}[Polynomial density]\label{lem:polynomial-density}
Under Assumptions~\ref{ass:stationarity and causality} and~\ref{ass:carleman},
real polynomials are dense in $L^2(P_m)$ for every finite $m$.
\end{lemma}
\begin{proof}
For each integer $r\geq1$,
$\|\boldsymbol{X}_0^{(m)}\|_2^r\leq
m^{r-1}\sum_{j=1}^m|Y_{-j}|^r$, so $P_m$ has moments of every order.
Stationarity and \eqref{eq:carleman} give the coordinate conditions in
\citet[Theorem~2.3]{de2003determinate}. Applying that theorem with the
standard basis vectors $\boldsymbol e_1,\ldots,\boldsymbol e_m$ of
$\Rbb^m$ gives density in $L^2(P_m)$.
\end{proof}
The next lemma shows how the finite-history approximation approaches the
optimal predictor as the number of retained lags increases.

\begin{lemma}[Finite-past conditional means]
\label{lem:finite-past-l2}
If $Y_0\in L^2$, then $\|f_m-f^\star\|_2\to0$ as $m\to\infty$.
For every $g\in L^2(\cF_m)$,
$$
\E(Y_0-g)^2-R(f^\star)=\E(g-f_m)^2+\E(f_m-f^\star)^2.
$$
\end{lemma}
\begin{proof}
The tower property gives $f_m=\E[f^\star\mid\cF_m]$. These conditional expectations form an $L^2$-bounded martingale. L\'evy's upward theorem and $L^2$ martingale convergence imply convergence to $\E[f^\star\mid\cF_\infty]=f^\star$ in $L^2$ as $m\to\infty$; see \citet[Appendix~B, Section~12.3, Theorems~12.25 and~12.28]{morters2010brownian}. In the identity
$$
Y_0-g=(Y_0-f^\star)+(f^\star-f_m)+(f_m-g),
$$
the first term is orthogonal to $L^2(\cF_\infty)$ and the second to $L^2(\cF_m)$. Expanding the squared norm therefore proves the risk decomposition.
\end{proof}
Polynomial density then yields the approximation result.

\begin{theorem}[Volterra approximation]
\label{prop:Volterrra-approx}
Under Assumptions~\ref{ass:stationarity and causality} and~\ref{ass:carleman},
$$
\lim_{m\to\infty}\inf_{p\in\mathbb N_0}\inf_{v\in\cH_{m,p}}\{R_m(v)-R(f^\star)\}=0.
$$
\end{theorem}
\begin{proof}
Fix $m\geq1$. By the Doob--Dynkin lemma, there exists a Borel-measurable
function $h_m:\Rbb^m\to\Rbb$ such that
$f_m=h_m(\boldsymbol{X}_0^{(m)})$ almost surely; see
\citet[Lemma~1.13]{kallenberg1997foundations}. Conditional Jensen's inequality
gives $\|h_m(\boldsymbol{X}_0^{(m)})\|_2^2\leq\mu_2$. By
Lemma~\ref{lem:polynomial-density}, for every $\varepsilon>0$ there are a
finite degree $p$ and $v\in\cH_{m,p}$ with
$\|v(\boldsymbol{X}_0^{(m)})-f_m\|_2<\varepsilon$.
By the triangle inequality, we have
$$
\|v(\boldsymbol{X}_0^{(m)})-f^\star\|_2
\leq\|v(\boldsymbol{X}_0^{(m)})-f_m\|_2
+\|f_m-f^\star\|_2
<\varepsilon+\|f_m-f^\star\|_2.
$$
Conditional-mean orthogonality, as in Lemma~\ref{lem:finite-past-l2}, gives
$R_m(v)-R(f^\star)=\|v(\boldsymbol{X}_0^{(m)})-f^\star\|_2^2$.
Taking the infimum over $p$ and $v$, and then letting
$\varepsilon\downarrow0$, we obtain
$$
0\leq\inf_{p\in\mathbb N_0}\inf_{v\in\cH_{m,p}}
\{R_m(v)-R(f^\star)\}\leq\|f_m-f^\star\|_2^2.
$$
The right-hand side tends to zero as $m\to\infty$ by
Lemma~\ref{lem:finite-past-l2}, proving the result.
\end{proof}
Theorem~\ref{prop:Volterrra-approx} consequently gives
\begin{equation}\label{eq:target-approximability}
\inf_{m\geq1,\,p\geq0}\inf_{v\in\cH_{m,p}}
\E\{f^\star-v(\boldsymbol{X}_0^{(m)})\}^2=0.
\end{equation}

The consistency proof requires that embedding a fixed approximating
polynomial into models with more lags or higher degree does not increase
its RKHS norm. The kernel in \eqref{e2} has this property.

\begin{lemma}[Nesting of the Volterra polynomial spaces]
\label{lem:volterra-embedding}
Let $v\in\cH_{m_0,p_0}$, $m\geq m_0$, and $p\geq p_0$. Define
$$
\widetilde v(x_1,\ldots,x_m)=v(x_1,\ldots,x_{m_0}).
$$
Then $\widetilde v\in\cH_{m,p}$ and its RKHS norm equals that of $v$.
\end{lemma}
\begin{proof}
Write
$v(\boldsymbol{x})=\sum_{|\boldsymbol{\alpha}|\leq p_0}
a_{\boldsymbol{\alpha}}\boldsymbol{x}^{\boldsymbol{\alpha}}$, where
$\boldsymbol{\alpha}=(\alpha_1,\ldots,\alpha_{m_0})$. For the model with
memory $m\geq m_0$, replace each exponent vector by
$\widetilde{\boldsymbol{\alpha}}=(\alpha_1,\ldots,\alpha_{m_0},0,\ldots,0)
\in\mathbb N_0^m$. The weight
$c_{\widetilde{\boldsymbol{\alpha}},m,p}
=|\boldsymbol{\alpha}|!/\prod_{j=1}^{m_0}\alpha_j!$ is therefore equal to
$c_{\boldsymbol{\alpha},m_0,p_0}$ and does not depend on the larger maximum
degree $p$. Hence \eqref{eq:volterra-coefficient-norm} gives
$$
\|\widetilde v\|_{\cH_{m,p}}^2
=\sum_{|\boldsymbol{\alpha}|\leq p_0}
a_{\boldsymbol{\alpha}}^2/c_{\boldsymbol{\alpha},m_0,p_0}
=\|v\|_{\cH_{m_0,p_0}}^2.
$$
\end{proof}

\begin{remark}[Moments of all orders do not guarantee polynomial density]
\label{rem:approximation-requirement}\label{rem:polynomial-obstruction}
The following counterexample shows that finite moments of all orders do not
ensure polynomial density, even for a stationary time series with
geometrically decaying dependence. Let $w$ be the lognormal density
$$
w(y)=\frac{1}{y\sqrt{2\pi}}\exp\left\{-\frac{(\log y)^2}{2}\right\},
\qquad y>0,
$$
and put $w(y)=0$ for $y\leq0$. Choose $0<\rho<1$ and define, for $x>0$,
$$
h(x)=\sin(2\pi\log x),\qquad
 g(x)=\mathbb{I}_{\{x>1\}}-\tfrac12,
$$
where $\mathbb{I}_{\{x>1\}}$ equals one when $x>1$ and zero otherwise.
Construct a positive-valued first-order Markov time series by drawing $Y_t$,
conditional on $Y_{t-1}=x$, from the density
\begin{equation}\label{eq:counterexample-transition}
p(y\mid x)=\{1+\rho h(x)g(y)\}w(y),\qquad x>0,\ y>0.
\end{equation}
Thus the conditional distribution of $Y_t$ given all preceding observations
depends only on $Y_{t-1}$. Take $Y_0$ to have density $w$ and generate the
chain using \eqref{eq:counterexample-transition}. The resulting process
$\{Y_t\}$ is a strictly stationary first-order Markov process with marginal
density $w$, and its strong-mixing coefficients decay geometrically; these
claims are verified in Appendix~\ref{app:obstruction}. Since $w$ is the
lognormal density, every $Y_t$ has finite moments of all orders. Its
conditional mean is
$$
f^\star=b+\rho c\sin(2\pi\log Y_{-1}),\qquad
b=e^{1/2},\quad c=e^{1/2}\{\Phi(1)-1/2\}>0,
$$
where $\Phi$ is the standard normal distribution function. For every memory
$m\geq1$, degree $p\geq0$ and polynomial $v\in\cH_{m,p}$,
\begin{equation}\label{eq:polynomial-obstruction}
R_m(v)-R(f^\star)\geq\frac{\rho^2c^2}{2}(1-e^{-8\pi^2})>0.
\end{equation}
Polynomials are not dense in $L^2(P_m)$, and the conditional mean cannot be
approximated arbitrarily closely in mean square by finite-memory polynomials.
Appendix~\ref{app:obstruction} proves \eqref{eq:polynomial-obstruction} using
the lognormal moment orthogonality discussed by
\citet{stoyanov2004stieltjes}. For this marginal distribution, the summands in
\eqref{eq:carleman} equal $e^{-j}$, so the Carleman condition fails.
\end{remark}

To control estimation error, we require a variance bound for averages of
functions of consecutive observations. Dependence between separated
observations is measured by the strong-mixing coefficients, defined for
$h\geq1$ by
\begin{equation}\label{def:mixing}
\alpha_Y(h)=\sup_{\substack{A\in\sigma(Y_t:t\leq0)\\
B\in\sigma(Y_t:t\geq h)}}
|\Pbb(A\cap B)-\Pbb(A)\Pbb(B)|.
\end{equation}
A process is strongly mixing when $\alpha_Y(h)\to0$ as $h\to\infty$.
For $q>2$, let $\eta=1-2/q$ and
$c_q=(\E|Z|^q)^{1/q}$ for $Z\sim N(0,1)$. For integers $n\geq1$ and
$m\geq0$, define
\begin{equation}\label{eq:finite-dependence-sum}
B_{n,m}(q)=1+2m+24c_q^2\sum_{h=1}^{n}\alpha_Y(h)^\eta.
\end{equation}
This quantity controls the variance of averages of functions of consecutive
observations. Specifically, for a Borel-measurable function
$g:\Rbb^{m+1}\to\Rbb$, let
$V_t=g(Y_t,Y_{t-1},\ldots,Y_{t-m})$. Under
Assumption~\ref{ass:stationarity and causality}, if $\|V_0\|_q<\infty$,
then
$$
\operatorname{Var}\left(\frac1n\sum_{t=1}^n V_t\right)
\leq\frac{B_{n,m}(q)}n\|V_0\|_q^2.
$$
For $1\leq h\leq m$, the blocks defining $V_t$ and $V_{t+h}$ can share
observations. For $h>m$, the two blocks are separated by $h-m$ time units,
and their covariance is controlled by $\alpha_Y(h-m)$. Appendix~\ref{app:covariance} proves the bound
using the covariance inequality of \citet{davydov1968}, in the form recorded
by \citet[inequality~(1.0)]{rio1993covariance}.\\[0.05 in]
Define
\begin{equation}\label{feature:moment}
\mathsf M_T(q)=\|Y_0^2+k^{(p_T)}(\boldsymbol{X}_0^{(m_T)},
\boldsymbol{X}_0^{(m_T)})\|_q,\qquad
\Delta_T(q)=\mathsf M_T(q)\sqrt{B_{n_T,m_T}(q)/n_T}.
\end{equation}
Here $\mathsf M_T(q)$ controls the moments of the response and kernel
features that enter the empirical-risk bound, while $\Delta_T(q)$ is the
corresponding sampling-error scale after accounting for temporal dependence
and the $n_T$ training pairs. For a fixed degree $p$, the condition
\begin{equation}\label{eq:polynomial-moment-sufficient}
\E|Y_0|^{2q\max(1,p)}<\infty
\end{equation}
is sufficient for $\mathsf M_T(q)<\infty$ when $p_T=p$ and $m_T$ is
finite. Assumption~\ref{ass:carleman} supplies these moments at every finite
degree. When the degree increases with $T$, the growth of the required moments
is reflected in $\mathsf M_T(q)$.

\label{subsec:assumptions-regimes}\label{subsec:moment-risk-consistency}
\label{subsec:fixed-regimes}\label{subsec:finite-dimensional-rate}
The following uniform bound is the link between the empirical criterion
and population risk.

\begin{proposition}[Uniform empirical-risk bound]
\label{prop:unbounded-uniform-bound}
Under Assumption~\ref{ass:stationarity and causality}, let $q>2$ and
$\mathsf M_T(q)<\infty$, with $\mathsf M_T(q)$ and $\Delta_T(q)$ defined in
\eqref{feature:moment}. For every $\Lambda\geq0$ and $u>0$,
$$
\Pbb\left(\sup_{f\in\cH_T:\,\|f\|_{\cH_T}\leq\Lambda}|\widehat R_T(f)-R_T(f)|>(1+\Lambda^2)u\right)\leq\frac{\Delta_T(q)^2}{u^2}.
$$
\end{proposition}
In particular, on any deterministic ball with radius $\Lambda_T$ satisfying
$(1+\Lambda_T^2)\Delta_T(q)\to0$, the empirical risk converges uniformly in
probability to the population risk. The proof is given in
Appendix~\ref{app:uniform-risk}.

\begin{riskratecorollary}[Uniform empirical-risk rate]
Suppose the conditions of Proposition~\ref{prop:unbounded-uniform-bound}
hold for every $T$. For any deterministic sequence $\Lambda_T\geq0$,
$$
\sup_{f\in\cH_T:\,\|f\|_{\cH_T}\leq\Lambda_T}|\widehat R_T(f)-R_T(f)|=O_p\{(1+\Lambda_T^2)\Delta_T(q)\}\quad\text{as }T\to\infty.
$$
\end{riskratecorollary}
\begin{proof}
Apply Proposition~\ref{prop:unbounded-uniform-bound} with
$u=C_2\Delta_T(q)$, where $C_2>0$ is arbitrary. The resulting
probability bound is $C_2^{-2}$, uniformly in $T$.
\end{proof}

For $\lambda>0$, let $\widehat f_{T,\lambda}$ minimise the empirical
criterion in \eqref{eq:ridge-estimator} with $\lambda_T=\lambda$, and let
$f_{T,\lambda}$ minimise the population counterpart
$R_T(f)+\lambda\|f\|_{\cH_T}^2$ over $f\in\cH_T$.
Define the regularised approximation error
\begin{equation}\label{eq:regularised-approximation-error}
\mathcal A_T(\lambda)=\inf_{f\in\cH_T}
\{R_T(f)-R(f^\star)+\lambda\|f\|_{\cH_T}^2\}.
\end{equation}
The next theorem bounds the fitted excess risk by this approximation term
and a sampling error.

\begin{theorem}[Finite-sample oracle inequality]
\label{prop:finite-sample-oracle}
Suppose Assumption~\ref{ass:stationarity and causality} holds and
$\mathsf M_T(q)<\infty$ for a fixed $q>2$. Use $\mathsf M_T(q)$ and
$\Delta_T(q)$ from \eqref{feature:moment} and $\mathcal A_T(\lambda)$ from
\eqref{eq:regularised-approximation-error}. For $\lambda>0$, $b>\mu_2$ and $u>0$,
$$
\Pbb\left\{R_T(\widehat f_{T,\lambda})-R(f^\star)>\mathcal A_T(\lambda)+2(1+b/\lambda)u\right\}
\leq\frac{\Delta_T(q)^2}{u^2}+\frac{\Delta_T(q)^2}{(b-\mu_2)^2}.
$$
\end{theorem}
The inequality bounds the fitted excess risk from above by the regularised
approximation error plus a sampling term, with the stated probability.
Its proof is given in Appendix~\ref{app:oracle-proof}.\\[0.05 in]
The next theorem establishes asymptotic risk consistency of the regularised
estimator $\widehat f_T=\widehat f_{T,\lambda_T}$.

\begin{theorem}[Risk consistency]\label{thm:unbounded-risk-consistency}
Suppose Assumptions~\ref{ass:stationarity and causality} and~\ref{ass:carleman}
hold. Let $\widehat f_T$ be the estimator in \eqref{eq:ridge-estimator},
and let $\Delta_T(q)$ be defined by \eqref{feature:moment} for a fixed
$q>2$. Suppose the deterministic memory lengths, degrees and penalties
satisfy, as $T\to\infty$,
$$
m_T\to\infty,\qquad p_T\to\infty,\qquad n_T=T-m_T\to\infty,
$$
$$
\lambda_T>0,\qquad\lambda_T\to0,\qquad
\frac{\Delta_T(q)}{\lambda_T}\to0.
$$
Then $R_T(\widehat f_T)-R(f^\star)$ converges in probability to $0$ as $T\to\infty$.
\end{theorem}
The condition $\Delta_T(q)/\lambda_T\to0$ restricts the rate
at which the penalty may decrease. Remark~\ref{rem:explicit-mixing-consistency}
gives explicit choices under polynomial decay of the strong-mixing
coefficients. The proof is in Appendix~\ref{app:prescribed-proof}.

\begin{remark}[Explicit choices under polynomial mixing]
\label{rem:explicit-mixing-consistency}
Polynomial decay of strong-mixing coefficients is a standard form of weak
dependence; see \citet[pp.~15--23]{Doukhan1994}.
Suppose Assumptions~\ref{ass:stationarity and causality} and~\ref{ass:carleman}
hold, and that for some $a>0$ and $K>0$,
$$
\alpha_Y(h)=O(h^{-a})\quad(h\to\infty),\qquad
\|Y_0\|_{2j}\leq K\sqrt j\quad(j\geq1).
$$
The second condition is a quantitative moment-growth assumption. It holds,
for example, for Gaussian observations and for bounded observations.
Appendix~\ref{app:nonlinear-benchmark} verifies both this moment bound and
geometric decay of the strong-mixing coefficients for the stationary models
$N_1,C_a$ and $J_a$ defined in Section~\ref{subsec:simulation-study}.\\[0.05 in]
For sufficiently large $T$, choose
\begin{equation}\label{eq:polynomial-mixing-choices}
m_T=p_T=\lfloor\log\log T\rfloor,\qquad
\lambda_T=T^{-\gamma_a},\qquad
\gamma_a=\min\{1/4,a/8\}.
\end{equation}
Then the sampling condition of Theorem~\ref{thm:unbounded-risk-consistency}
holds with $q=4$. In particular,
$$
\frac{\Delta_T(4)}{\lambda_T}\leq T^{-\gamma_a+o(1)}\longrightarrow0.
$$
Consequently, $R_T(\widehat f_T)-R(f^\star)$ converges in probability to
$0$ as $T\to\infty$.
For example, $a=1$ permits $\lambda_T=T^{-1/8}$, while $a\geq2$
permits $\lambda_T=T^{-1/4}$. Under exponential mixing,
$\alpha_Y(h)=O(e^{-Ch})$ for some $C>0$, the same memory and degree choices
give risk consistency with $\lambda_T=T^{-1/4}$.
\end{remark}

\subsection{Selection of the tuning parameters}
\label{subsec:tuning-parameter-selection}
We use a rolling-origin implementation of the prequential approach
\citep{Dawid1984,Tashman2000} to select the memory $m$, degree $p$ and
regularisation parameter $\lambda$ of the estimator in
\eqref{eq:ridge-estimator}.\\[0.05 in]
Let $\mathcal G_T\subset\mathbb N\times\mathbb N_0\times(0,\infty)$ be a
deterministic, finite, nonempty set of triples $\boldsymbol{\tau}=(m,p,\lambda)$.
Choose a deterministic integer $s_T$ as the first forecasting origin, with
$\max_{\boldsymbol{\tau}\in\mathcal G_T}m<s_T<T$, and put $N_T=T-s_T$.
The rolling-origin procedure is as follows.

\par\medskip
\noindent\rule{\linewidth}{0.5pt}\par\nobreak
\phantomsection\label{alg:rolling-origin}
\noindent\textbf{Rolling-origin selection of the Volterra predictor.}\par\nobreak
\noindent\rule{\linewidth}{0.5pt}\par\nobreak
\textbf{Input.} The observed series
$\boldsymbol y_{1:T}=(y_1,\ldots,y_T)^\top$, the candidate set
$\mathcal G_T$, and the first forecasting origin $s_T$.
\par
\textbf{Output.} The selected triple
$\widehat{\boldsymbol{\tau}}_T=(\widehat m_T,\widehat p_T,\widehat\lambda_T)$
and the full-sample predictor $\widehat f_T^{\mathrm{RO}}$.
\begin{enumerate}
\item \textbf{Construct the lag vectors.}
For each candidate memory $m$, form
$$
\boldsymbol x_t^{(m)}=(y_{t-1},\ldots,y_{t-m})^\top,
\qquad t=m+1,\ldots,T.
$$

\item \textbf{Obtain chronological predictions.}
For each candidate $\boldsymbol{\tau}=(m,p,\lambda)$ and each origin
$r=s_T,\ldots,T-1$, fit
\begin{equation}\label{eq:rolling-origin-estimator}
\widehat f_{r,\boldsymbol{\tau}}=
\argmin_{f\in\cH_{m,p}}
\left\{\frac1{r-m}\sum_{t=m+1}^{r}
\{y_t-f(\boldsymbol x_t^{(m)})\}^2
+\lambda\|f\|_{\cH_{m,p}}^2\right\}.
\end{equation}
Each fit uses only $y_1,\ldots,y_r$, and the candidate's penalty
$\lambda$ remains fixed across origins. Let $n_r=r-m$ and define
$$
\boldsymbol y_{r,m}=(y_{m+1},\ldots,y_r)^\top.
$$
Form the $n_r\times n_r$ Gram matrix
$\mathbf K_{r,\boldsymbol{\tau}}$ and the kernel-evaluation vector
$\boldsymbol\kappa_{r,\boldsymbol{\tau}}$ by
$$
[\mathbf K_{r,\boldsymbol{\tau}}]_{ij}
=k^{(p)}(\boldsymbol x_{m+i}^{(m)},\boldsymbol x_{m+j}^{(m)}),
\qquad
[\boldsymbol\kappa_{r,\boldsymbol{\tau}}]_i
=k^{(p)}(\boldsymbol x_{m+i}^{(m)},\boldsymbol x_{r+1}^{(m)}),
$$
for $1\leq i,j\leq n_r$. Then predict
\begin{equation}\label{eq:rolling-origin-kernel-prediction}
\widehat y_{r+1,\boldsymbol{\tau}}
=
\boldsymbol\kappa_{r,\boldsymbol{\tau}}^\top
\{\mathbf K_{r,\boldsymbol{\tau}}+n_r\lambda\mathbf I_{n_r}\}^{-1}
\boldsymbol y_{r,m}.
\end{equation}
Equivalently,
$\widehat y_{r+1,\boldsymbol{\tau}}
=\widehat f_{r,\boldsymbol{\tau}}(\boldsymbol x_{r+1}^{(m)})$.

\item \textbf{Calculate the prediction scores.}
For each candidate, calculate
\begin{equation}\label{eq:rolling-origin-score}
\operatorname{RO}_T(\boldsymbol{\tau})=
\frac1{N_T}\sum_{r=s_T}^{T-1}
\{y_{r+1}-\widehat f_{r,\boldsymbol{\tau}}
(\boldsymbol x_{r+1}^{(m)})\}^2.
\end{equation}
All candidates are assessed at the same prediction times.

\item \textbf{Select the tuning parameters.}
Choose
\begin{equation}\label{eq:selected-tuning-vector}
\widehat{\boldsymbol{\tau}}_T\in
\argmin_{\boldsymbol{\tau}\in\mathcal G_T}
\operatorname{RO}_T(\boldsymbol{\tau}).
\end{equation}
Resolve exact ties by smaller $D_{m,p}$, then smaller $m$, then smaller
$p$, and finally larger $\lambda$.

\item \textbf{Refit the selected model.}
Using $y_1,\ldots,y_T$, fit the selected candidate by
\eqref{eq:rolling-origin-estimator} with $r=T$. The resulting predictor is
$$
\widehat f_T^{\mathrm{RO}}
=\widehat f_{T,\widehat{\boldsymbol{\tau}}_T}.
$$
\end{enumerate}
\noindent\rule{\linewidth}{0.5pt}\par\medskip

The population counterpart of a candidate and its prediction risk are
\begin{equation}\label{eq:population-ridge-predictor}
f_{\boldsymbol{\tau}}=\argmin_{f\in\cH_{m,p}}\{R_m(f)+\lambda\|f\|_{\cH_{m,p}}^2\},
\qquad Q(\boldsymbol{\tau})=R_m(f_{\boldsymbol{\tau}}).
\end{equation}
For a candidate $\boldsymbol{\tau}=(m,p,\lambda)$ with finite feature
second moments, define
\begin{equation}\label{eq:candidate-feature-moments}
\begin{aligned}
\boldsymbol{\mathit{\Psi}}_{t,\boldsymbol{\tau}}
&=\boldsymbol{\psi}_{m,p}(\boldsymbol{X}_t^{(m)}),\\
\mathbf{G}_{\boldsymbol{\tau}}
&=\E[\boldsymbol{\mathit{\Psi}}_{0,\boldsymbol{\tau}}
\boldsymbol{\mathit{\Psi}}_{0,\boldsymbol{\tau}}^\top],\qquad
\boldsymbol{g}_{\boldsymbol{\tau}}
=\E[\boldsymbol{\mathit{\Psi}}_{0,\boldsymbol{\tau}}Y_0].
\end{aligned}
\end{equation}
For $r>m$, define the sample moments by
\begin{equation}\label{eq:empirical-feature-matrix}
\widehat{\mathbf G}_{r,\boldsymbol\tau}
=\frac1{r-m}\sum_{t=m+1}^r
\boldsymbol{\mathit\Psi}_{t,\boldsymbol\tau}
\boldsymbol{\mathit\Psi}_{t,\boldsymbol\tau}^{\top},\qquad
\widehat{\boldsymbol g}_{r,\boldsymbol\tau}
=\frac1{r-m}\sum_{t=m+1}^r
\boldsymbol{\mathit\Psi}_{t,\boldsymbol\tau}Y_t.
\end{equation}
Here the feature map is specified in \eqref{eq:scaled-monomials}, and
the lag vector is defined in \eqref{eq:lag-vector}.
The value $\min_{\boldsymbol\tau\in\mathcal G}Q(\boldsymbol\tau)$ is the
smallest population risk among the candidates. The next theorem shows that,
for a fixed finite grid, the rolling-origin procedure selects a candidate
attaining this minimum in the limit. The proof uses Birkhoff's ergodic
theorem \citep[Chapter~11]{EisnerEtAl2015}.

\begin{theorem}[Ergodic fixed-grid selection]
\label{thm:rolling-origin-selection-consistency}
Suppose the process is strictly stationary and ergodic, $\mathcal G$ is fixed, finite and nonempty, all penalties are positive, and
\begin{equation}
\E\{Y_0^2+k^{(p)}(\boldsymbol{X}_0^{(m)},\boldsymbol{X}_0^{(m)})\}<\infty\quad\text{for every }(m,p,\lambda)\in\mathcal G.
\label{eq:fixed-grid-moment}
\end{equation}
Take deterministic integer origins $\max_{\boldsymbol{\tau}\in\mathcal G}m<s_T<T$
and suppose $s_T/T\to\kappa\in[0,1)$ as $T\to\infty$. Use the scores
and selected refit defined in \eqref{eq:rolling-origin-score}--\eqref{eq:selected-tuning-vector}, and $Q$ from
\eqref{eq:population-ridge-predictor}. Then, almost surely as $T\to\infty$,
$$
\max_{\boldsymbol{\tau}\in\mathcal G}|\operatorname{RO}_T(\boldsymbol{\tau})-Q(\boldsymbol{\tau})|\to0,
\qquad \widehat{\boldsymbol{\tau}}_T\in\mathcal G^\star\ \text{eventually},
$$
$$
R_{\widehat m_T}(\widehat f_T^{\mathrm{RO}})-\min_{\boldsymbol{\tau}\in\mathcal G}Q(\boldsymbol{\tau})\to0,
$$
where $\mathcal G^\star=\argmin_{\boldsymbol{\tau}\in\mathcal G}Q(\boldsymbol{\tau})$. In particular, if $Q$ has a unique minimiser, that triple is eventually selected almost surely. A sufficient marginal moment condition is $\E|Y_0|^{2p_*}<\infty$, where $p_* =\max\{1,\max_{\boldsymbol{\tau}\in\mathcal G}p\}$.
\end{theorem}

\begin{remark}[Relation to the assumptions used for risk consistency]
Theorem~\ref{thm:rolling-origin-selection-consistency} has a different target from
Theorem~\ref{thm:unbounded-risk-consistency}. It compares finitely many
fixed candidates with positive penalties and proves convergence to
$\min_{\boldsymbol{\tau}\in\mathcal G}Q(\boldsymbol{\tau})$; it does not
claim convergence to $R(f^\star)$. Consequently, the Carleman condition and
the growing-model requirements used in Theorem~\ref{thm:unbounded-risk-consistency}
are not needed here. Ergodicity is used to obtain almost-sure convergence of
the time averages that define the feature moments and rolling-origin losses.
A strictly stationary strongly mixing process is ergodic, so strong mixing
is one sufficient dependence condition for this theorem, but no mixing rate
is required. The moment condition in \eqref{eq:fixed-grid-moment} guarantees
integrability for every candidate in the fixed finite grid. Under Assumptions~\ref{ass:stationarity and causality}
and~\ref{ass:carleman}, it is automatically satisfied
for every fixed finite $m$ and $p$, because
$k^{(p)}(\boldsymbol X_0^{(m)},\boldsymbol X_0^{(m)})$ is a polynomial in
finitely many lagged observations and all marginal moments are finite.
\end{remark}

The proof is in Appendix~\ref{app:fixed-grid}. The next remark explains
how the penalty $\lambda$ affects the population risk $Q(m,p,\lambda)$
used in Theorem~\ref{thm:rolling-origin-selection-consistency}.

\begin{remark}[Interpretation of the fixed-grid benchmark]
\label{rem:fixed-grid-benchmark}
For fixed memory $m$ and degree $p$, define
\begin{equation}\label{eq:population-ls-risk}
R_{\mathrm{LS}}(m,p)=\min_{f\in\cH_{m,p}}R_m(f)
=\min_{f\in\cH_{m,p}}
\E\bigl[\{Y_0-f(\boldsymbol{X}_0^{(m)})\}^2\bigr].
\end{equation}
This is the minimum unpenalised population prediction risk within the
Volterra polynomial space of memory $m$ and degree at most $p$. For $p>1$
the class $\cH_{m,p}$ contains nonlinear polynomial predictors; $p\leq1$
restricts the predictor to an affine function of the lagged observations.
For $\boldsymbol{\tau}=(m,p,\lambda)$, use the matrix
$\mathbf{G}_{\boldsymbol{\tau}}$ and vector $\boldsymbol{g}_{\boldsymbol{\tau}}$
in \eqref{eq:candidate-feature-moments}. Let $s_{\boldsymbol{\tau},j}$ be the $j$th eigenvalue of
$\mathbf{G}_{\boldsymbol{\tau}}$, in ascending order, and let
$\boldsymbol u_{\boldsymbol{\tau},j}$ be a corresponding unit eigenvector,
with the eigenvectors chosen orthonormal. Write
$g_{\boldsymbol{\tau},j}
=\boldsymbol u_{\boldsymbol{\tau},j}^{\top}\boldsymbol g_{\boldsymbol{\tau}}$.
Then
\begin{equation}\label{eq:population-ridge-risk-decomposition}
\begin{aligned}
R_{\mathrm{LS}}(m,p)
&=\mu_2-\sum_{j:s_{\boldsymbol{\tau},j}>0}
g_{\boldsymbol{\tau},j}^2/s_{\boldsymbol{\tau},j},\\
Q(m,p,\lambda)
&=R_{\mathrm{LS}}(m,p)+
\sum_{j:s_{\boldsymbol{\tau},j}>0}
g_{\boldsymbol{\tau},j}^2/s_{\boldsymbol{\tau},j}
\left\{\lambda/(s_{\boldsymbol{\tau},j}+\lambda)\right\}^2.
\end{aligned}
\end{equation}
Appendix~\ref{app:fixed-grid-risk-decomposition} derives
\eqref{eq:population-ridge-risk-decomposition}, including the case in which
$\mathbf{G}_{\boldsymbol{\tau}}$ is singular. Hence $Q(m,p,\lambda)$ is
nondecreasing in $\lambda$ for fixed $m$ and $p$. The fixed-grid
benchmark therefore uses the smallest available penalty, apart from ties. The frequent selections of the
smallest available penalty in Table~\ref{tab:simulation-selection}
are consistent with this population property, but do not establish a
finite-sample oracle result. The increasing-grid analysis below also
accounts for estimation while
memory, degree and penalty change.
\end{remark}

We now give two theorems on the prediction risk of the rolling-origin
selected estimator when the candidate grid changes with $T$. The first
provides finite-sample bounds for the rolling-origin scores and fitted risks.
The second gives uniform convergence rates and conditions for risk
consistency.

\label{subsec:growing-grid-selection}\label{subsec:growth-conditions}
For the finite candidate set $\mathcal G_T$, let $m_T^+$ and $p_T^+$ be its
largest memory and degree, and let $a_T$ be its smallest penalty:
\begin{equation}\label{eq:grid-extremes}
m_T^+=\max_{\boldsymbol{\tau}\in\mathcal G_T}m,\qquad
p_T^+=\max_{\boldsymbol{\tau}\in\mathcal G_T}p,\qquad
a_T=\min_{\boldsymbol{\tau}\in\mathcal G_T}\lambda>0,\qquad
D_T=D_{m_T^+,p_T^+}.
\end{equation}
Using the feature map from \eqref{eq:scaled-monomials}, define
\begin{equation}\label{eq:common-feature-vector}
\boldsymbol{\psi}_T=\boldsymbol{\psi}_{m_T^+,p_T^+},\qquad
\boldsymbol{\mathit{\Psi}}_{t,T}
=\boldsymbol{\psi}_T(\boldsymbol{X}_t^{(m_T^+)})\in\Rbb^{D_T}.
\end{equation}
Set
\begin{equation}\label{eq:leverage}
\mathbf G_T=\E[\boldsymbol{\mathit{\Psi}}_{0,T}
\boldsymbol{\mathit{\Psi}}_{0,T}^{\top}],\qquad
\mathbf C_T=\mathbf G_T+a_T\mathbf I,\qquad
\ell_T(\boldsymbol x)=\boldsymbol\psi_T(\boldsymbol x)^\top
\mathbf C_T^{-1}\boldsymbol\psi_T(\boldsymbol x).
\end{equation}
Its expectation is
\begin{equation}\label{eq:effective-dimension}
d_T=\E\ell_T(\boldsymbol X_0^{(m_T^+)})
=\operatorname{tr}(\mathbf G_T\mathbf C_T^{-1})\leq D_T.
\end{equation}
Let $\sigma_{T,j}$ be the $j$th eigenvalue, in ascending order, of
$\mathbf G_T$, for $j=1,\ldots,D_T$. Then
$d_T=\sum_{j=1}^{D_T}\sigma_{T,j}/(\sigma_{T,j}+a_T)$.
Thus $d_T$ measures the number of feature directions that remain
substantial after regularisation; it is the effective dimension
\citep{caponnetto2007optimal,sutherland2017fixing}.

For $q>2$, define the following quantities:
\begin{equation}\label{eq:grid-moments}
L_T(q)=\|Y_0^2+\ell_T(\boldsymbol X_0^{(m_T^+)})\|_q,\qquad
B_T(q)=B_{T,m_T^+}(q),\qquad
\nu_T=\min\{s_T-m_T^+,T-s_T\},
\end{equation}
Here $L_T(q)$ is a moment bound, $B_T(q)$ is the dependence
factor defined in \eqref{eq:finite-dependence-sum}, and $\nu_T$ is
the smaller of the fitting and validation sample lengths. Define
\begin{equation}\label{eq:grid-rate}
J_T=1+\lceil\log_2T\rceil,\qquad
\rho_T=\frac{L_T(q)J_T\sqrt{TB_T(q)}}{\nu_T}.
\end{equation}
The factor $J_T$ comes from bounding the sample averages over all
forecasting origins. The quantity $\rho_T$ is the resulting error scale.
These quantities enter the risk bounds; computing the selector
in \eqref{eq:selected-tuning-vector} does not require them.\\[0.05 in]
For $0<\epsilon<1$, write
\begin{equation}\label{eq:perturbation-functions}
A(\epsilon)=\frac{\epsilon(1+\sqrt{\mu_2})}{1-\epsilon},\qquad
V(\epsilon)=A(\epsilon)^2+2\sqrt{\mu_2}A(\epsilon).
\end{equation}

\begin{theorem}[Finite-sample growing-grid bounds]
\label{prop:growing-grid-finite}
Suppose Assumption~\ref{ass:stationarity and causality} holds and
$\mathcal G_T$ is deterministic, finite and nonempty. Assume
$\E\|\boldsymbol{\mathit{\Psi}}_{0,T}\|_2^2<\infty$ for the common
feature vector in \eqref{eq:common-feature-vector}, and $L_T(q)<\infty$
for one fixed $q>2$, with $L_T(q)$ defined in \eqref{eq:grid-moments}.
Use $d_T$ from \eqref{eq:effective-dimension}, $\rho_T$ from
\eqref{eq:grid-rate}, and $Q$ from \eqref{eq:population-ridge-predictor}.
For $0<\epsilon<1$, let $A(\epsilon)$ and $V(\epsilon)$ be given by
\eqref{eq:perturbation-functions}, and define
\begin{equation}\label{eq:score-error-bound}
U_T(\epsilon)=(1+\mu_2)\epsilon+A(\epsilon)^2(d_T+\epsilon)
+2A(\epsilon)\sqrt{(d_T+\epsilon)\{\mu_2+(1+\mu_2)\epsilon\}}.
\end{equation}
With probability at least $1-5\rho_T^2/\epsilon^2$, the following four bounds hold simultaneously:
$$
\sup_{\boldsymbol{\tau}\in\mathcal G_T}|\operatorname{RO}_T(\boldsymbol{\tau})-Q(\boldsymbol{\tau})|\leq U_T(\epsilon),\qquad
\sup_{\boldsymbol{\tau}=(m,p,\lambda)\in\mathcal G_T}|R_m(\widehat f_{T,\boldsymbol{\tau}})-Q(\boldsymbol{\tau})|\leq V(\epsilon),
$$
$$
\left|R_{\widehat m_T}(\widehat f_T^{\mathrm{RO}})-\min_{\boldsymbol{\tau}\in\mathcal G_T}Q(\boldsymbol{\tau})\right|\leq2U_T(\epsilon)+V(\epsilon),
$$
$$
0\leq R_{\widehat m_T}(\widehat f_T^{\mathrm{RO}})-\min_{\boldsymbol{\tau}=(m,p,\lambda)\in\mathcal G_T}R_m(\widehat f_{T,\boldsymbol{\tau}})\leq2U_T(\epsilon)+2V(\epsilon).
$$

\end{theorem}
Appendix~\ref{app:growing-grid-proof} proves the result by bounding a common
second-moment error over the fitting and validation intervals. The same event
controls the normal equations for every candidate and both oracle comparisons.

\Needspace{12\baselineskip}
\begin{theorem}[Uniform consistency and oracle bounds for Volterra selection]
\label{thm:growing-grid}
\label{thm:main-consistency}
\label{cor:selected-bayes-consistency}
Suppose Assumption~\ref{ass:stationarity and causality} holds.
For each $T$, let $\mathcal G_T$ be deterministic, finite, and nonempty,
and use the rolling-origin selector in \eqref{eq:selected-tuning-vector}.
Use the common feature vector $\boldsymbol{\mathit{\Psi}}_{0,T}$ in
\eqref{eq:common-feature-vector}, $L_T(q)$ in \eqref{eq:grid-moments},
$d_T$ in \eqref{eq:effective-dimension}, and $\rho_T$ in
\eqref{eq:grid-rate}. Assume $\E\|\boldsymbol{\mathit{\Psi}}_{0,T}\|_2^2<\infty$
and $L_T(q)<\infty$ for one fixed $q>2$. The population candidate risk
$Q$ is defined in \eqref{eq:population-ridge-predictor}.
All limits and rates in (i)--(iii) are taken as $T\to\infty$.\\[0.05 in]
(i) If $\rho_T\to0$, the candidate-specific feature matrices in
\eqref{eq:candidate-feature-moments} and \eqref{eq:empirical-feature-matrix}
satisfy the following bound in Frobenius norm:
\begin{equation}\label{eq:relative-operator-consistency}
\max_{\boldsymbol{\tau}=(m,p,\lambda)\in\mathcal G_T}\max_{s_T\leq r\leq T}
\left\|(\mathbf{G}_{\boldsymbol{\tau}}+\lambda \mathbf{I})^{-1/2}
(\widehat{\mathbf{G}}_{r,\boldsymbol{\tau}}-\mathbf{G}_{\boldsymbol{\tau}})
(\mathbf{G}_{\boldsymbol{\tau}}+\lambda \mathbf{I})^{-1/2}\right\|_{\mathrm F}=O_p(\rho_T),
\end{equation}
and the final refits satisfy
\begin{equation}\label{eq:uniform-refit-rate}
\sup_{\boldsymbol{\tau}=(m,p,\lambda)\in\mathcal G_T}
|R_m(\widehat f_{T,\boldsymbol{\tau}})-Q(\boldsymbol{\tau})|=O_p(\rho_T).
\end{equation}

(ii) If $\rho_T\sqrt{1+d_T}\to0$, then
\begin{equation}\label{eq:uniform-score-rate}
\sup_{\boldsymbol{\tau}\in\mathcal G_T}|\operatorname{RO}_T(\boldsymbol{\tau})-Q(\boldsymbol{\tau})|
=O_p\{\rho_T\sqrt{1+d_T}\},
\end{equation}
\begin{equation}\label{eq:selected-population-oracle}
\left|R_{\widehat m_T}(\widehat f_T^{\mathrm{RO}})
-\min_{\boldsymbol{\tau}\in\mathcal G_T}Q(\boldsymbol{\tau})\right|
=O_p\{\rho_T\sqrt{1+d_T}\},
\end{equation}
and
\begin{equation}\label{eq:selected-fitted-oracle}
0\leq R_{\widehat m_T}(\widehat f_T^{\mathrm{RO}})
-\min_{\boldsymbol{\tau}=(m,p,\lambda)\in\mathcal G_T}R_m(\widehat f_{T,\boldsymbol{\tau}})
=O_p\{\rho_T\sqrt{1+d_T}\}.
\end{equation}

(iii) In addition to (ii), suppose
Assumption~\ref{ass:carleman} holds and $\mathcal G_T$ contains
$(m_T^\circ,p_T^\circ,\lambda_T^\circ)$ with
$m_T^\circ,p_T^\circ\to\infty$ and $\lambda_T^\circ\to0$.
For the target $f^\star$ and its risk in \eqref{eq:prediction-targets},
\begin{equation}\label{eq:selected-bayes-limit}
R_{\widehat m_T}(\widehat f_T^{\mathrm{RO}})-R(f^\star)\longrightarrow0
\qquad\text{in probability as }T\to\infty.
\end{equation}

\end{theorem}
The proof is in Appendix~\ref{app:main-theorem-proof}.
When $s_T/T\to\kappa\in(0,1)$ and $m_T^+=o(T)$, the condition
in part (ii) is equivalent to
\begin{equation}\label{eq:explicit-growth}
\frac{L_T(q)^2 B_T(q)(1+d_T)(\log T)^2}{T}\to0\quad\text{as }T\to\infty.
\end{equation}
It permits the dimension and polynomial moments to grow, provided that the
sampling error still decreases. The following corollaries give explicit
sufficient choices.

\begin{corollary}[Polynomial-order growth bounds]
\label{cor:power-growth}
Suppose the stationarity, moment and candidate-grid hypotheses stated before
part~(i) of Theorem~\ref{thm:growing-grid} hold. Assume
$s_T/T\to\kappa\in(0,1)$ and $m_T^+=o(T)$ as $T\to\infty$.
Let $\ell,b,\delta\geq0$ be fixed exponents bounding the growth of the
moment quantity $L_T(q)$, the dependence factor $B_T(q)$ and the effective
dimension $1+d_T$, respectively; these quantities are defined in
\eqref{eq:grid-moments} and \eqref{eq:effective-dimension}.
Specifically, suppose that, as $T\to\infty$,
\begin{equation}\label{eq:power-growth-exponents}
L_T(q)=O(T^\ell),\qquad B_T(q)=O(T^b),\qquad 1+d_T=O(T^\delta).
\end{equation}
If $2\ell+b<1$, the uniform matrix and fitted-risk conclusions in
part~(i) of Theorem~\ref{thm:growing-grid} hold.
If $2\ell+b+\delta<1$, the rolling-origin scores converge uniformly,
and both oracle risk differences in part~(ii) are
$$
O_p\{T^{\ell+(b+\delta-1)/2}\log T\}\quad\text{as }T\to\infty.
$$
Under this latter inequality, if Assumption~\ref{ass:carleman} also holds
and $\mathcal G_T$ contains a sequence
$(m_T^\circ,p_T^\circ,\lambda_T^\circ)$ with
$m_T^\circ,p_T^\circ\to\infty$ and $\lambda_T^\circ\to0$, then the selected
full-sample predictor is risk consistent, as in
\eqref{eq:selected-bayes-limit}.
\end{corollary}
\begin{proof}
Since $s_T/T\to\kappa\in(0,1)$ and $m_T^+=o(T)$, the definition
of $\nu_T$ in \eqref{eq:grid-moments} gives $\nu_T\asymp T$.
Also $J_T=O(\log T)$. Substitution of
\eqref{eq:power-growth-exponents} into \eqref{eq:grid-rate} therefore yields
$$
\rho_T=O\{T^{\ell+(b-1)/2}\log T\},\qquad
\rho_T\sqrt{1+d_T}
=O\{T^{\ell+(b+\delta-1)/2}\log T\}.
$$
If $2\ell+b<1$, the first bound tends to zero, so part~(i) of
Theorem~\ref{thm:growing-grid} applies. If $2\ell+b+\delta<1$,
the second bound tends to zero. Part~(ii) then gives the asserted
$O_p$ rates for the scores and the two oracle-risk differences.
Finally, the additional hypotheses in the corollary are precisely those
required in part~(iii) of Theorem~\ref{thm:growing-grid}, which gives
risk consistency of the selected predictor.
\end{proof}

\Needspace{17\baselineskip}
\begin{corollary}[A Gaussian feasibility result]
\label{cor:gaussian-feasible}
Suppose $\{Y_t\}$ is a strictly stationary Gaussian process with
$0<\mu_2<\infty$ and $\alpha_Y(h)\leq C_3e^{-C_4h}$ for positive constants
$C_3,C_4$. Let $s_T=\lfloor\kappa T\rfloor$, $0<\kappa<1$, and, for
sufficiently large $T$, set
$$
r_T=\left\lfloor\frac{\log T}{10}\right\rfloor,\qquad
\mathcal G_T=\{1,\ldots,r_T\}\times\{0,\ldots,r_T\}\times\Lambda_T,
$$
where $\Lambda_T$ is any deterministic finite positive set containing
$T^{-\gamma}$ for a fixed $\gamma>0$. Then all three parts of
Theorem~\ref{thm:growing-grid} apply. In particular,
$D_T\leq4^{r_T}\leq T^{(\log4)/10}$.\\[0.05 in]
Appendix~\ref{app:gaussian-feasible} shows that this choice makes
$\rho_T\sqrt{1+d_T}=O\{T^{-1/20}(\log T)^{3/2}\}$. Hence both
oracle-risk differences in \eqref{eq:selected-population-oracle} and
\eqref{eq:selected-fitted-oracle} are
$O_p\{T^{-1/20}(\log T)^{3/2}\}$ as $T\to\infty$.
This rate concerns comparison with the corresponding oracle benchmark;
total excess risk relative to $R(f^\star)$ also contains approximation error.
\end{corollary}
The proof in Appendix~\ref{app:gaussian-feasible} bounds Gaussian polynomial
moments. For example, even for independent standard Gaussian observations,
the variance of the empirical second moment of the degree-$p$ feature is
$n^{-1}\{\E Y_0^{4p}-(\E Y_0^{2p})^2\}$. Polynomial moments therefore
enter the analysis along with dimension. The next condition also covers
non-Gaussian series with controlled marginal moment growth.

\begin{corollary}[Slowly growing grids with controlled marginal moments]
\label{cor:loglog-grid}
Suppose $\{Y_t\}$ is strictly stationary and that, for positive
constants $C_5,C_6,C_7$,
$$
\alpha_Y(h)\leq C_5e^{-C_6h}\quad(h\geq1),\qquad
\|Y_0\|_{2j}\leq C_7\sqrt j\quad(j\geq1).
$$
Let $r_T$ be positive integers satisfying $r_T\to\infty$ and
$r_T=O(\log\log T)$. Take $s_T=\lfloor\kappa T\rfloor$
with $0<\kappa<1$ and define
\begin{equation}\label{eq:loglog-grid-general}
\mathcal G_T=\{1,\ldots,r_T\}\times\{0,\ldots,r_T\}
\times T^{-\xi}\mathcal C,
\qquad 0<\xi<\tfrac12,
\end{equation}
where $\mathcal C\subset(0,\infty)$ is fixed, finite and nonempty.
For all sufficiently large $T$, this grid satisfies all three parts of
Theorem~\ref{thm:growing-grid}. In particular, its selected refit is
risk consistent.
\end{corollary}
Appendix~\ref{app:loglog-grid} gives the proof. This corollary applies,
for example, to the stationary models $N_1,C_a$ and $J_a$
defined in Section~\ref{subsec:simulation-study}.
Appendix~\ref{app:nonlinear-benchmark} verifies their geometric mixing
and marginal moment bound. Taking $r_T=\lfloor\log\log T\rfloor$,
$\xi=1/4$, $\mathcal C=\{0.1,1,10\}$ and $s_T=\lfloor0.60T\rfloor$
in \eqref{eq:loglog-grid-general} therefore gives a risk-consistent
selected predictor for these stationary models.
Table~\ref{tab:assumptions} summarises the different regimes.

\begin{table}[H]
\centering
\caption{Candidate regimes, conclusions and sufficient conditions for the
principal asymptotic results in Section~\ref{sec:theoretical-results}. The
limits and rates are as $T\to\infty$.}
\label{tab:assumptions}
\footnotesize
\begin{tabularx}{\textwidth}{@{}p{0.14\textwidth}p{0.19\textwidth}p{0.27\textwidth}Y@{}}
\toprule
Regime & Candidate structure & Conclusion & Sufficient conditions \\
\midrule
Prespecified models
& $m_T,p_T,n_T\to\infty$; $\lambda_T\to0$
& Risk consistency: $R_T(\widehat f_T)-R(f^\star)$ converges in probability to $0$
& Stationarity, Carleman condition, and $\Delta_T(q)/\lambda_T\to0$ \\
Fixed grid
& $\mathcal G$ fixed; all $\lambda>0$
& Uniform score convergence, eventual selection in $\mathcal G^\star$, and selected-refit risk convergence to the fixed-grid population oracle
& Stationarity, ergodicity, finite feature second moments, and $s_T/T\to\kappa<1$ \\
Growing grid
& Deterministic finite $\mathcal G_T$
& Uniform score and refit oracle bounds; additionally, risk consistency when the grid contains an approximating sequence
& Stationarity and $\rho_T\sqrt{1+d_T}\to0$; for risk consistency also the Carleman condition and a candidate with increasing memory and degree and vanishing penalty \\
\bottomrule
\end{tabularx}
\end{table}

\section{Simulation study and data analysis}\label{sim}
We evaluate the one-step-ahead prediction performance of the closed-loop
Volterra method on simulated time series and compare its prediction errors
with those of alternative forecasting methods. We also evaluate the method
in a real-data analysis involving annual sunspot numbers. Prediction accuracy is
measured by the root mean squared error (RMSE),
\begin{equation}\label{eq:forecast-rmse}
\operatorname{RMSE}=\left\{\frac{1}{N_{\mathrm{ev}}}
\sum_{t\in\mathcal I_{\mathrm{ev}}}(Y_t-\widehat Y_t)^2\right\}^{1/2},
\end{equation}
where $\mathcal I_{\mathrm{ev}}$ is the set of evaluation times and
$N_{\mathrm{ev}}$ its cardinality.

\subsection{Simulation study}\label{subsec:simulation-study}
\noindent\textbf{Generating models.}
Let $Z_t$ be independent standard normal random variables, set
$\sigma=0.3$, and write $\varepsilon_t=\sigma Z_t$ for additive
innovations. The linear autoregressive (AR) and autoregressive
moving-average (ARMA) models considered here are
$$
L_1:\quad Y_t=0.60Y_{t-1}+\varepsilon_t,\qquad
L_2:\quad Y_t=0.55Y_{t-1}-0.25Y_{t-2}+\varepsilon_t,
$$
$$
L_3:\quad Y_t=0.50Y_{t-1}-0.20Y_{t-2}+0.10Y_{t-3}
+\varepsilon_t+0.40\varepsilon_{t-1}+0.20\varepsilon_{t-2}.
$$
For $(a,b)\in\{2,3\}^2$, the polynomial autoregressive models are
$$
I_{a,b}:\quad Y_t=0.35Y_{t-1}-0.10Y_{t-2}
+0.10Y_{t-1}^{a}Y_{t-2}^{b}+\varepsilon_t,
$$
$$
A_{a,b}:\quad Y_t=0.35Y_{t-1}+0.10Y_{t-2}^{a}
-0.10Y_{t-3}^{b}+\varepsilon_t.
$$
The autoregressive conditional heteroskedastic (ARCH) and generalised
ARCH (GARCH) models considered here are
$$
\begin{aligned}
H_1:&\quad Y_t=\sqrt{H_t}Z_t,\qquad
H_t=0.75\sigma^2+0.25Y_{t-1}^2,\\
H_2:&\quad Y_t=\sqrt{H_t}Z_t,\qquad
H_t=0.05\sigma^2+0.10Y_{t-1}^2+0.85H_{t-1}.
\end{aligned}
$$
We also consider the following nonlinear models:
$$
N_1:\quad Y_t=0.50\tanh(Y_{t-1})+\varepsilon_t,
$$
and, for $a\in\{0.30,0.60,0.90\}$,
$$
C_a:\quad Y_t=a\cos(2Y_{t-1})+\varepsilon_t,
$$
$$
J_a:\quad Y_t=a\tanh(2Y_{t-1})\tanh(2Y_{t-2})+\varepsilon_t.
$$

Exponential smooth-transition autoregressive (ESTAR) models are a
class of nonlinear autoregressive models whose coefficients vary
smoothly with the state of the process. \citet{terasvirta1994smooth}
studied the specification, estimation and evaluation of such models.
We consider the following restricted ESTAR specification:
$$
\mathrm{ESTAR}:\quad
Y_t=Y_{t-1}\exp(-2Y_{t-1}^2)+\varepsilon_t.
$$

The conditional mean is approximately linear near zero and attenuates
large lagged values.\\[0.05 in]
The stationary versions of $L_1,L_2,L_3,N_1,C_a,J_a$ and ESTAR satisfy
Assumptions~\ref{ass:stationarity and causality} and~\ref{ass:carleman}.
Appendix~\ref{app:nonlinear-benchmark} verifies the dependence and moment
conditions for the nonlinear models. The ARCH and GARCH models have
stationary versions with finite variance but fail the all-moments
condition. Together with the polynomial recursions, for which stationarity
is not assumed, they provide stress tests outside the verified assumptions.

\noindent\textbf{Simulation procedure.}
For each generating model and each training sample size
$T\in\{100,200,500\}$, we perform $100$ independent replications.
The first $1000$ observations are discarded as burn-in for each series.
Each replication uses a training series of length $T$ and an independent
realisation of the same model for evaluating $500$ one-step-ahead
predictions.\\[0.05 in]
For each Volterra estimator, the tuning grid is
$$
(m,p,\lambda)\in\{1,\ldots,6\}\times\{0,\ldots,6\}
\times\{10^{-6},10^{-5},\ldots,10^1\}.
$$
We select the tuning triple by the rolling-origin procedure in
Section~\ref{subsec:tuning-parameter-selection}, with first origin
$\lfloor0.60T\rfloor$.\\[0.05 in]
We compare six prediction methods: penalised-intercept Volterra,
unpenalised-intercept Volterra, linear ridge regression
\citep{hoerlKennard1970Ridge}, Gaussian radial-basis-function (RBF)
kernel ridge regression \citep{girosiJonesPoggio1995Regularization},
nearest-neighbour regression \citep{stone1977NonparametricRegression},
and autoregressive fractionally integrated moving-average (ARFIMA)
prediction \citep{hosking1981fractional}.
As explained in Remark~\ref{rem:scale-penalty}, the estimator studied
in this paper penalises the intercept. The unpenalised-intercept
variant is included to assess the effect of that penalty.\\[0.05 in]
Linear ridge uses $p=1$ and the same memory and penalty grids.
The RBF kernel is
$$
k_\ell(\boldsymbol x,\boldsymbol z)
=\exp\{-\|\boldsymbol x-\boldsymbol z\|_2^2/(2m\ell^2)\},
$$
with $m\in\{1,\ldots,6\}$,
$\ell\in\{0.05,0.10,0.25,0.50,1,2,4,8,16\}$ and the same penalty
grid; its intercept is unpenalised. Nearest-neighbour regression
averages the responses at the $k$ closest lag vectors, with
$m\in\{1,\ldots,6\}$ and $k\in\{3,5,10,20,40,80,160\}$ whenever the fitting sample contains
at least $k$ lag vectors. To account for the temporal ordering of the
observations, the tuning parameters of these three competing methods
are selected by the
rolling-origin procedure in Section~\ref{subsec:tuning-parameter-selection}.
ARFIMA is fitted using the \texttt{arfima} function in the
\texttt{forecast} package of R software, which selects the AR and
moving-average orders and estimates the fractional differencing
parameter over $[0,0.5]$.\\[0.05 in]
Section~\ref{subsec:simulation-selection} examines recovery of the
generating memory and polynomial degree in models where both have a
specific meaning, namely the linear and polynomial autoregressive models.
Section~\ref{subsec:simulation-results} compares the mean RMSEs of the
six methods over the $100$ replications.

\subsubsection{Selection of tuning parameters}
\label{subsec:simulation-selection}
Table~\ref{tab:simulation-selection} reports the tuning-parameter
selections at $T=500$ for the closed-loop Volterra method with penalised
and unpenalised intercepts, under two linear and two polynomial
autoregressive generating models. The generating degree $p$ is four for
$I_{2,2}$, because its interaction is $Y_{t-1}^2Y_{t-2}^2$, and two for
$A_{2,2}$.

\begin{table}[htbp]
\centering
\caption{Tuning-parameter selection for the closed-loop Volterra method
at $T=500$ over $100$ replications, under penalised and unpenalised
intercept specifications. The third column gives the generating
memory--degree pair $(m,p)$. The fourth column gives the most frequently
selected triple and its count. The last three columns give the
percentages selecting the correct memory, degree and pair.}
\label{tab:simulation-selection}
\footnotesize
\setlength{\tabcolsep}{4pt}
\begin{tabular}{@{}llccrrr@{}}
\toprule
Model & Intercept & \shortstack{True\\$(m,p)$} & \shortstack{Most selected triple\\$(m,p,\lambda)$ (count)}
& \shortstack{Correct\\$m$ (\%)} & \shortstack{Correct\\$p$ (\%)} & \shortstack{Correct\\$(m,p)$ (\%)} \\
\midrule
$L_1$ & Penalised & $(1,1)$ & $(1,1,10^{-6})$ (22) & 59 & 52 & 34 \\
 & Unpenalised & $(1,1)$ & $(1,1,10^{-6})$ (22) & 59 & 52 & 34 \\
\addlinespace[2pt]
$L_2$ & Penalised & $(2,1)$ & $(2,1,10^{-6})$ (14) & 53 & 61 & 32 \\
 & Unpenalised & $(2,1)$ & $(2,1,10^{-6})$ (14) & 54 & 61 & 32 \\
\addlinespace[2pt]
$I_{2,2}$ & Penalised & $(2,4)$ & $(2,1,10^{-6})$ (13) & 36 & 10 & 2 \\
 & Unpenalised & $(2,4)$ & $(2,1,10^{-6})$ (13) & 36 & 11 & 3 \\
\addlinespace[2pt]
$A_{2,2}$ & Penalised & $(3,2)$ & $(1,1,10^{-6})$ (13) & 12 & 27 & 5 \\
 & Unpenalised & $(3,2)$ & $(1,1,10^{-6})$ (13) & 12 & 27 & 5 \\
\bottomrule
\end{tabular}
\end{table}

For $L_1$ and $L_2$, the most frequently selected memory and degree
coincide with those of the generating models. For $I_{2,2}$ and
$A_{2,2}$, the most frequent selections have degree one, indicating
that rolling-origin validation favours lower-degree predictors for
these models at the considered sample size. The two intercept
implementations give similar selection frequencies across all four
models. Interestingly, the smallest penalisation parameter in the grid,
$\lambda=10^{-6}$, appears in the most frequently selected triple in
every case reported in Table~\ref{tab:simulation-selection}.

\subsubsection{Prediction accuracy}\label{subsec:simulation-results}
Table~\ref{tab:simulation-rmse-long} in
Appendix~\ref{app:complete-simulation-rmse} reports mean RMSE over the
$100$ replications, calculated from $500$ evaluation observations per
replication. Results are grouped by $T=100,200,500$ and rounded to two
decimal places. Across the generating models, the Volterra method remains competitive with the alternative prediction methods, particularly at $T=500$, where its mean RMSEs are generally comparable to or lower than those of the competing methods. \\[0.05 in]
At $T=500$, the mean RMSEs of Volterra and linear ridge are similar
for the linear and polynomial autoregressive models. For the highly
nonlinear models $C_{0.60}$, $C_{0.90}$, $J_{0.60}$ and $J_{0.90}$,
both Volterra implementations have lower mean RMSEs than linear ridge
and ARFIMA. RBF ridge has comparable results for the cosine models and
lower mean RMSEs for the interaction models. The two Volterra intercept
implementations have identical mean RMSEs to two decimal places at
$T=500$ for all reported models.

\subsection{Data analysis of annual sunspot numbers}\label{subsec:real-data}
We analyse annual mean total sunspot numbers for $1749$--$2024$
from the International Sunspot Number Version 2.0, published by the
World Data Center for the Sunspot Index and Long-term Solar Observations
(SILSO), Royal Observatory of Belgium \citep{silso2015v2}.
Although the yearly series starts in 1700, we begin in 1749 because
observations are sparser in the earlier period \citep{clette2016corrections}.
Sunspot forecasting has previously been studied with autoregressive
models \citep{lombAndersen1980Sunspot}.\\[0.05 in]
Writing $s_t$ for the annual mean in year $t$, we consider the time
series of scaled annual changes $\{y_t\}$ for our analysis, where
\begin{equation}\label{eq:sunspot-difference}
y_t=\frac{s_t-s_{t-1}}{100},\qquad t=1750,\ldots,2024.
\end{equation}
The first $200$ samples of this series ($1750$--$1949$) form the training
sample, and the remaining $75$ samples ($1950$--$2024$) form the
evaluation sample. The annual changes are divided by 100 before fitting the prediction models. A forecast $\widehat y_t$ gives the annual-mean forecast
$\widehat s_t=s_{t-1}+100\widehat y_t$, and prediction errors are
reported in sunspot-number units.\\[0.05 in]
We first check whether the stationarity assumption is reasonable for
the training sample using the augmented Dickey--Fuller (ADF) test and
the Kwiatkowski--Phillips--Schmidt--Shin (KPSS) test
\citep{dickey1979,kpss1992}. The null hypothesis of the ADF test is
that the series has a unit root, whereas the null hypothesis of the
KPSS test is level stationarity. For the ADF regression with an
intercept, the number of lagged differences is selected by the Akaike
information criterion (AIC) from $0,\ldots,20$; AIC selects seven lags.
The resulting ADF statistic is $-9.96$ ($p<0.001$). The KPSS statistic
is $0.027$ ($p>0.10$), with truncation lag four. Thus the unit-root
null is rejected, while the level-stationarity null is not rejected.
Both diagnostics therefore support modelling $\{y_t\}$ as a
stationary series.\\[0.05 in]
For both Volterra estimators we use
$m\in\{1,\ldots,6\}$, $p\in\{0,\ldots,6\}$ and the logarithmic
penalty grid $10^{-6},\ldots,10$. To give the competing methods a broad
tuning range, linear ridge uses $m=1,\ldots,20$ and the logarithmic
penalty grid $10^{-8},\ldots,10^2$, while RBF kernel ridge uses the
same memory and penalty grids together with
$$
\ell\in\{0.025,0.05,0.10,0.25,0.50,1,2,4,8,16,32\}.
$$
Nearest-neighbour regression uses $m\in\{1,\ldots,20\}$ and every
feasible number of neighbours at the first validation origin, and the
ordinary autoregressive candidates have orders $1,\ldots,10$.
Volterra, linear ridge, RBF ridge, nearest-neighbour regression and the
ordinary autoregressive models are tuned by the
rolling-origin procedure defined in Section~\ref{subsec:tuning-parameter-selection}.
The first origin is $r=120$, so that $60\%$ of the $200$ training observations
are available at the first fit; validation then proceeds through
$r=199$.\\[0.05 in]
For ARMA, we fit a model for each pair of autoregressive and
moving-average orders $(p,q)\in\{0,\ldots,10\}^2$. Each order pair is
fitted both with an estimated mean and with the mean fixed at zero.
We select the combination of orders and mean specification that minimises
the small-sample-corrected Akaike information criterion (AICc), computed
using the training sample.
ARFIMA is fitted in R software using \texttt{forecast::arfima}; its
order search allows autoregressive and moving-average orders up to ten
and estimates the fractional differencing parameter over $[0,0.5]$.
After model selection, the fitted models and their parameters are kept
fixed throughout the evaluation period. Each one-step-ahead forecast uses
the observed samples of $\{y_t\}$ available before the corresponding year.\\[0.05 in]
The selected Volterra triple is $(m,p,\lambda)=(5,3,10^{-2})$ for
both intercept specifications. Linear ridge selects
$(m,\lambda)=(9,10^{-2})$, RBF ridge selects
$(m,\ell,\lambda)=(8,0.25,10^{-2})$, and nearest-neighbour regression
selects $(m,k)=(3,6)$. The ordinary autoregressive model selects order
nine. AICc selects an ARMA$(5,2)$ model without a mean.
The ARFIMA fit gives
$(p,d,q)=(5,4.6\times10^{-5},2)$, so the estimated fractional
differencing parameter is essentially zero. Table~\ref{tab:sunspot-prediction}
reports root mean squared error (RMSE) and mean absolute error (MAE)
over the $75$ evaluation years.

\begin{table}[H]
\centering
\caption{One-step-ahead forecasts of annual sunspot numbers for
$1950$--$2024$, using the $200$ training samples from $1750$--$1949$. Errors are in sunspot-number units.}
\label{tab:sunspot-prediction}
\small
\begin{tabular}{@{}lrr@{}}
\toprule
Method & RMSE & MAE \\
\midrule
Volterra, penalised intercept & 24.32 & 18.38 \\
Volterra, unpenalised intercept & 24.31 & 18.37 \\
Linear ridge & 26.57 & 20.96 \\
RBF kernel ridge & 26.84 & 20.35 \\
Nearest neighbour & 26.11 & 19.15 \\
AR(9) & 26.84 & 21.32 \\
ARMA(5,2) & 26.00 & 20.74 \\
ARFIMA & 26.06 & 20.89 \\
\bottomrule
\end{tabular}
\end{table}

\begin{figure}[H]
\centering
\includegraphics[width=0.94\textwidth]{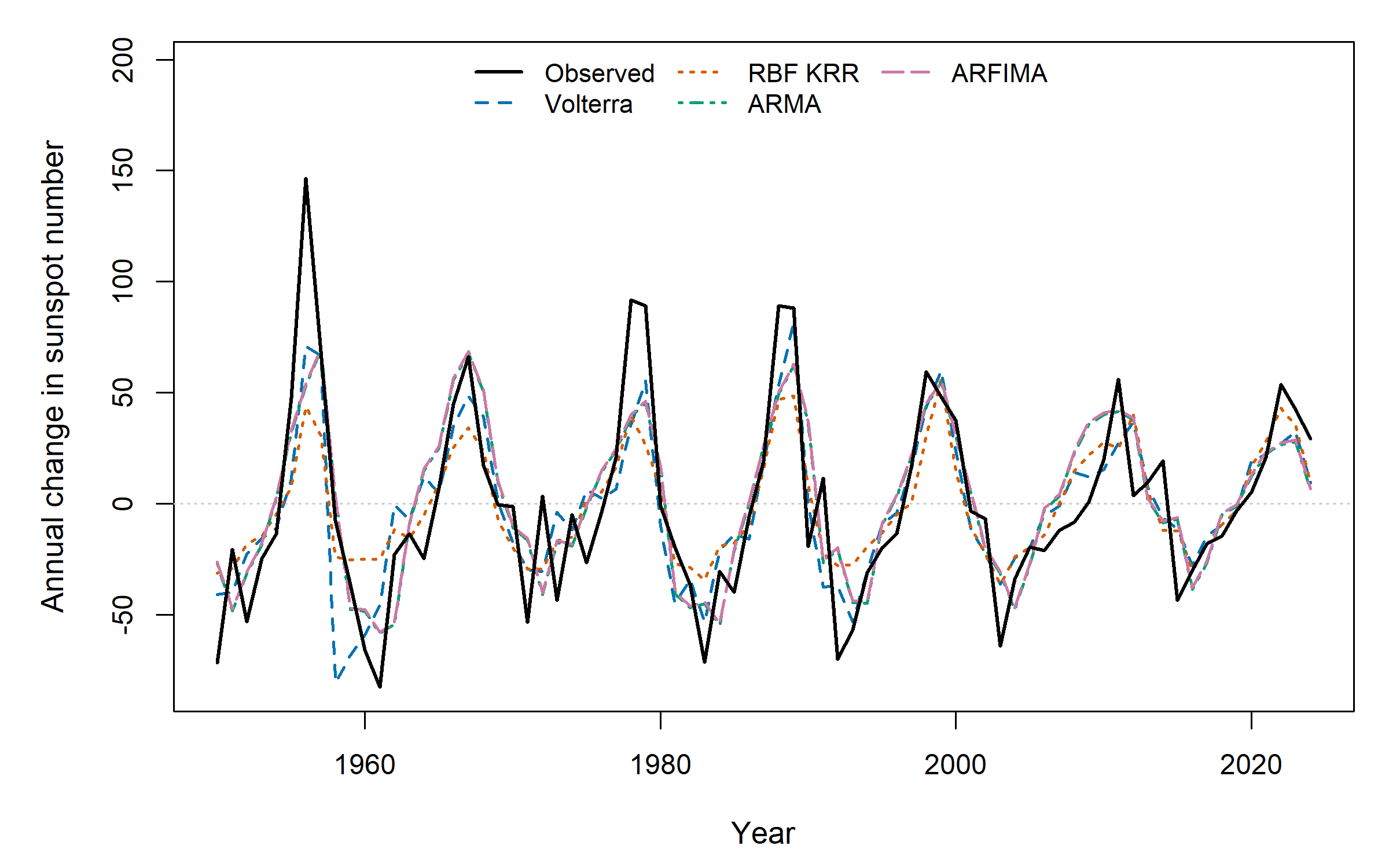}
\caption{Observed annual changes in sunspot numbers and one-step-ahead
forecasts, 1950--2024. The four forecasting methods are the
penalised-intercept Volterra estimator, radial-basis-function kernel
ridge regression (RBF KRR), ARMA and ARFIMA. All models are fitted to
the $200$ training samples from 1750--1949; each forecast uses
observations available before the corresponding year. Changes are
shown in sunspot-number units.}
\label{fig:sunspot-change-forecasts}
\end{figure}

The two Volterra estimators give nearly identical results and attain
the smallest RMSE and MAE among the fitted methods. Among the
competing methods, ARMA, ARFIMA and nearest-neighbour regression give
the smallest RMSEs, all close to $26$. Linear ridge, RBF ridge and
AR(9) have RMSEs between $26.57$ and $26.84$. Thus the advantage of Volterra
in this application remains after allowing the competing methods the
larger tuning ranges described above, although several alternatives
produce prediction errors of similar magnitude.\\[0.05 in]
Figure~\ref{fig:sunspot-change-forecasts} compares the observed annual
changes with forecasts from Volterra, RBF ridge, ARMA and ARFIMA. The
largest positive changes, particularly those in 1956 and the late
1970s, are underpredicted by all four methods.

\Needspace{18\baselineskip}
\section{Discussion}\label{conclusion}
We develop a statistical framework for the closed-loop Volterra method for
analysing and forecasting stationary time series. The main contribution is
the theoretical development of regularised Volterra prediction under
temporal dependence. By representing finite-memory Volterra predictors
through the polynomial kernel
$k^{(p)}(\boldsymbol{x},\boldsymbol{z})
=\sum_{n=0}^{p}(\boldsymbol{x}^{\top}\boldsymbol{z})^n$, the method combines
the classical Volterra representation with regularised kernel estimation,
while preserving the coefficient structure of the underlying Volterra
expansion. This formulation makes it possible to study approximation,
estimation and data-driven selection of the memory ($m$), polynomial degree ($p$) and
regularisation parameter ($\lambda$) within a common framework.\\[0.05 in]
The theoretical results provide a rigorous justification for this
approach. Under strict stationarity and a marginal Carleman condition, there exists a sequence of finite-memory Volterra polynomials whose squared-error prediction risks converge to the optimal prediction risk. For the regularised estimator,
we derive a finite-sample oracle inequality under temporal dependence and
establish risk consistency when $m$ and $p$ increase
and $\lambda$ decreases at suitable rates. We further
analyse rolling-origin selection of the tuning parameters. For fixed
candidate sets, the selected predictor asymptotically attains the smallest
population risk among the candidates, while for growing candidate sets we
obtain finite-sample oracle bounds and sufficient conditions for risk
consistency of the selected predictor. These results connect the
approximation capability of Volterra expansions with statistical
estimation and model selection, and provide a theoretical basis for using
closed-loop Volterra methods as data-driven nonlinear predictors for
dependent time series.\\[0.05 in]
The numerical experiments support the practical relevance of this
framework. The Volterra estimator remains competitive across a broad
collection of linear and nonlinear generating mechanisms and becomes
particularly useful when the conditional mean exhibits pronounced
nonlinear dependence. The application to annual sunspot numbers further
illustrates that the method can provide accurate forecasts in a real
time-series problem while retaining a structured polynomial
representation of the dependence on past observations.\\[0.05 in]
A natural continuation of this work is to sharpen the connection between
the approximation properties of Volterra expansions and the statistical
rates of the resulting estimator. In particular, under additional
regularity or fading-memory assumptions on the prediction functional, one
may derive explicit rates at which finite-memory, finite-degree Volterra
polynomials approximate the optimal predictor. Combining such
approximation rates with the finite-sample estimation bounds developed
here could lead to explicit excess-risk rates and provide further guidance
for the joint choice of the memory, polynomial degree and regularisation
level as the sample size increases.\\[0.05 in]
Overall, the results establish closed-loop Volterra prediction as a
theoretically justified and practically useful framework for nonlinear
time-series forecasting, combining the interpretability of polynomial
Volterra representations with regularised estimation and data-driven
model selection.

\clearpage
\appendix
\section*{Appendix}
\addcontentsline{toc}{section}{Appendix}
Appendices A--E provide proofs and supporting arguments for the main
results. Appendix F verifies the stationarity, mixing and moment
claims used for the simulated models. Appendix G reports the
simulation prediction errors.
\section{Proof of the polynomial-density counterexample}
\label{app:approximation}\label{app:obstruction}
For probability measures $\mu,\nu$, let
$\|\mu-\nu\|_{\mathrm{TV}}=\sup_A|\mu(A)-\nu(A)|$.
This appendix proves the counterexample stated in
Remark~\ref{rem:polynomial-obstruction}, including stationarity, geometric
mixing and the failure of polynomial density for the constructed process.

Using the density $w$ and functions $h,g$ defined in
Remark~\ref{rem:polynomial-obstruction}, let $Z\sim N(0,1)$. For every
integer $k\geq0$,
$$
\int_0^\infty x^kh(x)w(x)\,dx
=\E[e^{kZ}\sin(2\pi Z)]
=\operatorname{Im}\exp\{(k+2\pi i)^2/2\}
=e^{(k^2-4\pi^2)/2}\sin(2\pi k)=0.
$$
In particular, $\int_0^\infty h(x)w(x)\,dx=0$. Symmetry of $Z$ also gives
$\int_0^\infty g(y)w(y)\,dy=0$. Hence the transition density satisfies
$$
\int_0^\infty p(y\mid x)\,dy=1,\qquad
\int_0^\infty p(y\mid x)w(x)\,dx=w(y).
$$
The second identity proves that $w$ is invariant. Moreover,
$$
p(y\mid x)=(1-\rho/2)w(y)+(\rho/2)\widetilde p(y\mid x),\qquad
\widetilde p(y\mid x)=\{1+2h(x)g(y)\}w(y).
$$
Since $|h(x)|\leq1$ and $|g(y)|\leq1/2$, $\widetilde p(\,\cdot\mid x)$ is
also a probability density. Let $W$ denote the probability measure with
density $w$ and let $\widetilde P$ denote the transition kernel with density
$\widetilde p$. Thus $P=(1-\rho/2)W+(\rho/2)\widetilde P$, where
$W(x,A)=W(A)$ for every $x>0$. For probability measures $\mu,\nu$,
$$
\|\mu P-\nu P\|_{\mathrm{TV}}
=\frac{\rho}{2}\|\mu\widetilde P-\nu\widetilde P\|_{\mathrm{TV}}
\leq\frac{\rho}{2}\|\mu-\nu\|_{\mathrm{TV}}.
$$
The inequality follows directly from the definition of total variation:
for $|f|\leq1$, also $|\widetilde Pf|\leq1$.
Iteration and invariance of $W$ give, for every $x>0$ and $k\geq1$,
$$
\|P^k(x,\cdot)-W\|_{\mathrm{TV}}
=\|\delta_xP^k-WP^k\|_{\mathrm{TV}}
\leq(\rho/2)^k.
$$
This is the usual contraction argument under a uniform minorisation
condition; see \citet[Section~4.3]{RobertsRosenthal2004}.
It implies $\alpha_Y(k)\leq(\rho/2)^k$ for the stationary
Markov chain. It also proves uniqueness of the invariant distribution.
Its two-sided stationary extension is the process used in
Remark~\ref{rem:polynomial-obstruction}.\\[0.05 in]
The lognormal marginal distribution satisfies
$\E Y_0^r=e^{r^2/2}<\infty$ for every $r>0$.
For $Z\sim N(0,1)$, completing the square gives
$e^z\varphi(z)=e^{1/2}\varphi(z-1)$, where $\varphi$ is the standard
normal density. Hence
$$
\begin{aligned}
b&=\int_0^\infty y w(y)\,dy=e^{1/2},\\
c&=\int_0^\infty y g(y)w(y)\,dy
=e^{1/2}\{\Phi(1)-1/2\}>0.
\end{aligned}
$$
Integrating $y$ against \eqref{eq:counterexample-transition} therefore
gives
$$
\E(Y_0\mid Y_{-1})=b+\rho c h(Y_{-1}).
$$
To prove orthogonality, fix $m\geq1$ and integers
$a_1,\ldots,a_m\geq0$. In the joint density of
$(Y_{-m},\ldots,Y_{-1})$, write $x_j$ for $Y_{-j}$.
For every nonnegative integer $a$,
$$
\begin{aligned}
\int_0^\infty x^a p(y\mid x)w(x)\,dx
&=w(y)\int_0^\infty x^a\{1+\rho h(x)g(y)\}w(x)\,dx\\
&=w(y)\E Y_0^a,
\end{aligned}
$$
where the last equality follows from the polynomial orthogonality of $h$
proved above. Apply this identity successively to the integrations with
respect to $x_m,x_{m-1},\ldots,x_2$. The remaining integral contains
$$
\int_0^\infty x_1^{a_1}h(x_1)w(x_1)\,dx_1=0.
$$
Thus
$$
\E\left[h(Y_{-1})\prod_{j=1}^mY_{-j}^{a_j}\right]=0,
$$
and the same identity holds with the product replaced by any polynomial
$v(\boldsymbol X_0^{(m)})$. The exchanges of integration are justified by
absolute integrability: $p(y\mid x)\leq(1+\rho/2)w(y)$ and all
lognormal moments are finite.
Finally, by conditional-mean orthogonality and the preceding identity,
$$
\begin{aligned}
R_m(v)-R(f^\star)
&=\E\left[\{v(\boldsymbol X_0^{(m)})-b
-\rho c h(Y_{-1})\}^2\right]\\
&=\E\{v(\boldsymbol X_0^{(m)})-b\}^2
+\rho^2c^2\E h(Y_{-1})^2.
\end{aligned}
$$
Since $\E h(Y_{-1})^2=\{1-e^{-8\pi^2}\}/2$,
\eqref{eq:polynomial-obstruction} follows. Equality holds for $v=b$.

\section{Covariance and empirical-risk bounds}
\label{app:covariance}

This appendix proves Proposition~\ref{prop:unbounded-uniform-bound},
Theorem~\ref{prop:finite-sample-oracle}, and
Theorem~\ref{thm:unbounded-risk-consistency}. Before those proofs, we establish
two auxiliary lemmas that provide the covariance and block second-moment
bounds needed to control empirical risks under temporal dependence. The
constants and dependence sums are those defined in
\eqref{eq:finite-dependence-sum}.

\subsection{Auxiliary covariance bounds for the prescribed-model results}
The following two lemmas are used in the proofs of
Proposition~\ref{prop:unbounded-uniform-bound} and
Theorems~\ref{prop:finite-sample-oracle}--\ref{thm:unbounded-risk-consistency}.
For two sigma-fields $\mathcal A,\mathcal B$, write
$$
\alpha(\mathcal A,\mathcal B)=\sup_{A\in\mathcal A,\,B\in\mathcal B}
|\Pbb(A\cap B)-\Pbb(A)\Pbb(B)|.
$$
For a measurable lag functional $V_{t,T}=g_T(Y_t,\ldots,Y_{t-m_T})$,
its past and future sigma-fields are contained in those of $Y$ at the
corresponding times. Hence
$\alpha_{V,T}(h)\leq\alpha_Y(h-m_T)$ for $h>m_T$;
see \citet[Section~1.1]{bradley2005basic}.
\label{rem:mixing-and-moments}
The moment requirement is separate: \eqref{eq:polynomial-moment-sufficient}
controls the products arising from degree-$p$ features. The proofs below
retain both their moments and the shifted dependence coefficients.

\begin{lemma}[A dimension-free covariance bound]
\label{lem:alpha-covariance}
Let $\boldsymbol U,\boldsymbol V$ be $\Rbb^d$-valued random vectors,
for some integer $d\geq1$, measurable with respect to
$\mathcal A,\mathcal B$, respectively, and satisfying
$\E\|\boldsymbol U\|_2^q<\infty$ and
$\E\|\boldsymbol V\|_2^q<\infty$ for some $q>2$,
where $\|\cdot\|_2$ is the Euclidean norm.
With $\eta=1-2/q$ and $c_q=(\E|Z|^q)^{1/q}$ for $Z\sim N(0,1)$,
$$
|\E\langle \boldsymbol{U},\boldsymbol{V}\rangle
-\langle\E \boldsymbol{U},\E \boldsymbol{V}\rangle|
\leq12c_q^2\alpha(\mathcal A,\mathcal B)^\eta
\{\E\|\boldsymbol U\|_2^q\}^{1/q}
\{\E\|\boldsymbol V\|_2^q\}^{1/q}.
$$
\end{lemma}
\begin{proof}
Use the standard orthonormal basis of $\Rbb^d$ and
let $\gamma_d$ denote the standard Gaussian probability measure on
$\Rbb^d$. For every fixed deterministic $\boldsymbol a\in\Rbb^d$,
Davydov's scalar covariance inequality \citep{davydov1968} gives
$$
|\operatorname{Cov}(\boldsymbol a^\top\boldsymbol U,
\boldsymbol a^\top\boldsymbol V)|
\leq12\alpha(\mathcal A,\mathcal B)^\eta
\|\boldsymbol a^\top\boldsymbol U\|_q
\|\boldsymbol a^\top\boldsymbol V\|_q.
$$
Integrate this inequality with respect to $\gamma_d$. By Fubini's
theorem we may interchange the expectation over $\boldsymbol U,\boldsymbol V$
with the integral over $\boldsymbol a$. The identity
$\int \boldsymbol a\boldsymbol a^\top\,\gamma_d(d\boldsymbol a)=\mathbf I_d$
then gives
$$
\int \operatorname{Cov}(\boldsymbol a^\top\boldsymbol U,
\boldsymbol a^\top\boldsymbol V)\,\gamma_d(d\boldsymbol a)
=\E\langle\boldsymbol U,\boldsymbol V\rangle
-\langle\E\boldsymbol U,\E\boldsymbol V\rangle.
$$
The required absolute integrability follows from
$$
\E\int |(\boldsymbol a^\top\boldsymbol U)
(\boldsymbol a^\top\boldsymbol V)|\,\gamma_d(d\boldsymbol a)
\leq \E(\|\boldsymbol U\|_2\,\|\boldsymbol V\|_2)
\leq (\E\|\boldsymbol U\|_2^2)^{1/2}(\E\|\boldsymbol V\|_2^2)^{1/2}<\infty,
$$
and the corresponding term involving the means is finite as well. Hence
$$
\begin{aligned}
&|\E\langle\boldsymbol U,\boldsymbol V\rangle
-\langle\E\boldsymbol U,\E\boldsymbol V\rangle|\\
&\quad\leq12\alpha(\mathcal A,\mathcal B)^\eta
\int\|\boldsymbol a^\top\boldsymbol U\|_q
\|\boldsymbol a^\top\boldsymbol V\|_q\,\gamma_d(d\boldsymbol a).
\end{aligned}
$$
The Cauchy--Schwarz inequality for integration with respect to $\gamma_d$
bounds the last integral by the
product of the square roots of
$\int\|\boldsymbol a^\top\boldsymbol U\|_q^2\,d\gamma_d$ and its
analogue for $\boldsymbol V$. Since $2/q<1$, concavity of $x^{2/q}$ gives
$$
\begin{aligned}
\int\|\boldsymbol a^\top\boldsymbol U\|_q^2\,\gamma_d(d\boldsymbol a)
&\leq
\left\{\E\int|\boldsymbol a^\top\boldsymbol U|^q
\,\gamma_d(d\boldsymbol a)\right\}^{2/q}\\
&=c_q^2\{\E\|\boldsymbol U\|_2^q\}^{2/q},
\end{aligned}
$$
because, for every fixed $\boldsymbol u$,
$\boldsymbol a^\top\boldsymbol u$ is $N(0,\|\boldsymbol u\|_2^2)$ under
$\gamma_d$. The same bound holds for $\boldsymbol V$, and substitution
completes the proof.
\end{proof}

\begin{lemma}[Lagged second-moment matrices]
\label{lem:lag-variance}
Let $q>2$, let $\boldsymbol{Z}_t$ be a stationary vector-valued measurable
function of $(Y_t,\ldots,Y_{t-m})$, and define
$$
K(q)=\left\|\|\boldsymbol Z_0\|_2^2\right\|_q<\infty,
\qquad
\mathbf{\Gamma}=\E(\boldsymbol Z_0\boldsymbol Z_0^\top).
$$
For a block of $n$ consecutive time indices
$\mathcal I=\{r+1,\ldots,r+n\}$, define its empirical second-moment matrix by
$$
\widehat{\mathbf\Gamma}_{\mathcal I}
=\frac1n\sum_{t\in\mathcal I}\boldsymbol Z_t\boldsymbol Z_t^\top.
$$
With $B_{n,m}(q)$ defined in \eqref{eq:finite-dependence-sum},
\begin{equation}
\E\|\widehat{\mathbf\Gamma}_{\mathcal I}-\mathbf\Gamma\|_{\mathrm F}^2
\leq \frac{B_{n,m}(q)K(q)^2}{n}.
\label{eq:matrix-block-bound}
\end{equation}
The same argument applies to a scalar stationary lag functional $V_t$,
with $K(q)$ replaced by $\|V_0\|_q$, and bounds the variance of its
block average.
\end{lemma}
\begin{proof}
Put $\mathbf A_t=\boldsymbol Z_t\boldsymbol Z_t^\top$ and
$\mathbf W_t=\mathbf A_t-\E\mathbf A_t$. Since
$\|\mathbf A_t\|_{\mathrm F}=\|\boldsymbol Z_t\|_2^2$,
$$
\E\|\mathbf W_t\|_{\mathrm F}^2
=\E\|\mathbf A_t\|_{\mathrm F}^2-\|\E\mathbf A_t\|_{\mathrm F}^2
\leq \E\|\mathbf A_t\|_{\mathrm F}^2
\leq K(q)^2.
$$
For $1\leq h\leq m$, the Cauchy--Schwarz inequality for the Frobenius
inner product \citep[Section~5.7]{hornjohnson2012matrix}, followed by the
scalar Cauchy--Schwarz inequality, gives
$$
|\E\langle\mathbf W_0,\mathbf W_h\rangle_{\mathrm F}|
\leq
\{\E\|\mathbf W_0\|_{\mathrm F}^2\}^{1/2}
\{\E\|\mathbf W_h\|_{\mathrm F}^2\}^{1/2}
\leq K(q)^2.
$$
For $h>m$, $\mathbf A_0$ and $\mathbf A_h$ are measurable with respect to
sigma-fields separated by $h-m$. Applying Lemma~\ref{lem:alpha-covariance}
in the Hilbert space of matrices with the Frobenius inner product yields
$$
|\E\langle\mathbf W_0,\mathbf W_h\rangle_{\mathrm F}|
\leq12c_q^2\alpha_Y(h-m)^\eta K(q)^2.
$$
For $\mathcal I=\{r+1,\ldots,r+n\}$, expanding the squared Frobenius norm
and grouping cross-products by their lag gives, by stationarity,
$$
\E\left\|\sum_{t\in\mathcal I}\mathbf W_t\right\|_{\mathrm F}^2
=n\E\|\mathbf W_0\|_{\mathrm F}^2
+2\sum_{h=1}^{n-1}(n-h)
\E\langle\mathbf W_0,\mathbf W_h\rangle_{\mathrm F},
$$
because a block of length $n$ contains exactly $n-h$ pairs of indices
separated by lag $h$. Splitting the last sum at $m$ and using the two
preceding bounds gives
$$
\E\left\|\sum_{t\in\mathcal I}\mathbf W_t\right\|_{\mathrm F}^2
\leq nB_{n,m}(q)K(q)^2.
$$
Division by $n^2$ proves \eqref{eq:matrix-block-bound}. The scalar assertion
follows from the same stationary covariance-sum expansion.
\end{proof}

\subsection{Proofs of the prescribed-model risk bounds}
\label{app:uniform-risk}
\label{app:oracle-proof}
\label{app:prescribed-proof}
This subsection now uses Lemmas~\ref{lem:alpha-covariance} and
\ref{lem:lag-variance} to prove Proposition~\ref{prop:unbounded-uniform-bound},
Theorem~\ref{prop:finite-sample-oracle}, and
Theorem~\ref{thm:unbounded-risk-consistency}, in that order.

\begin{proof}[Proof of Proposition~\ref{prop:unbounded-uniform-bound}]
For fixed $T$, define the augmented column vector
$$
\boldsymbol{Z}_{t,T}=\begin{pmatrix}
Y_t\\
\boldsymbol{\psi}_{m_T,p_T}(\boldsymbol{X}_t^{(T)})
\end{pmatrix}\in\Rbb^{D_{m_T,p_T}+1}.
$$
Its population second-moment matrix and empirical second-moment matrix over
the training block $\mathcal I_T=\{m_T+1,\ldots,T\}$ are, respectively,
$$
\mathbf{\Gamma}_T=\E(\boldsymbol Z_{0,T}\boldsymbol Z_{0,T}^\top),
\qquad
\widehat{\mathbf{\Gamma}}_T
=\frac1{n_T}\sum_{t=m_T+1}^T
\boldsymbol Z_{t,T}\boldsymbol Z_{t,T}^\top.
$$
For $f_{\boldsymbol{\theta}}=\boldsymbol{\theta}^\top\boldsymbol{\psi}_{m_T,p_T}$ and
$\boldsymbol{a}_{\boldsymbol{\theta}}=(1,-\boldsymbol{\theta}^\top)^\top$,
$$
\widehat R_T(f_{\boldsymbol{\theta}})-R_T(f_{\boldsymbol{\theta}})=\boldsymbol{a}_{\boldsymbol{\theta}}^\top(\widehat{\mathbf{\Gamma}}_T-\mathbf{\Gamma}_T)\boldsymbol{a}_{\boldsymbol{\theta}}.
$$
The supremum is therefore at most $(1+\Lambda^2)\|\widehat{\mathbf{\Gamma}}_T-\mathbf{\Gamma}_T\|_{\mathrm F}$. By Lemma~\ref{lem:feature-representation},
$\|\boldsymbol{Z}_{0,T}\|_2^2
=Y_0^2+k^{(p_T)}(\boldsymbol{X}_0^{(T)},
\boldsymbol{X}_0^{(T)})$.
Lemma~\ref{lem:lag-variance} therefore bounds the second moment of
$\widehat{\mathbf\Gamma}_T-\mathbf\Gamma_T$ by $\Delta_T(q)^2$. Markov's inequality completes the proof.
\end{proof}

\begin{proof}[Proof of Theorem~\ref{prop:finite-sample-oracle}]
Let $f_{T,\lambda}$ minimise the population penalised criterion. Since the
zero function belongs to $\cH_T$,
$$
R_T(f_{T,\lambda})+\lambda\|f_{T,\lambda}\|_{\cH_T}^2
\leq R_T(0)=\mu_2,
$$
so $\lambda\|f_{T,\lambda}\|_{\cH_T}^2\leq\mu_2<b$.
On the event $A=\{\widehat R_T(0)\leq b\}$, the definition of
$\widehat f_{T,\lambda}$ gives
$$
\widehat R_T(\widehat f_{T,\lambda})
+\lambda\|\widehat f_{T,\lambda}\|_{\cH_T}^2
\leq \widehat R_T(0)\leq b,
$$
and hence $\lambda\|\widehat f_{T,\lambda}\|_{\cH_T}^2\leq b$.
Thus both functions lie in the ball of radius $\sqrt{b/\lambda}$. Define
$$
\delta_T=\sup_{f\in\cH_T:\,\|f\|_{\cH_T}\leq\sqrt{b/\lambda}}
|\widehat R_T(f)-R_T(f)|.
$$
On $A$,
$$
\begin{aligned}
R_T(\widehat f_{T,\lambda})
+\lambda\|\widehat f_{T,\lambda}\|_{\cH_T}^2
&\leq \widehat R_T(\widehat f_{T,\lambda})
+\lambda\|\widehat f_{T,\lambda}\|_{\cH_T}^2+\delta_T\\
&\leq \widehat R_T(f_{T,\lambda})
+\lambda\|f_{T,\lambda}\|_{\cH_T}^2+\delta_T\\
&\leq R_T(f_{T,\lambda})
+\lambda\|f_{T,\lambda}\|_{\cH_T}^2+2\delta_T\\
&=R(f^\star)+\mathcal A_T(\lambda)+2\delta_T.
\end{aligned}
$$
Since $\lambda\|\widehat f_{T,\lambda}\|_{\cH_T}^2\geq0$,
the preceding display implies, on $A$,
$$
R_T(\widehat f_{T,\lambda})-R(f^\star)
\leq\mathcal A_T(\lambda)+2\delta_T.
$$
Proposition~\ref{prop:unbounded-uniform-bound} yields
$$
\Pbb\{\delta_T>(1+b/\lambda)u\}\leq\frac{\Delta_T(q)^2}{u^2}.
$$
Also,
$A^c=\{\widehat R_T(0)-R_T(0)>b-\mu_2\}$, so the same proposition with
radius zero gives
$$
\Pbb(A^c)\leq\frac{\Delta_T(q)^2}{(b-\mu_2)^2}.
$$
The union bound proves the theorem.
\end{proof}

\begin{proof}[Proof of Theorem~\ref{thm:unbounded-risk-consistency}]
Fix $\epsilon>0$. By Theorem~\ref{prop:Volterrra-approx} there is a
polynomial $v$, depending on finitely many lags and having finite degree,
such that its prediction risk is less than $R(f^\star)+\epsilon$.
Since $m_T,p_T\to\infty$, this polynomial belongs to $\cH_T$ for
all sufficiently large $T$. By Lemma~\ref{lem:volterra-embedding}, its
RKHS norm is unchanged when considered as an element of $\cH_T$.
Consequently,
$$
0\leq\mathcal A_T(\lambda_T)
\leq\epsilon+\lambda_T\|v\|_{\cH_T}^2.
$$
The norm of this fixed polynomial is finite and $\lambda_T\to0$.
Since $\epsilon$ is arbitrary, $\mathcal A_T(\lambda_T)\to0$.\\[0.05 in]
Take $u_T=\sqrt{\lambda_T\Delta_T(q)}+\lambda_T^2$ and fix
$b>\mu_2$; the second term ensures $u_T>0$ even if $\Delta_T(q)=0$. The assumptions give
$$
\frac{u_T}{\lambda_T}\to0,
\qquad \frac{\Delta_T(q)}{u_T}\to0,
\qquad\Delta_T(q)\to0.
$$
Theorem~\ref{prop:finite-sample-oracle} gives
$$
\Pbb\{R_T(\widehat f_T)-R(f^\star)
>\mathcal A_T(\lambda_T)+2(1+b/\lambda_T)u_T\}
\longrightarrow0.
$$
The threshold on the left tends to zero. For any fixed $\varepsilon>0$,
it is eventually smaller than $\varepsilon$, and hence
$\Pbb\{R_T(\widehat f_T)-R(f^\star)>\varepsilon\}\to0$.
\end{proof}

\begin{proof}[Proof of Remark~\ref{rem:explicit-mixing-consistency}]
\label{app:explicit-mixing-consistency}
Take $q=4$ and first use the explicit choice
$r_T=m_T=p_T=\lfloor\log\log T\rfloor$ in
\eqref{eq:polynomial-mixing-choices}. Since $m_T=r_T$, the training-sample
size defined in Section~\ref{subsec:ridge-estimation} satisfies
$n_T=T-r_T\sim T$.
The moment-growth assumption gives
$\|Y_0\|_{8r_T}\leq2K\sqrt{r_T}$. By stationarity and Minkowski's
inequality,
$$
\left\|\|\boldsymbol X_0^{(r_T)}\|_2^2\right\|_{4r_T}
\leq r_T\|Y_0\|_{8r_T}^2
\leq4K^2r_T^2.
$$
For $u\geq0$, $\sum_{d=0}^{r_T}u^d\leq(1+u)^{r_T}$. Hence, using
$\mathsf M_T(4)$ from \eqref{feature:moment},
$$
\mathsf M_T(4)
\leq\|Y_0\|_8^2+
\left\|\{1+\|\boldsymbol X_0^{(r_T)}\|_2^2\}^{r_T}\right\|_4
\leq C+\{1+4K^2r_T^2\}^{r_T}
=T^{o(1)}.
$$
Since $\alpha_Y(h)=O(h^{-a})$, there is a finite constant $C_\alpha$ such
that $\alpha_Y(h)\leq C_\alpha h^{-a}$ for all $h\geq1$. Therefore, for
$q=4$,
$$
B_{n_T,r_T}(4)\leq1+2r_T+
24c_4^2C_\alpha^{1/2}\sum_{h=1}^{n_T}h^{-a/2}.
$$
Integral comparison bounds the sum by a constant when $a>2$, by
$1+\log n_T$ when $a=2$, and by a constant multiple of
$n_T^{1-a/2}$ when $0<a<2$. Consequently,
$$
\sqrt{\frac{B_{n_T,r_T}(4)}{n_T}}
\leq T^{-\min\{1/2,a/4\}+o(1)}
=T^{-2\gamma_a+o(1)}.
$$
Combining the last two displays with $\lambda_T=T^{-\gamma_a}$ gives
$$
\frac{\Delta_T(4)}{\lambda_T}
\leq T^{-\gamma_a+o(1)}\longrightarrow0.
$$
Together with $m_T,p_T\to\infty$ and $\lambda_T\to0$, this verifies the
conditions of Theorem~\ref{thm:unbounded-risk-consistency}. Under exponential
mixing, $\sum_{h\geq1}\alpha_Y(h)^{1/2}<\infty$, so
$B_{n_T,r_T}(4)=O(r_T+1)$ and the same argument with
$\lambda_T=T^{-1/4}$ gives $\Delta_T(4)/\lambda_T\to0$.\\[0.05 in]
For completeness, the quantitative bound
$\|Y_0\|_{2j}\leq K\sqrt j$ can be removed at the cost of a
population-moment-dependent sequence. Under Assumption~\ref{ass:carleman},
define, for sufficiently large $T$,
\begin{equation}\label{eq:moment-adapted-memory}
r_T=\max\left\{r\in\mathbb N:
 r\leq\lfloor\log\log T\rfloor,\quad
 \|Y_0\|_{8r}\leq\log T\right\},
\qquad m_T=p_T=r_T.
\end{equation}
The defining set is eventually nonempty, and every fixed positive integer
$r$ eventually belongs to it because $\|Y_0\|_{8r}<\infty$ while
$\log T\to\infty$. Thus $r_T\to\infty$. Since $m_T=r_T$, the previously
defined training-sample size again satisfies $n_T=T-r_T\sim T$.
Moreover,
$$
\left\|\|\boldsymbol X_0^{(r_T)}\|_2^2\right\|_{4r_T}
\leq r_T(\log T)^2,
$$
and therefore
$$
\mathsf M_T(4)
\leq\|Y_0\|_8^2+
\{1+r_T(\log T)^2\}^{r_T}=T^{o(1)}.
$$
The preceding mixing calculation is unchanged, so the same choice
$\lambda_T=T^{-\gamma_a}$ yields risk consistency. This construction depends
on unknown population moments and serves only as a theoretical feasibility
argument under the weaker Carleman condition.
\end{proof}

\section{Proof of fixed-grid selection}
\label{app:fixed-grid}
We first prove Theorem~\ref{thm:rolling-origin-selection-consistency}.
The population-risk formula in Remark~\ref{rem:fixed-grid-benchmark}
is derived afterwards.

\subsection{Proof of Theorem~\ref{thm:rolling-origin-selection-consistency}}
The notation for the candidate features and population risk is given in
\eqref{eq:candidate-feature-moments} and
\eqref{eq:population-ridge-predictor}.
\begin{proof}
Fix $\boldsymbol{\tau}=(m,p,\lambda)\in\mathcal G$ and use the
sample moments defined in \eqref{eq:empirical-feature-matrix}.
The moment condition \eqref{eq:fixed-grid-moment} and the Cauchy--Schwarz inequality
imply integrability of every entry. Birkhoff's ergodic theorem \citep[Chapter~11]{EisnerEtAl2015} gives
$$
\widehat{\mathbf G}_{r,\boldsymbol{\tau}}\to\mathbf G_{\boldsymbol{\tau}},
\qquad
\widehat{\boldsymbol g}_{r,\boldsymbol{\tau}}\to\boldsymbol g_{\boldsymbol{\tau}}
\qquad\text{almost surely as }r\to\infty,
$$
where the population quantities are defined in
\eqref{eq:candidate-feature-moments}. Since $\lambda>0$,
\begin{equation}\label{eq:fixed-grid-coefficient-convergence}
\widehat{\boldsymbol\theta}_{r,\boldsymbol{\tau}}
=(\widehat{\mathbf G}_{r,\boldsymbol{\tau}}+\lambda\mathbf I)^{-1}
\widehat{\boldsymbol g}_{r,\boldsymbol{\tau}}
\longrightarrow
\boldsymbol\theta_{\boldsymbol{\tau}}
=(\mathbf G_{\boldsymbol{\tau}}+\lambda\mathbf I)^{-1}
\boldsymbol g_{\boldsymbol{\tau}}
\qquad\text{almost surely}.
\end{equation}
The population predictor $f_{\boldsymbol{\tau}}$ and
$Q(\boldsymbol{\tau})=R_m(f_{\boldsymbol{\tau}})$ are defined in
\eqref{eq:population-ridge-predictor}. Put
$$
Z_{r,\boldsymbol{\tau}}=
\{Y_{r+1}-f_{\boldsymbol{\tau}}(\boldsymbol X_{r+1}^{(m)})\}^2.
$$
Then $\{Z_{r,\boldsymbol{\tau}}\}$ is stationary and ergodic and
$\E Z_{0,\boldsymbol{\tau}}=Q(\boldsymbol{\tau})<\infty$. For $n\geq1$, define
$$
S_n=\sum_{r=1}^{n}\{Z_{r,\boldsymbol\tau}-Q(\boldsymbol\tau)\},
\qquad S_0=0.
$$
Birkhoff's theorem gives $S_n/n\to0$ almost surely. It follows that
$\max_{0\leq n\leq T}|S_n|/T\to0$ almost surely: for any
$\epsilon>0$, there is an integer $n_0$ such that
$|S_n|\leq\epsilon n$ for $n\geq n_0$, while the maximum of the
finitely many $|S_n|$ with $n<n_0$ is finite. Since $T-s_T\sim
(1-\kappa)T$, we obtain
$$
\begin{aligned}
&\left|\frac1{T-s_T}\sum_{r=s_T}^{T-1}
\{Z_{r,\boldsymbol\tau}-Q(\boldsymbol\tau)\}\right|\\
&\qquad=\left|\frac{S_{T-1}-S_{s_T-1}}{T-s_T}\right|
\leq\frac{2\max_{0\leq n\leq T}|S_n|}{T-s_T}\longrightarrow0
\qquad\text{almost surely}.
\end{aligned}
$$
Therefore,
\begin{equation}\label{eq:fixed-grid-population-loss-average}
\frac1{T-s_T}\sum_{r=s_T}^{T-1}Z_{r,\boldsymbol\tau}
\longrightarrow Q(\boldsymbol\tau)
\qquad\text{almost surely}.
\end{equation}

Let
$$
A_{r,\boldsymbol{\tau}}=
\|\widehat{\boldsymbol\theta}_{r,\boldsymbol{\tau}}
-\boldsymbol\theta_{\boldsymbol{\tau}}\|_2,
\quad
e_{r,\boldsymbol{\tau}}=Y_{r+1}-f_{\boldsymbol{\tau}}(\boldsymbol X_{r+1}^{(m)}),
$$
and
$$
d_{r,\boldsymbol{\tau}}=
(\widehat{\boldsymbol\theta}_{r,\boldsymbol{\tau}}
-\boldsymbol\theta_{\boldsymbol{\tau}})^\top
\boldsymbol{\mathit{\Psi}}_{r+1,\boldsymbol{\tau}}.
$$
Then
$$
\{Y_{r+1}-\widehat f_{r,\boldsymbol{\tau}}(\boldsymbol X_{r+1}^{(m)})\}^2
-Z_{r,\boldsymbol{\tau}}
=d_{r,\boldsymbol{\tau}}^2-2e_{r,\boldsymbol{\tau}}d_{r,\boldsymbol{\tau}},
$$
and
$$
|d_{r,\boldsymbol{\tau}}|
\leq A_{r,\boldsymbol{\tau}}
\|\boldsymbol{\mathit{\Psi}}_{r+1,\boldsymbol{\tau}}\|_2.
$$
Take an event of probability one on which both the coefficient limit
\eqref{eq:fixed-grid-coefficient-convergence} and the ergodic averages
of $Z_{r,\boldsymbol\tau}$, $\|\boldsymbol{\mathit{\Psi}}_{r+1,\boldsymbol\tau}\|_2^2$,
and $\|\boldsymbol{\mathit{\Psi}}_{r+1,\boldsymbol\tau}\|_2|e_{r,\boldsymbol\tau}|$
converge. Such an event exists because the intersection of these four
probability-one events has probability one.
On this event, $A_{r,\boldsymbol{\tau}}\to0$.
Fix $\varepsilon>0$ and choose
$r_\varepsilon$ so that $A_{r,\boldsymbol{\tau}}\leq\varepsilon$ for
$r\geq r_\varepsilon$. Then
\begin{align*}
&\left|\operatorname{RO}_T(\boldsymbol{\tau})
-\frac1{T-s_T}\sum_{r=s_T}^{T-1}Z_{r,\boldsymbol{\tau}}\right|\\
&\leq \frac1{T-s_T}\sum_{\substack{s_T\leq r<r_\varepsilon\\r\leq T-1}}
|d_{r,\boldsymbol{\tau}}^2-2e_{r,\boldsymbol{\tau}}d_{r,\boldsymbol{\tau}}|\\
&\quad+\varepsilon^2\frac1{T-s_T}
\sum_{r=\max(s_T,r_\varepsilon)}^{T-1}
\|\boldsymbol{\mathit{\Psi}}_{r+1,\boldsymbol{\tau}}\|_2^2\\
&\quad+2\varepsilon\frac1{T-s_T}
\sum_{r=\max(s_T,r_\varepsilon)}^{T-1}
\|\boldsymbol{\mathit{\Psi}}_{r+1,\boldsymbol{\tau}}\|_2
|e_{r,\boldsymbol{\tau}}|.
\end{align*}
The first term tends to zero because it contains only finitely many fixed
summands. By \eqref{eq:fixed-grid-moment} and the Cauchy--Schwarz inequality,
$$
\E\|\boldsymbol{\mathit{\Psi}}_{0,\boldsymbol{\tau}}\|_2^2<\infty,
\qquad
\E\!\left[\|\boldsymbol{\mathit{\Psi}}_{0,\boldsymbol{\tau}}\|_2
|Y_0-f_{\boldsymbol{\tau}}(\boldsymbol X_0^{(m)})|\right]<\infty.
$$
Applying the same tail-average argument to these stationary ergodic
sequences gives, almost surely,
\begin{align*}
\limsup_{T\to\infty}
\left|\operatorname{RO}_T(\boldsymbol{\tau})
-\frac1{T-s_T}\sum_{r=s_T}^{T-1}Z_{r,\boldsymbol{\tau}}\right|
\leq{}&\varepsilon^2
\E\|\boldsymbol{\mathit{\Psi}}_{0,\boldsymbol{\tau}}\|_2^2\\
&+2\varepsilon\E\!\left[
\|\boldsymbol{\mathit{\Psi}}_{0,\boldsymbol{\tau}}\|_2
|Y_0-f_{\boldsymbol{\tau}}(\boldsymbol X_0^{(m)})|\right].
\end{align*}
Letting $\varepsilon\downarrow0$ and using
\eqref{eq:fixed-grid-population-loss-average},
$$
\operatorname{RO}_T(\boldsymbol{\tau})-Q(\boldsymbol{\tau})\to0
\qquad\text{almost surely}.
$$
Since $\mathcal G$ is finite,
\begin{equation}\label{eq:fixed-grid-uniform-score}
\max_{\boldsymbol{\tau}\in\mathcal G}
|\operatorname{RO}_T(\boldsymbol{\tau})-Q(\boldsymbol{\tau})|
\longrightarrow0\qquad\text{almost surely}.
\end{equation}

For each candidate, \eqref{eq:fixed-grid-coefficient-convergence}
with $r=T$ gives
$\widehat{\boldsymbol\theta}_{T,\boldsymbol\tau}
\to\boldsymbol\theta_{\boldsymbol\tau}$ almost surely.
Expanding the squared-error risk and substituting this limit gives
\begin{align*}
R_m(\widehat f_{T,\boldsymbol{\tau}})
&=\mu_2-2\widehat{\boldsymbol\theta}_{T,\boldsymbol{\tau}}^\top
\boldsymbol g_{\boldsymbol{\tau}}
+\widehat{\boldsymbol\theta}_{T,\boldsymbol{\tau}}^\top
\mathbf G_{\boldsymbol{\tau}}
\widehat{\boldsymbol\theta}_{T,\boldsymbol{\tau}}\\
&\longrightarrow
\mu_2-2\boldsymbol\theta_{\boldsymbol{\tau}}^\top
\boldsymbol g_{\boldsymbol{\tau}}
+\boldsymbol\theta_{\boldsymbol{\tau}}^\top
\mathbf G_{\boldsymbol{\tau}}\boldsymbol\theta_{\boldsymbol{\tau}}
=Q(\boldsymbol{\tau})
\qquad\text{almost surely}
\end{align*}
Thus
\begin{equation}\label{eq:fixed-grid-uniform-refit}
\max_{\boldsymbol{\tau}\in\mathcal G}
|R_m(\widehat f_{T,\boldsymbol{\tau}})-Q(\boldsymbol{\tau})|
\longrightarrow0\qquad\text{almost surely}.
\end{equation}

Let
$$
Q^\star=\min_{\boldsymbol{\tau}\in\mathcal G}Q(\boldsymbol{\tau}),
\qquad
\mathcal G^\star=\argmin_{\boldsymbol{\tau}\in\mathcal G}Q(\boldsymbol{\tau}).
$$
If $\mathcal G\setminus\mathcal G^\star\neq\varnothing$, put
$$
d=\min_{\boldsymbol{\tau}\in\mathcal G\setminus\mathcal G^\star}
\{Q(\boldsymbol{\tau})-Q^\star\}>0.
$$
Whenever the maximum in \eqref{eq:fixed-grid-uniform-score} is less than
$d/3$, for $\boldsymbol{\tau}\notin\mathcal G^\star$ and
$\boldsymbol{\tau}^\star\in\mathcal G^\star$,
$$
\operatorname{RO}_T(\boldsymbol{\tau})>Q^\star+\frac{2d}{3},
\qquad
\operatorname{RO}_T(\boldsymbol{\tau}^\star)<Q^\star+\frac d3.
$$
Hence $\widehat{\boldsymbol{\tau}}_T\in\mathcal G^\star$ eventually almost surely.
If $\mathcal G^\star=\mathcal G$, this conclusion is immediate. Therefore,
using \eqref{eq:fixed-grid-uniform-refit},
$$
\left|R_{\widehat m_T}(\widehat f_T^{\mathrm{RO}})-Q^\star\right|
\leq
\max_{\boldsymbol{\tau}\in\mathcal G}
|R_m(\widehat f_{T,\boldsymbol{\tau}})-Q(\boldsymbol{\tau})|
\longrightarrow0
\qquad\text{almost surely}
$$
Finally, Lemma~\ref{lem:feature-representation} and the kernel definition
\eqref{e2} give
$\|\boldsymbol\psi_{m,p}(\boldsymbol X_0^{(m)})\|_2^2
=k^{(p)}(\boldsymbol X_0^{(m)},\boldsymbol X_0^{(m)})
=\sum_{j=0}^p\|\boldsymbol X_0^{(m)}\|_2^{2j}$.
For a fixed finite $m$ and $p$, this sum has finite expectation when
$\E|Y_0|^{2p_*}<\infty$, by stationarity and
$\|\boldsymbol X_0^{(m)}\|_2^{2j}
\leq m^{j-1}\sum_{i=1}^m|Y_{-i}|^{2j}$ for $j\geq1$.
The term $\E Y_0^2$ is finite under the same assumption.
Thus \eqref{eq:fixed-grid-moment} holds.
\end{proof}

\subsection{Derivation of the population-risk formula}
\label{app:fixed-grid-risk-decomposition}
Fix $\boldsymbol{\tau}=(m,p,\lambda)$ and write
$f_{\boldsymbol{\theta}}(\boldsymbol{x})
=\boldsymbol{\theta}^{\top}\boldsymbol{\psi}_{m,p}(\boldsymbol{x})$.
Using \eqref{eq:candidate-feature-moments},
\begin{equation}\label{eq:population-risk-quadratic}
R_m(f_{\boldsymbol{\theta}})
=\mu_2-2\boldsymbol{\theta}^{\top}\boldsymbol g_{\boldsymbol{\tau}}
+\boldsymbol{\theta}^{\top}\mathbf G_{\boldsymbol{\tau}}\boldsymbol{\theta}.
\end{equation}
Let $(s_{\boldsymbol{\tau},j},\boldsymbol u_{\boldsymbol{\tau},j})$ be an
orthonormal eigenbasis of $\mathbf G_{\boldsymbol{\tau}}$, and put
$\theta_j=\boldsymbol u_{\boldsymbol{\tau},j}^{\top}\boldsymbol{\theta}$ and
$g_{\boldsymbol{\tau},j}
=\boldsymbol u_{\boldsymbol{\tau},j}^{\top}\boldsymbol g_{\boldsymbol{\tau}}$.
If $s_{\boldsymbol{\tau},j}=0$, then
$\E(\boldsymbol u_{\boldsymbol{\tau},j}^{\top}
\boldsymbol{\mathit{\Psi}}_{0,\boldsymbol{\tau}})^2=0$, so
$\boldsymbol u_{\boldsymbol{\tau},j}^{\top}
\boldsymbol{\mathit{\Psi}}_{0,\boldsymbol{\tau}}=0$ almost surely and hence
$g_{\boldsymbol{\tau},j}=0$. Therefore \eqref{eq:population-risk-quadratic}
reduces, in the eigenbasis, to
$$
\mu_2+\sum_{j:s_{\boldsymbol{\tau},j}>0}
\{s_{\boldsymbol{\tau},j}\theta_j^2
-2g_{\boldsymbol{\tau},j}\theta_j\}.
$$
Minimising each term gives
$\theta_j=g_{\boldsymbol{\tau},j}/s_{\boldsymbol{\tau},j}$ and therefore
$$
R_{\mathrm{LS}}(m,p)
=\mu_2-\sum_{j:s_{\boldsymbol{\tau},j}>0}
g_{\boldsymbol{\tau},j}^2/s_{\boldsymbol{\tau},j}.
$$
For the penalised population criterion, the coefficient in each positive
eigendirection is
$\theta_j=g_{\boldsymbol{\tau},j}/(s_{\boldsymbol{\tau},j}+\lambda)$.
Substitution into \eqref{eq:population-risk-quadratic} gives
$$
Q(m,p,\lambda)
=R_{\mathrm{LS}}(m,p)+
\sum_{j:s_{\boldsymbol{\tau},j}>0}
g_{\boldsymbol{\tau},j}^2/s_{\boldsymbol{\tau},j}
\left\{\lambda/(s_{\boldsymbol{\tau},j}+\lambda)\right\}^2,
$$
which proves \eqref{eq:population-ridge-risk-decomposition}.

\section{Proofs for the growing-grid selection results}
\label{app:growing-grid}
This appendix proves Theorems~\ref{prop:growing-grid-finite} and
\ref{thm:growing-grid}. We first control the sample feature moments
uniformly over fitting and validation periods. We then use that bound
in the normal equations and the rolling-origin prediction errors.

\subsection{Proof of Theorem~\ref{prop:growing-grid-finite}}
\label{app:growing-grid-proof}
\begin{proof}
Let
$$
\boldsymbol w_{t,T}=\begin{pmatrix}
Y_t\\ \mathbf C_T^{-1/2}\boldsymbol{\mathit\Psi}_{t,T}
\end{pmatrix},\qquad
\mathbf M_{t,T}=\boldsymbol w_{t,T}\boldsymbol w_{t,T}^{\top}
-\E(\boldsymbol w_{0,T}\boldsymbol w_{0,T}^{\top}).
$$
By \eqref{eq:leverage} and \eqref{eq:grid-moments},
$\|\|\boldsymbol w_{0,T}\|_2^2\|_q=L_T(q)$. For any consecutive
interval $I=\{a,a+1,\ldots,b\}\subseteq\{1,\ldots,T\}$, let
$|I|=b-a+1$. Since $B_{n,m}(q)$ is nondecreasing in $n$,
Lemma~\ref{lem:lag-variance} gives
\begin{equation}\label{eq:selection-block-second-moment}
\E\left\|\sum_{t=a}^b\mathbf M_{t,T}\right\|_{\mathrm F}^2
\leq |I|B_T(q)L_T(q)^2.
\end{equation}

Put $\mathbf S_{0,T}=\mathbf 0$ and
$\mathbf S_{k,T}=\sum_{t=1}^k\mathbf M_{t,T}$ for $1\leq k\leq T$.
For each $j=0,\ldots,J_T-1$, divide $\{1,\ldots,T\}$ into
consecutive blocks of length $2^j$, with a shorter final block if
necessary. Denote these blocks by $\mathcal B_j$.
For every $1\leq k\leq T$, the binary expansion of $k$ partitions
$\{1,\ldots,k\}$ into disjoint blocks $I_1,\ldots,I_s$, with
$s\leq J_T$ and at most one block of each length $2^j$,
$j=0,\ldots,J_T-1$. The Cauchy--Schwarz inequality gives
$$
\begin{aligned}
\|\mathbf S_{k,T}\|_{\mathrm F}^2
&=\left\|\sum_{i=1}^s\sum_{t\in I_i}\mathbf M_{t,T}\right\|_{\mathrm F}^2
\leq s\sum_{i=1}^s\left\|\sum_{t\in I_i}\mathbf M_{t,T}\right\|_{\mathrm F}^2\\
&\leq J_T\sum_{j=0}^{J_T-1}\sum_{I\in\mathcal B_j}
\left\|\sum_{t\in I}\mathbf M_{t,T}\right\|_{\mathrm F}^2.
\end{aligned}
$$
The blocks in each $\mathcal B_j$ partition $\{1,\ldots,T\}$, so
$\sum_{I\in\mathcal B_j}|I|=T$. Taking expectations and applying
\eqref{eq:selection-block-second-moment} yields
\begin{equation}\label{eq:dyadic-maximal-bound}
\E\max_{1\leq k\leq T}\|\mathbf S_{k,T}\|_{\mathrm F}^2
\leq J_T\sum_{j=0}^{J_T-1}\sum_{I\in\mathcal B_j}
|I|B_T(q)L_T(q)^2
=J_T^2TB_T(q)L_T(q)^2.
\end{equation}

For every consecutive interval $I=\{a,\ldots,b\}$ with
$|I|\geq\nu_T$,
$$
\left\|\frac1{|I|}\sum_{t\in I}\mathbf M_{t,T}\right\|_{\mathrm F}
\leq\frac{\|\mathbf S_{b,T}\|_{\mathrm F}
+\|\mathbf S_{a-1,T}\|_{\mathrm F}}{|I|}
\leq\frac2{\nu_T}\max_{0\leq k\leq T}
\|\mathbf S_{k,T}\|_{\mathrm F}.
$$
By Markov's inequality and \eqref{eq:dyadic-maximal-bound},
$$
\Pbb\left(\frac{2}{\nu_T}\max_{0\leq k\leq T}\|\mathbf S_{k,T}\|_{\mathrm F}>\epsilon\right)
\leq\frac{4\rho_T^2}{\epsilon^2}.
$$
Every fitting interval $\{m+1,\ldots,r\}$, for
$\boldsymbol\tau=(m,p,\lambda)\in\mathcal G_T$ and $s_T\leq r\leq T$,
and the validation interval $\{s_T+1,\ldots,T\}$ have length at least
$\nu_T$. Thus, with probability at least $1-4\rho_T^2/\epsilon^2$,
the corresponding sample moment matrices satisfy the stated bound
simultaneously for all candidates and forecasting origins.\\[0.05 in]
Fix $\boldsymbol\tau=(m,p,\lambda)\in\mathcal G_T$ and
$s_T\leq r\leq T$. The feature vector $\boldsymbol{\mathit\Psi}_{t,\boldsymbol\tau}$
uses the monomials in $\boldsymbol{\mathit\Psi}_{t,T}$ involving at
most $m$ lags and having degree at most $p$.
Let $\mathbf P_{\boldsymbol\tau}$ select those entries of
$\boldsymbol{\mathit\Psi}_{t,T}$ that appear in
$\boldsymbol{\mathit\Psi}_{t,\boldsymbol\tau}$, so that
$\boldsymbol{\mathit\Psi}_{t,\boldsymbol\tau}
=\mathbf P_{\boldsymbol\tau}\boldsymbol{\mathit\Psi}_{t,T}$.
Since $\lambda\geq a_T$,
$$
\mathbf P_{\boldsymbol\tau}\mathbf C_T\mathbf P_{\boldsymbol\tau}^{\top}
=\mathbf G_{\boldsymbol\tau}+a_T\mathbf I
\preceq\mathbf G_{\boldsymbol\tau}+\lambda\mathbf I.
$$
Consequently,
$\| (\mathbf G_{\boldsymbol\tau}+\lambda\mathbf I)^{-1/2}
\mathbf P_{\boldsymbol\tau}\mathbf C_T^{1/2}\|_{\mathrm{op}}\leq1$.
The lower-right block of
$|I|^{-1}\sum_{t\in I}\mathbf M_{t,T}$ is the matrix
$\mathbf C_T^{-1/2}(\widehat{\mathbf G}_{I,T}-\mathbf G_T)
\mathbf C_T^{-1/2}$, where $\widehat{\mathbf G}_{I,T}$ is the
sample second-moment matrix of the common feature vector over $I$.
Using $\|\mathbf A\mathbf D\mathbf B\|_{\mathrm F}
\leq\|\mathbf A\|_{\mathrm{op}}
\|\mathbf D\|_{\mathrm F}\|\mathbf B\|_{\mathrm{op}}$,
we obtain, for every candidate and fitting interval,
$$
\left\| (\mathbf G_{\boldsymbol\tau}+\lambda\mathbf I)^{-1/2}
(\widehat{\mathbf G}_{r,\boldsymbol\tau}-\mathbf G_{\boldsymbol\tau})
(\mathbf G_{\boldsymbol\tau}+\lambda\mathbf I)^{-1/2}
\right\|_{\mathrm F}\leq\epsilon.
$$
The same bound for the response--feature entries yields, for any
$\boldsymbol u,\boldsymbol v\in\Rbb^{D_{m,p}}$,
\begin{equation}\label{eq:selection-moment-inequalities}
\begin{aligned}
|\boldsymbol u^\top(\widehat{\mathbf G}_{r,\boldsymbol\tau}
-\mathbf G_{\boldsymbol\tau})\boldsymbol v|
&\leq\epsilon
\|\boldsymbol u\|_{\mathbf G_{\boldsymbol\tau}+\lambda\mathbf I}
\|\boldsymbol v\|_{\mathbf G_{\boldsymbol\tau}+\lambda\mathbf I},\\
|\boldsymbol u^\top(\widehat{\boldsymbol g}_{r,\boldsymbol\tau}
-\boldsymbol g_{\boldsymbol\tau})|
&\leq\epsilon
\|\boldsymbol u\|_{\mathbf G_{\boldsymbol\tau}+\lambda\mathbf I},
\end{aligned}
\end{equation}
where $\|\boldsymbol u\|_{\mathbf H}^2=\boldsymbol u^\top\mathbf H\boldsymbol u$
for a positive definite matrix $\mathbf H$.
Thus
$\widehat{\mathbf G}_{r,\boldsymbol\tau}+\lambda\mathbf I\succeq
(1-\epsilon)(\mathbf G_{\boldsymbol\tau}+\lambda\mathbf I)$.

The population and sample coefficient vectors are
$$
\boldsymbol\theta_{\boldsymbol\tau}
=(\mathbf G_{\boldsymbol\tau}+\lambda\mathbf I)^{-1}
\boldsymbol g_{\boldsymbol\tau},\qquad
\widehat{\boldsymbol\theta}_{r,\boldsymbol\tau}
=(\widehat{\mathbf G}_{r,\boldsymbol\tau}+\lambda\mathbf I)^{-1}
\widehat{\boldsymbol g}_{r,\boldsymbol\tau}.
$$
The population normal equation and the Cauchy--Schwarz inequality yield
$\|\boldsymbol\theta_{\boldsymbol\tau}\|_{\mathbf G_{\boldsymbol\tau}+\lambda\mathbf I}
\leq\sqrt{\mu_2}$.
Subtracting the normal equations and applying
\eqref{eq:selection-moment-inequalities} gives
\begin{equation}\label{eq:selection-coefficient-bound}
\|\widehat{\boldsymbol\theta}_{r,\boldsymbol\tau}
-\boldsymbol\theta_{\boldsymbol\tau}\|_{\mathbf G_{\boldsymbol\tau}+\lambda\mathbf I}
\leq\frac{\epsilon(1+\sqrt{\mu_2})}{1-\epsilon}=A(\epsilon).
\end{equation}
Expanding the squared population loss and using
$\boldsymbol g_{\boldsymbol\tau}
=(\mathbf G_{\boldsymbol\tau}+\lambda\mathbf I)
\boldsymbol\theta_{\boldsymbol\tau}$ shows that
$$
\begin{aligned}
&R_m(\widehat f_{r,\boldsymbol\tau})-Q(\boldsymbol\tau)\\
&=(\widehat{\boldsymbol\theta}_{r,\boldsymbol\tau}
-\boldsymbol\theta_{\boldsymbol\tau})^\top
\mathbf G_{\boldsymbol\tau}
(\widehat{\boldsymbol\theta}_{r,\boldsymbol\tau}
-\boldsymbol\theta_{\boldsymbol\tau})
-2\lambda
(\widehat{\boldsymbol\theta}_{r,\boldsymbol\tau}
-\boldsymbol\theta_{\boldsymbol\tau})^\top
\boldsymbol\theta_{\boldsymbol\tau}.
\end{aligned}
$$
By \eqref{eq:selection-coefficient-bound}, its absolute value is at most
$V(\epsilon)$. In particular, this holds for all refits at $r=T$.\\[0.05 in]
On the validation interval, the same matrix bound gives
\begin{equation}\label{eq:selection-population-score}
\begin{aligned}
\left|\frac1{N_T}\sum_{r=s_T}^{T-1}
\{Y_{r+1}-f_{\boldsymbol\tau}(\boldsymbol X_{r+1}^{(m)})\}^2
-Q(\boldsymbol\tau)\right|&\leq(1+\mu_2)\epsilon,\\
\frac1{N_T}\sum_{r=s_T}^{T-1}
\{Y_{r+1}-f_{\boldsymbol\tau}(\boldsymbol X_{r+1}^{(m)})\}^2
&\leq\mu_2+(1+\mu_2)\epsilon.
\end{aligned}
\end{equation}
By stationarity and Lemma~\ref{lem:lag-variance},
$$
\E\left[\frac1{N_T}\sum_{r=s_T}^{T-1}
\ell_T(\boldsymbol X_{r+1}^{(m_T^+)})\right]=d_T,\qquad
\operatorname{Var}\left(\frac1{N_T}\sum_{r=s_T}^{T-1}
\ell_T(\boldsymbol X_{r+1}^{(m_T^+)})\right)
\leq\frac{B_T(q)L_T(q)^2}{N_T}.
$$
As $\nu_T^2\leq TN_T$, Chebyshev's inequality shows that this
average is at most $d_T+\epsilon$ except on an event of probability
at most $\rho_T^2/\epsilon^2$. Hence all the preceding bounds hold
simultaneously with probability at least $1-5\rho_T^2/\epsilon^2$.\\[0.05 in]
By $\lambda\geq a_T$ and the inclusion of candidate monomials in
$\boldsymbol{\mathit\Psi}_{t,T}$,
$$
\boldsymbol{\mathit\Psi}_{r+1,\boldsymbol\tau}^{\top}
(\mathbf G_{\boldsymbol\tau}+\lambda\mathbf I)^{-1}
\boldsymbol{\mathit\Psi}_{r+1,\boldsymbol\tau}
\leq \ell_T(\boldsymbol X_{r+1}^{(m_T^+)}).
$$
Together with \eqref{eq:selection-coefficient-bound}, this gives
$$
|\widehat f_{r,\boldsymbol\tau}(\boldsymbol X_{r+1}^{(m)})
-f_{\boldsymbol\tau}(\boldsymbol X_{r+1}^{(m)})|^2
\leq A(\epsilon)^2\ell_T(\boldsymbol X_{r+1}^{(m_T^+)}).
$$
Expanding the squared forecasting errors and applying the
Cauchy--Schwarz inequality, together with
\eqref{eq:selection-population-score}, gives
$$
\begin{aligned}
|\operatorname{RO}_T(\boldsymbol\tau)-Q(\boldsymbol\tau)|
&\leq (1+\mu_2)\epsilon+A(\epsilon)^2(d_T+\epsilon)\\
&\quad+2A(\epsilon)\sqrt{(d_T+\epsilon)
\{\mu_2+(1+\mu_2)\epsilon\}}=U_T(\epsilon).
\end{aligned}
$$
The selected candidate minimises $\operatorname{RO}_T$, so
$$
0\leq Q(\widehat{\boldsymbol\tau}_T)
-\min_{\boldsymbol\tau\in\mathcal G_T}Q(\boldsymbol\tau)
\leq2U_T(\epsilon).
$$
Since the refitted risk of every candidate differs from $Q$ by at most
$V(\epsilon)$, the remaining assertions follow:
$$
\left|R_{\widehat m_T}(\widehat f_T^{\mathrm{RO}})
-\min_{\boldsymbol\tau\in\mathcal G_T}Q(\boldsymbol\tau)\right|
\leq2U_T(\epsilon)+V(\epsilon),
$$
$$
0\leq R_{\widehat m_T}(\widehat f_T^{\mathrm{RO}})
-\min_{\boldsymbol\tau=(m,p,\lambda)\in\mathcal G_T}
R_m(\widehat f_{T,\boldsymbol\tau})
\leq2U_T(\epsilon)+2V(\epsilon).
$$
\end{proof}

\subsection{Proof of Theorem~\ref{thm:growing-grid}}
\label{app:main-theorem-proof}
\begin{proof}
The proof of Theorem~\ref{prop:growing-grid-finite} gives, for every
$0<\epsilon<1$, the Frobenius-norm bound established immediately before
\eqref{eq:selection-moment-inequalities} and the four finite-sample
bounds, except on an event of probability at most
$5\rho_T^2/\epsilon^2$.
Specifically, the Frobenius-norm bound yields
$$
\left\|(\mathbf G_{\boldsymbol\tau}+\lambda\mathbf I)^{-1/2}
(\widehat{\mathbf G}_{r,\boldsymbol\tau}-\mathbf G_{\boldsymbol\tau})
(\mathbf G_{\boldsymbol\tau}+\lambda\mathbf I)^{-1/2}
\right\|_{\mathrm F}\leq\epsilon
$$
uniformly in the grid and in $s_T\leq r\leq T$.\\[0.05 in]
Assume $\rho_T\to0$. For each fixed $M>0$, take
$\epsilon=M\rho_T$; this lies in $(0,1)$ for sufficiently large $T$.
The probability of failure is at most $5/M^2$. As $M$ can be chosen
arbitrarily large, the matrix error is $O_p(\rho_T)$.
Also, $V(\epsilon)=O(\epsilon)$ as $\epsilon\downarrow0$, so the
uniform fitted-risk error is $O_p(\rho_T)$. This proves (i).\\[0.05 in]
Under the stronger condition $\rho_T\sqrt{1+d_T}\to0$, the score
bound in \eqref{eq:score-error-bound} satisfies, for fixed $M$,
$$
U_T(M\rho_T)=O\{\rho_T\sqrt{1+d_T}\},
$$
because the term $\rho_T^2(1+d_T)$ is of smaller order. The two
oracle bounds follow from Theorem~\ref{prop:growing-grid-finite} in
the same way. This proves (ii).\\[0.05 in]
For (iii), fix $\delta>0$. Theorem~\ref{prop:Volterrra-approx}
gives a polynomial $v_\delta\in\cH_{m_\delta,p_\delta}$ with
$$
R_{m_\delta}(v_\delta)-R(f^\star)<\delta.
$$
Set $C_\delta=\|v_\delta\|_{\cH_{m_\delta,p_\delta}}^2<\infty$.
For sufficiently large $T$, the candidate
$(m_T^\circ,p_T^\circ,\lambda_T^\circ)$ contains $v_\delta$ and
preserves its norm by Lemma~\ref{lem:volterra-embedding}.
By the definition of $f_{\boldsymbol\tau}$,
$$
0\leq\min_{\boldsymbol\tau\in\mathcal G_T}Q(\boldsymbol\tau)
-R(f^\star)
\leq\delta+\lambda_T^\circ C_\delta.
$$
Let $T\to\infty$ with $\delta$ fixed and then let
$\delta\downarrow0$. Since $\lambda_T^\circ\to0$,
$$
\min_{\boldsymbol\tau\in\mathcal G_T}Q(\boldsymbol\tau)
\longrightarrow R(f^\star).
$$
Combining this limit with the selected-population oracle bound from
(ii) proves \eqref{eq:selected-bayes-limit}.
\end{proof}

\section{Proofs of the explicit growing-grid corollaries}
\label{app:gaussian-feasible}
This appendix verifies the sufficient growth conditions used in
Corollary~\ref{cor:gaussian-feasible} and
Corollary~\ref{cor:loglog-grid}.

\subsection{Proof of Corollary~\ref{cor:gaussian-feasible}}
\begin{proof}
We first record the Gaussian polynomial moment bound used in the proof. Let
$\boldsymbol G$ be a finite-dimensional Gaussian random vector and let $H$ be
a polynomial of total degree at most $p$. Then, for every $q\geq1$,
\begin{equation}\label{eq:gaussian-hypercontractive}
\|H(\boldsymbol G)\|_{2q}
\leq(2q-1)^{p/2}\|H(\boldsymbol G)\|_2.
\end{equation}
This is the Gaussian hypercontractivity inequality for polynomials; see
\citet[Theorem~5.10 and Remark~5.11]{janson1997gaussian}. The result also
applies when the covariance matrix of $\boldsymbol G$ is singular. Indeed, if
$\boldsymbol G$ has mean $\boldsymbol\mu$ and covariance matrix of rank $s$,
then $\boldsymbol G=\boldsymbol\mu+\mathbf A\boldsymbol Z$
for some matrix $\mathbf A$ and $\boldsymbol Z\sim N_s(\boldsymbol0,\mathbf I_s)$.
Since $H(\boldsymbol\mu+\mathbf A\boldsymbol z)$ is again a polynomial of
degree at most $p$, \eqref{eq:gaussian-hypercontractive} applies.\\[0.05 in]
For the grid in Corollary~\ref{cor:gaussian-feasible},
$m_T^+=p_T^+=r_T$. Hence $\boldsymbol X_0^{(r_T)}=(Y_{-1},\ldots,Y_{-r_T})^\top$
is a Gaussian random vector. Let $\boldsymbol e_j$ denote the $j$th standard
basis vector of $\Rbb^{D_T}$ and define
$$
H_{j,T}(\boldsymbol x)
=\boldsymbol e_j^\top\mathbf C_T^{-1/2}\boldsymbol\psi_T(\boldsymbol x),
\qquad j=1,\ldots,D_T.
$$
Every component of $\boldsymbol\psi_T(\boldsymbol x)$ is a monomial of total
degree at most $r_T$. Therefore $H_{j,T}$, being a linear combination of
these components, is a polynomial of total degree at most $r_T$. Applying
\eqref{eq:gaussian-hypercontractive} gives
\begin{equation}\label{eq:normalized-coordinate-moment}
\|H_{j,T}(\boldsymbol X_0^{(r_T)})\|_{2q}^2
\leq(2q-1)^{r_T}\|H_{j,T}(\boldsymbol X_0^{(r_T)})\|_2^2.
\end{equation}

By the definition of $\ell_T$,
$$
\ell_T(\boldsymbol X_0^{(r_T)})
=\|\mathbf C_T^{-1/2}\boldsymbol\psi_T(\boldsymbol X_0^{(r_T)})\|_2^2
=\sum_{j=1}^{D_T}H_{j,T}(\boldsymbol X_0^{(r_T)})^2.
$$
Minkowski's inequality and \eqref{eq:normalized-coordinate-moment} yield
$$
\begin{aligned}
\|\ell_T(\boldsymbol X_0^{(r_T)})\|_q
&\leq\sum_{j=1}^{D_T}\|H_{j,T}(\boldsymbol X_0^{(r_T)})^2\|_q\\
&=\sum_{j=1}^{D_T}\|H_{j,T}(\boldsymbol X_0^{(r_T)})\|_{2q}^2\\
&\leq(2q-1)^{r_T}\sum_{j=1}^{D_T}\|H_{j,T}(\boldsymbol X_0^{(r_T)})\|_2^2.
\end{aligned}
$$
The last sum equals
$$
\begin{aligned}
\sum_{j=1}^{D_T}\E H_{j,T}(\boldsymbol X_0^{(r_T)})^2
&=\E\|\mathbf C_T^{-1/2}\boldsymbol\Psi_{0,T}\|_2^2\\
&=\operatorname{tr}(\mathbf C_T^{-1/2}\mathbf G_T\mathbf C_T^{-1/2})\\
&=\operatorname{tr}(\mathbf G_T\mathbf C_T^{-1})=d_T.
\end{aligned}
$$
Consequently,
\begin{equation}\label{eq:gaussian-leverage-bound}
\|\ell_T(\boldsymbol X_0^{(r_T)})\|_q\leq(2q-1)^{r_T}d_T.
\end{equation}
Applying \eqref{eq:gaussian-hypercontractive} to the degree-one polynomial
$Y_0$ gives
$$
\|Y_0^2\|_q=\|Y_0\|_{2q}^2\leq(2q-1)\mu_2.
$$
Therefore, from the definition of $L_T(q)$ and \eqref{eq:gaussian-leverage-bound},
\begin{equation}\label{eq:gaussian-LT-bound}
L_T(q)\leq(2q-1)\mu_2+(2q-1)^{r_T}d_T.
\end{equation}

We now take $q=4$. Since $D_T=D_{r_T,r_T}=\binom{2r_T}{r_T}\leq4^{r_T}$
and $d_T\leq D_T$, \eqref{eq:gaussian-LT-bound} gives
$$
L_T(4)\leq7\mu_2+7^{r_T}4^{r_T}\leq C 28^{r_T}
$$
for a constant $C>0$ independent of $T$. We use $C$ for positive
constants whose values may change between inequalities. The geometric mixing assumption implies
$$
\sum_{h=1}^{T}\alpha_Y(h)^{1/2}
\leq C_3^{1/2}\sum_{h=1}^{\infty}e^{-C_4h/2}<\infty.
$$
Hence $B_T(4)=O(r_T+1)$. Moreover,
$J_T=O(\log T)$ and $\nu_T\asymp T$
because $s_T=\lfloor\kappa T\rfloor$ and $r_T=o(T)$. Also $\sqrt{1+d_T}\leq\sqrt{1+4^{r_T}}\leq\sqrt2\,2^{r_T}$.
Using the definition of $\rho_T$, we therefore obtain
$$
\rho_T\sqrt{1+d_T}
=\frac{L_T(4)J_T\sqrt{T B_T(4)}}{\nu_T}\sqrt{1+d_T}
\leq C 56^{r_T}\sqrt{\frac{r_T+1}{T}}\log T.
$$
Since $r_T=\lfloor(\log T)/10\rfloor$, we have
$56^{r_T}\leq T^{9/20}$ and $r_T=O(\log T)$,
since $\log56<9/2$. Consequently,
$$
\rho_T\sqrt{1+d_T}=O\{T^{-1/20}(\log T)^{3/2}\}\longrightarrow0.
$$
In particular, $\rho_T\to0$, so parts (i) and (ii) of Theorem~\ref{thm:growing-grid} apply.\\[0.05 in]
Finally, applying \eqref{eq:gaussian-hypercontractive} to $Y_0$ gives, for every $j\geq1$,
$$
\|Y_0\|_{2j}\leq\sqrt{2j-1}\,\|Y_0\|_2.
$$
Hence
$$
\sum_{j=1}^{\infty}\{\E|Y_0|^{2j}\}^{-1/(2j)}
=\sum_{j=1}^{\infty}\|Y_0\|_{2j}^{-1}
\geq\frac{1}{\|Y_0\|_2}\sum_{j=1}^{\infty}(2j-1)^{-1/2}=\infty.
$$
Thus Assumption~\ref{ass:carleman} holds. Since $r_T\to\infty$ and the grid
contains $(r_T,r_T,T^{-\gamma})$, with $T^{-\gamma}\to0$, part (iii) of
Theorem~\ref{thm:growing-grid} also applies.
\end{proof}

\subsection{Proof of Corollary~\ref{cor:loglog-grid}}
\label{app:loglog-grid}
\begin{proof}
For the grid in \eqref{eq:loglog-grid-general},
$m_T^+=p_T^+=r_T$ and $a_T=T^{-\xi}c_*$, where
$c_*=\min\mathcal C>0$. We verify the conditions of
Theorem~\ref{thm:growing-grid} with $q=4$.
By stationarity, Minkowski's inequality, and the marginal moment bound,
$$
\bigl\|\|\boldsymbol X_0^{(r_T)}\|_2^2\bigr\|_{4r_T}
\leq\sum_{j=1}^{r_T}\|Y_{-j}^2\|_{4r_T}
=r_T\|Y_0\|_{8r_T}^2\leq 4C_7^2r_T^2.
$$
For $u\geq0$, $\sum_{j=0}^{r_T}u^j\leq(1+u)^{r_T}$.
Using the kernel definition \eqref{e2} and Minkowski's inequality,
$$
\begin{aligned}
\bigl\|k^{(r_T)}(\boldsymbol X_0^{(r_T)},\boldsymbol X_0^{(r_T)})\bigr\|_4
&\leq\bigl\|1+\|\boldsymbol X_0^{(r_T)}\|_2^2\bigr\|_{4r_T}^{r_T}\\
&\leq(1+4C_7^2r_T^2)^{r_T}.
\end{aligned}
$$
Since $\mathbf C_T\succeq a_T\mathbf I$,
$\ell_T(\boldsymbol x)\leq a_T^{-1}k^{(r_T)}(\boldsymbol x,\boldsymbol x)$.
Consequently,
$$
L_T(4)\leq\|Y_0\|_8^2
+\frac{T^\xi}{c_*}(1+4C_7^2r_T^2)^{r_T}.
$$
The assumption $r_T=O(\log\log T)$ implies
$r_T\log(1+4C_7^2r_T^2)=o(\log T)$, and hence
$L_T(4)\leq T^{\xi+o(1)}$.
Furthermore, by \eqref{eq:finite-dependence-sum},
$$
B_T(4)\leq1+2r_T+24c_4^2C_5^{1/2}
\sum_{h=1}^\infty e^{-C_6h/2}=O(1+r_T).
$$
The feature dimension satisfies
$d_T\leq D_T=\binom{2r_T}{r_T}\leq4^{r_T}=T^{o(1)}$.
Finally $J_T=O(\log T)$ and $\nu_T\asymp T$ because
$s_T=\lfloor\kappa T\rfloor$ and $r_T=o(T)$. Substituting these
bounds into \eqref{eq:grid-rate} gives
$$
\rho_T\sqrt{1+d_T}
=O\{T^{\xi-1/2+o(1)}\}\longrightarrow0,
$$
where the powers of $\log T$ and $r_T$ are absorbed into $T^{o(1)}$.
Thus parts (i) and (ii) of Theorem~\ref{thm:growing-grid} apply.\\[0.05 in]
The marginal moment bound also gives
$\|Y_0\|_{2j}^{-1}\geq C_7^{-1}j^{-1/2}$ for $j\geq1$
(with the convention in Assumption~\ref{ass:carleman}),
so the Carleman series diverges. Since $r_T\to\infty$ and
$cT^{-\xi}\to0$ for every fixed $c\in\mathcal C$,
$(r_T,r_T,cT^{-\xi})\in\mathcal G_T$ supplies the
sequence required in part (iii) of Theorem~\ref{thm:growing-grid}.
This proves risk consistency of the selected refit.
\end{proof}

\section{Verification of the assumptions for the stationary simulation models}
\label{app:nonlinear-benchmark}

\subsection{The models \texorpdfstring{$N_1$, $C_a$ and ESTAR}{N\_1, C\_a and ESTAR}}
Consider first a process of the form
\begin{equation}\label{eq:F-bounded-regression}
Y_t=g(Y_{t-1})+\sigma Z_t,
\end{equation}
where $\{Z_t\}$ are independent $N(0,1)$ variables, $\sigma>0$, and
$$
|g(x)|\leq b\qquad\text{for every }x\in\Rbb.
$$
Its transition density is
$$
p(x,y)=\varphi_\sigma\{y-g(x)\},
$$
where
$$
\varphi_\sigma(u)=\frac{1}{\sqrt{2\pi}\sigma}
\exp\left(-\frac{u^2}{2\sigma^2}\right).
$$
Since $|g(x)|\leq b$ and $\varphi_\sigma(u)$ decreases as $|u|$ increases,
\begin{equation}\label{eq:F-minorization-one}
p(x,y)\geq q_b(y):=\varphi_\sigma(|y|+b)
\qquad\text{for all }x,y\in\Rbb.
\end{equation}
Moreover,
$$
\epsilon_b:=\int_{\Rbb}q_b(y)\,dy=2\Phi(-b/\sigma)>0.
$$
Define the probability measure
$$
\nu_b(dy)=\epsilon_b^{-1}q_b(y)\,dy.
$$
Equation~\eqref{eq:F-minorization-one} implies
\begin{equation}\label{eq:F-mixture-one}
P(x,\cdot)=\epsilon_b\nu_b(\cdot)+(1-\epsilon_b)Q_x(\cdot)
\end{equation}
for a probability kernel $Q_x$.\\[0.05 in]
Let $\pi_1$ and $\pi_2$ be two probability measures. Since a Markov kernel
does not increase total variation distance,
$$
\begin{aligned}
\|\pi_1P-\pi_2P\|_{\mathrm{TV}}
&=(1-\epsilon_b)\|\pi_1Q-\pi_2Q\|_{\mathrm{TV}}\\
&\leq(1-\epsilon_b)\|\pi_1-\pi_2\|_{\mathrm{TV}}.
\end{aligned}
$$
The map $\mu\mapsto\mu P$ is therefore a contraction on the complete space
of probability measures under total variation distance. Hence it has a
unique fixed point $\pi_b$, and iteration gives
\begin{equation}\label{eq:F-uniform-TV-one}
\sup_x\|P^h(x,\cdot)-\pi_b\|_{\mathrm{TV}}
\leq(1-\epsilon_b)^h.
\end{equation}
For a stationary Markov chain, this uniform total-variation bound implies
\begin{equation}\label{eq:F-mixing-one}
\alpha_Y(h)\leq(1-\epsilon_b)^h;
\end{equation}
see \citet[Section~3]{bradley2005basic}. Thus the stationary process is
geometrically strongly mixing.\\[0.05 in]
Under stationarity, Minkowski's inequality gives, for every integer $j\geq1$,
\begin{equation}\label{eq:F-moment-one}
\begin{aligned}
\|Y_0\|_{2j}
&=\|g(Y_{-1})+\sigma Z_0\|_{2j}\\
&\leq b+\sigma\|Z_0\|_{2j}\\
&\leq b+\sigma\sqrt{2j}.
\end{aligned}
\end{equation}
Consequently, all moments are finite and
$$
\sum_{j=1}^{\infty}\|Y_0\|_{2j}^{-1}
\geq\sum_{j=1}^{\infty}\frac{1}{b+\sigma\sqrt{2j}}=\infty.
$$
Therefore the marginal Carleman condition holds.\\[0.05 in]
For $N_1$,
$$
g(x)=0.50\tanh(x),\qquad b=0.50.
$$
For $C_a$,
$$
g(x)=a\cos(2x),\qquad b=a.
$$
For ESTAR,
$$
g(x)=x\exp(-2x^2),\qquad b=\tfrac12 e^{-1/2},
$$
since $\sup_{x\in\Rbb}|x|\exp(-2x^2)=\tfrac12 e^{-1/2}$.
Hence all three models have unique stationary versions satisfying geometric
strong mixing and the marginal moment bound used in
Corollary~\ref{cor:loglog-grid}.

\subsection{The model \texorpdfstring{$J_a$}{J\_a}}
For
$$
J_a:\qquad Y_t=a\tanh(2Y_{t-1})\tanh(2Y_{t-2})+\sigma Z_t,
$$
define the Markov state
$$
\boldsymbol S_t=(Y_t,Y_{t-1})^\top.
$$
For every $x,z\in\Rbb$,
$$
|a\tanh(2x)\tanh(2z)|\leq a.
$$
Let
$$
q_a(y)=\varphi_\sigma(|y|+a),\qquad
\epsilon_a=\int_{\Rbb}q_a(y)\,dy=2\Phi(-a/\sigma)>0.
$$
Starting from $\boldsymbol S_0=(x,z)$, let $u=Y_1$ and $v=Y_2$. Their joint
conditional density is
$$
\begin{aligned}
p_{x,z}(u,v)
&=\varphi_\sigma\{u-a\tanh(2x)\tanh(2z)\}\\
&\qquad\times\varphi_\sigma\{v-a\tanh(2u)\tanh(2x)\}.
\end{aligned}
$$
Both conditional means belong to $[-a,a]$. Therefore
\begin{equation}\label{eq:F-two-step-minorization}
p_{x,z}(u,v)\geq q_a(u)q_a(v).
\end{equation}
Let $\nu_a^{(2)}$ be the probability measure on $\Rbb^2$ with density
$$
\frac{q_a(u)q_a(v)}{\epsilon_a^2}.
$$
Then \eqref{eq:F-two-step-minorization} gives
$$
P^2(s,\cdot)\geq\epsilon_a^2\nu_a^{(2)}(\cdot)
\qquad\text{for every }s\in\Rbb^2.
$$
Hence, for any probability measures $\mu_1,\mu_2$ on $\Rbb^2$,
\begin{equation}\label{eq:F-two-step-contraction}
\|\mu_1P^{2k}-\mu_2P^{2k}\|_{\mathrm{TV}}
\leq(1-\epsilon_a^2)^k\|\mu_1-\mu_2\|_{\mathrm{TV}}.
\end{equation}
The contraction argument used above therefore gives a unique invariant
probability measure $\pi$ for $P^2$.\\[0.05 in]
Since $\pi P^2=\pi$,
$$
(\pi P)P^2=\pi P^3=(\pi P^2)P=\pi P.
$$
Therefore $\pi P$ is also invariant for $P^2$. Uniqueness of the
$P^2$-invariant distribution gives
$$
\pi P=\pi.
$$
Thus $\pi$ is invariant for $P$. It is also the unique invariant
distribution for $P$, because every $P$-invariant distribution is
$P^2$-invariant.\\[0.05 in]
From \eqref{eq:F-two-step-contraction} and the Markov-chain mixing bound cited
above,
$$
\alpha_S(h)\leq(1-\epsilon_a^2)^{\lfloor h/2\rfloor}.
$$
Since $Y_t$ is a coordinate of $\boldsymbol S_t$,
$$
\alpha_Y(h)\leq\alpha_S(h)
\leq(1-\epsilon_a^2)^{\lfloor h/2\rfloor}.
$$
Hence the stationary $J_a$ process is geometrically strongly mixing.
Finally,
$$
|a\tanh(2Y_{-1})\tanh(2Y_{-2})|\leq a,
$$
so, exactly as in \eqref{eq:F-moment-one},
$$
\|Y_0\|_{2j}\leq a+\sigma\sqrt{2j},\qquad j\geq1.
$$
Thus all moments are finite and the marginal Carleman condition holds.\\[0.05 in]
For both $C_a$ and $J_a$, the innovation $Z_t$ is independent of the past and
has mean zero. Hence their conditional means are the regression functions
defined in Section~\ref{subsec:simulation-study}, and
$$
R(f^\star)=\E(\sigma Z_0)^2=\sigma^2.
$$

\subsection{The models \texorpdfstring{$H_1$ and $H_2$}{H\_1 and H\_2}}
For $H_1$,
$$
Y_t=\sqrt{H_t}Z_t,\qquad
H_t=\omega+0.25Y_{t-1}^2,
\qquad \omega=0.75\sigma^2>0.
$$
Suppose, for contradiction, that a stationary version satisfies
$\E|Y_0|^{12}<\infty$. Since $Z_t$ is independent of $H_t$,
$$
\E Y_0^{12}=\E Z_0^{12}\,\E H_0^6
>0.25^6\E Z_0^{12}\,\E Y_{-1}^{12}.
$$
Stationarity gives $\E Y_{-1}^{12}=\E Y_0^{12}$, whereas
$$
0.25^6\E Z_0^{12}=\frac{10395}{4096}>1.
$$
This is impossible. Therefore
\begin{equation}\label{eq:F-H1-infinite-moment}
\E|Y_0|^{12}=\infty.
\end{equation}

For $H_2$,
$$
Y_t=\sqrt{H_t}Z_t,\qquad
H_t=\omega+(0.10Z_{t-1}^2+0.85)H_{t-1},
\qquad \omega=0.05\sigma^2.
$$
Put
$$
A_t=0.10Z_{t-1}^2+0.85.
$$
Then $A_t$ is independent of $H_{t-1}$ and
$$
H_t>A_tH_{t-1}.
$$
If a stationary version had $\E H_0^6<\infty$, then
$$
\begin{aligned}
\E H_0^6=\E H_t^6
&>\E(A_t^6H_{t-1}^6)\\
&=\E A_t^6\,\E H_{t-1}^6.
\end{aligned}
$$
However,
$$
\E A_t^6=\sum_{j=0}^{6}\binom{6}{j}(0.10)^j(0.85)^{6-j}(2j-1)!!>1.
$$
This is a contradiction. Hence
$$
\E H_0^6=\infty.
$$
Since $Z_0$ is independent of $H_0$,
$$
\E|Y_0|^{12}=\E H_0^6\E Z_0^{12}=\infty.
$$

Thus the degree-six moment condition required by \eqref{eq:fixed-grid-moment}
fails for both $H_1$ and $H_2$. Indeed, the degree-six kernel diagonal
contains twelfth-order lag terms. For example,
$$
v(\boldsymbol x)=x_1^6
$$
belongs to the degree-six polynomial class and
$$
R_m(v)=\E(Y_0-Y_{-1}^6)^2.
$$
Using
$$
(a-b)^2\geq\frac12 b^2-a^2,
$$
we obtain
$$
R_m(v)\geq\frac12\E Y_{-1}^{12}-\E Y_0^2=\infty.
$$
Hence finite moments observed on a simulated finite path do not imply that
every degree-six candidate has finite population risk.

\subsection{The polynomial recursions}
The polynomial recursions in Section~\ref{subsec:simulation-study}
are evaluated over finite horizons; no stationary version is assumed.
\clearpage
\section{Simulation prediction errors}
\label{app:complete-simulation-rmse}
Table~\ref{tab:simulation-rmse-long} reports the mean RMSE over $100$
replications for the $21$ generating models in
Section~\ref{subsec:simulation-study}. Each replication is evaluated
on $500$ observations. Volterra (P) and Volterra (U) denote penalised-
and unpenalised-intercept Volterra estimation, respectively; NN denotes
nearest-neighbour regression.

\begingroup
\small
\setlength{\tabcolsep}{5pt}
\renewcommand{\arraystretch}{0.95}
\begin{longtable}{@{}lrrrrrr@{}}
\caption{Mean one-step-ahead prediction RMSE over $100$ replications,
grouped by training sample size.}\label{tab:simulation-rmse-long}\\
\toprule
Model & Volterra (P) & Volterra (U) & Linear ridge & RBF KRR & NN & ARFIMA \\
\midrule
\endfirsthead
\caption[]{Mean one-step-ahead prediction RMSE (continued).}\\
\toprule
Model & Volterra (P) & Volterra (U) & Linear ridge & RBF KRR & NN & ARFIMA \\
\midrule
\endhead
\midrule
\multicolumn{7}{r}{\footnotesize Continued on next page}\\
\endfoot
\bottomrule
\endlastfoot
\multicolumn{7}{c}{\textbf{Training sample size $T=100$}}\\*
\midrule
$L_1$ & 0.33 & 0.33 & 0.31 & 0.32 & 0.33 & 0.30 \\
$L_2$ & 0.32 & 0.32 & 0.31 & 0.32 & 0.33 & 0.31 \\
$L_3$ & 0.32 & 0.32 & 0.31 & 0.32 & 0.35 & 0.31 \\
$I_{2,2}$ & 0.33 & 0.33 & 0.31 & 0.32 & 0.32 & 0.31 \\
$I_{2,3}$ & 0.33 & 0.33 & 0.31 & 0.32 & 0.32 & 0.31 \\
$I_{3,2}$ & 0.34 & 0.34 & 0.31 & 0.32 & 0.32 & 0.31 \\
$I_{3,3}$ & 0.33 & 0.33 & 0.31 & 0.32 & 0.32 & 0.31 \\
$A_{2,2}$ & 0.31 & 0.32 & 0.31 & 0.32 & 0.32 & 0.31 \\
$A_{2,3}$ & 0.32 & 0.32 & 0.31 & 0.32 & 0.32 & 0.31 \\
$A_{3,2}$ & 0.33 & 0.33 & 0.31 & 0.33 & 0.32 & 0.31 \\
$A_{3,3}$ & 0.32 & 0.32 & 0.31 & 0.33 & 0.32 & 0.31 \\
$H_1$ & 0.35 & 0.35 & 0.30 & 0.32 & 0.31 & 0.31 \\
$H_2$ & 0.37 & 0.37 & 0.30 & 0.33 & 0.31 & 0.30 \\
$N_1$ & 0.35 & 0.34 & 0.31 & 0.32 & 0.32 & 0.31 \\
$C_{0.30}$ & 0.32 & 0.32 & 0.31 & 0.33 & 0.32 & 0.32 \\
$C_{0.60}$ & 0.38 & 0.38 & 0.36 & 0.32 & 0.33 & 0.36 \\
$C_{0.90}$ & 0.35 & 0.34 & 0.54 & 0.33 & 0.35 & 0.54 \\
$J_{0.30}$ & 0.33 & 0.33 & 0.31 & 0.32 & 0.32 & 0.32 \\
$J_{0.60}$ & 0.36 & 0.36 & 0.35 & 0.33 & 0.34 & 0.35 \\
$J_{0.90}$ & 0.41 & 0.42 & 0.46 & 0.39 & 0.40 & 0.45 \\
ESTAR & 0.33 & 0.33 & 0.32 & 0.32 & 0.33 & 0.32 \\
\midrule
\multicolumn{7}{c}{\textbf{Training sample size $T=200$}}\\*
\midrule
$L_1$ & 0.31 & 0.31 & 0.30 & 0.31 & 0.32 & 0.30 \\
$L_2$ & 0.31 & 0.31 & 0.31 & 0.31 & 0.32 & 0.31 \\
$L_3$ & 0.31 & 0.31 & 0.31 & 0.31 & 0.33 & 0.30 \\
$I_{2,2}$ & 0.33 & 0.33 & 0.30 & 0.31 & 0.31 & 0.30 \\
$I_{2,3}$ & 0.31 & 0.31 & 0.31 & 0.31 & 0.31 & 0.31 \\
$I_{3,2}$ & 0.32 & 0.32 & 0.30 & 0.31 & 0.31 & 0.30 \\
$I_{3,3}$ & 0.32 & 0.32 & 0.30 & 0.31 & 0.31 & 0.30 \\
$A_{2,2}$ & 0.31 & 0.31 & 0.31 & 0.31 & 0.31 & 0.30 \\
$A_{2,3}$ & 0.31 & 0.31 & 0.30 & 0.31 & 0.31 & 0.30 \\
$A_{3,2}$ & 0.32 & 0.32 & 0.31 & 0.31 & 0.31 & 0.31 \\
$A_{3,3}$ & 0.31 & 0.31 & 0.30 & 0.31 & 0.31 & 0.30 \\
$H_1$ & 0.31 & 0.32 & 0.30 & 0.31 & 0.31 & 0.30 \\
$H_2$ & 0.30 & 0.31 & 0.30 & 0.30 & 0.30 & 0.30 \\
$N_1$ & 0.31 & 0.31 & 0.30 & 0.31 & 0.31 & 0.30 \\
$C_{0.30}$ & 0.31 & 0.31 & 0.31 & 0.31 & 0.31 & 0.31 \\
$C_{0.60}$ & 0.32 & 0.32 & 0.35 & 0.31 & 0.32 & 0.35 \\
$C_{0.90}$ & 0.37 & 0.37 & 0.53 & 0.32 & 0.33 & 0.53 \\
$J_{0.30}$ & 0.31 & 0.31 & 0.31 & 0.31 & 0.31 & 0.31 \\
$J_{0.60}$ & 0.32 & 0.32 & 0.34 & 0.32 & 0.33 & 0.35 \\
$J_{0.90}$ & 0.40 & 0.40 & 0.41 & 0.33 & 0.34 & 0.40 \\
ESTAR & 0.32 & 0.32 & 0.32 & 0.31 & 0.31 & 0.31 \\
\midrule
\pagebreak
\multicolumn{7}{c}{\textbf{Training sample size $T=500$}}\\*
\midrule
$L_1$ & 0.30 & 0.30 & 0.30 & 0.30 & 0.31 & 0.30 \\
$L_2$ & 0.30 & 0.30 & 0.30 & 0.30 & 0.31 & 0.30 \\
$L_3$ & 0.31 & 0.31 & 0.30 & 0.30 & 0.32 & 0.30 \\
$I_{2,2}$ & 0.30 & 0.30 & 0.30 & 0.30 & 0.30 & 0.30 \\
$I_{2,3}$ & 0.30 & 0.30 & 0.30 & 0.30 & 0.31 & 0.30 \\
$I_{3,2}$ & 0.30 & 0.30 & 0.30 & 0.30 & 0.31 & 0.30 \\
$I_{3,3}$ & 0.30 & 0.30 & 0.30 & 0.30 & 0.31 & 0.30 \\
$A_{2,2}$ & 0.30 & 0.30 & 0.30 & 0.30 & 0.31 & 0.30 \\
$A_{2,3}$ & 0.30 & 0.30 & 0.30 & 0.30 & 0.31 & 0.30 \\
$A_{3,2}$ & 0.30 & 0.30 & 0.30 & 0.30 & 0.31 & 0.30 \\
$A_{3,3}$ & 0.30 & 0.30 & 0.30 & 0.30 & 0.31 & 0.30 \\
$H_1$ & 0.31 & 0.31 & 0.30 & 0.31 & 0.30 & 0.30 \\
$H_2$ & 0.31 & 0.31 & 0.30 & 0.30 & 0.30 & 0.30 \\
$N_1$ & 0.30 & 0.30 & 0.30 & 0.30 & 0.31 & 0.30 \\
$C_{0.30}$ & 0.31 & 0.31 & 0.31 & 0.31 & 0.31 & 0.31 \\
$C_{0.60}$ & 0.31 & 0.31 & 0.35 & 0.31 & 0.31 & 0.35 \\
$C_{0.90}$ & 0.30 & 0.30 & 0.53 & 0.30 & 0.31 & 0.53 \\
$J_{0.30}$ & 0.31 & 0.31 & 0.31 & 0.31 & 0.31 & 0.31 \\
$J_{0.60}$ & 0.32 & 0.32 & 0.34 & 0.31 & 0.31 & 0.34 \\
$J_{0.90}$ & 0.34 & 0.34 & 0.40 & 0.31 & 0.32 & 0.40 \\
ESTAR & 0.31 & 0.31 & 0.31 & 0.30 & 0.31 & 0.31 \\
\end{longtable}
\endgroup
\clearpage
\bibliography{myendref}
\end{document}